\documentclass[twocolumn,showpacs,preprintnumbers,nofootinbib,prd,superscriptaddress,groupedaddress,10pt]{revtex4-1}

\usepackage{graphicx,amssymb,amsmath,amsthm,amsfonts,epsfig}
\usepackage[linktocpage]{hyperref}
\usepackage[usenames]{color}
\usepackage{epstopdf}
\usepackage{bm}
\usepackage{dcolumn}
\usepackage[latin1]{inputenc}
\usepackage{latexsym}
\usepackage{rotating}
\usepackage{hyperref}
\usepackage{color}
\usepackage{longtable}
\usepackage{enumerate}
\usepackage{mathtools}
\usepackage{url}
\def\non{\nonumber}
\def\p{\partial}
\def\na{\nabla}

\def\dif{\textrm{d}}

\renewcommand{\vec}[1]{\boldsymbol{#1}}
\def\kwave{\mathfrak{p}}

\newcommand{\ben}{\begin{enumerate}}
\newcommand{\een}{\end{enumerate}}

\def\be{\begin{equation}}
\def\ee{\end{equation}}
\def\bea{\begin{eqnarray}}
\def\eea{\end{eqnarray}}
\newcommand{\beq}{\begin{eqnarray}}
\newcommand{\eeq}{\end{eqnarray}} 
\newcommand{\ba}{\begin{align}}
\newcommand{\ea}{\end{align}}

\def\F{\mathcal{F}}

\def\eps{\epsilon}

\def\si{\sigma}

\def\ba{\bar{a}}

\definecolor{darkorange}{rgb}{1,0.549,0}

\begin{document}
{\hfill KCL-PH-TH/2018-65}

\title{
Axionic instabilities and new black hole solutions
}

\author{
Mateja Bo\v{s}kovi\'{c}$^{1,2}$,
Richard Brito$^{3}$,
Vitor Cardoso$^{1,4}$,
Taishi Ikeda$^{1}$,
Helvi Witek$^{5}$
}
\affiliation{${^1}$ CENTRA, Departamento de F\'{\i}sica, Instituto Superior T\'ecnico -- IST, Universidade de Lisboa -- UL,
Avenida Rovisco Pais 1, 1049 Lisboa, Portugal}
\affiliation{${^2}$ Faculty of Physics, University of Belgrade, Studentski trg 12, 11000 Beograd, Serbia}
\affiliation{${^3}$ Max Planck Institute for Gravitational Physics (Albert Einstein Institute), Am M\"{u}hlenberg 1, Potsdam-Golm, 14476, Germany}
\affiliation{${^4}$ Theoretical Physics Department, CERN 1 Esplanade des Particules, Geneva 23, CH-1211, Switzerland}
\affiliation{${^5}$ Department of Physics, King's College London, Strand, London, WC2R 
2LS, United Kingdom}

\begin{abstract}
The coupling between scalar and vector fields has a long and interesting history. Axions are one key possibility to solve the strong CP problem and axion-like particles
could be one solution to the dark matter puzzle. Extensive experimental and observational efforts are actively looking for ``axionic'' imprints. Given the nature of the coupling, 
and the universality of free fall, nontrivial important effects are expected in regions where gravity is strong. Rotating black holes (immersed, or not in magnetic fields) are a prime example of such regions. Here, we show that

\noindent {\bf (i)} A background electromagnetic field induces an axionic instability in flat space, for electric fields above a certain threshold value. Conversely, a homogeneous
harmonic axion field induces an instability in the Maxwell sector. When carried over to curved spacetime, this phenomena translates into generic instabilities of charged black holes.
We describe the instability and its likely final state, new black hole solutions.

\noindent {\bf (ii)} In the presence of charge, black hole uniqueness results are lost. We find solutions which are small deformations of the Kerr-Newman geometry {\it and}
hairy stationary solutions without angular momentum but which are ``dragged'' by the axion. Axion fields must exist around spinning black holes if these are immersed in external magnetic fields. The axion profile can be obtained perturbatively from the electro-vacuum solution derived by Wald.

\noindent {\bf (iii)} Ultralight axions trigger superradiant instabilities of spinning black holes and form an axionic cloud in the exterior geometry. The superradiant growth
can be interrupted or suppressed through couplings such as ${\bf E}\cdot {\bf B}$ (typical axionic coupling) but also more generic terms such as direct couplings to the invariant ${\bf E}^2- {\bf B}^2$. These couplings lead to periodic {\it bursts} of light, which occur throughout the history of energy extraction from the black hole. We provide numerical and simple analytical estimates for the rates of these processes.

\noindent {\bf (iv)} Finally, we discuss how plasma effects can affect the evolution of superradiant instabilities.

\end{abstract}

\maketitle

\tableofcontents

\section{Introduction}

Peccei and Quinn first introduced the axion, a pseudo-Goldstone scalar field, in order to solve the strong CP problem~\cite{Peccei:1977hh}. More recently it was shown that a plenitude of ultralight axion-like bosons might arise from moduli compactification in string theory. In this ``axiverse'' scenario, a landscape of light axion-like fields can populate a mass range down to the Hubble scale $m_H \sim 10^{-33}$ eV,
with possible implications at astrophysical and cosmological scales~\cite{Arvanitaki:2009fg}. In particular, axions and axion-like particles are also strong candidates for both cold and non-cold dark matter~\cite{Bergstrom:2009ib,Fairbairn:2014zta, Marsh:2015xka,Hui:2016ltb}.

The photon-axion mixing in the presence of an external magnetic or electric field, can be used to impose strong constraints on axion-like particles due to intergalactic magnetic fields~(see e.g. Ref.\cite{Mirizzi:2006zy} for a review) or can lead to a detectable signature in the spectra of high-energy gamma ray sources~\cite{Hooper:2007bq}. In addition, even in the absence of 
electromagnetic (EM) fields, superradiant instabilities can be triggered, through which the axion grows and ``condensates'' around massive, spinning black holes (BHs)~\cite{Detweiler:1980uk,Cardoso:2005vk,Dolan:2007mj,East:2017ovw,East:2018glu,Brito:2015oca}. 
The instability extracts rotational energy away from the spinning BH and deposits it into an axion cloud with high occupation number~\cite{Brito:2015oca}. Eventually, gravitational-wave (GW) emission dominates over the superradiant growth, leading to a secular spin-down and decay of the cloud. Such systems are a promising source of GWs that can be detected with current and future detectors~\cite{Arvanitaki:2010sy,Arvanitaki:2014wva,Brito:2014wla,Arvanitaki:2016qwi,Baryakhtar:2017ngi,Brito:2017wnc,Brito:2017zvb,Hannuksela:2018izj}.

In an astrophysically relevant situation, BHs are often surrounded by a plasma in an accretion disk, which generates its own EM field. In addition, galactic magnetic fields and background EM radiation is present. The presence of magnetic fields in regions where gravity is strong may give rise to new phenomena, for example the triggering of instabilities or the induction of nontrivial BH hair. In the context of axions or axion-like particles, such scenarios have hardly been considered. One of the purposes of this work is to understand possible new BH configurations in the context of axionic physics.

It has also been argued recently that the coupling of superradiant axion clouds with photons can lead to bursts of radiation which in the quantum version resemble laser-like emission~\cite{Rosa:2017ury,Sen:2018cjt}. Thus, the evolution of superradiant instabilities would produce periodic emission of light. These arguments are order-of-magnitude, highly approximate and partially inconsistent, but have very recently been put on a firmer ground through the full numerical solution of the relevant equations~\cite{Ikeda:2018nhb}. More generally, the study of axion electrodynamics in curved spacetime has been the topic of some but few studies, with some results in the Schwarzschild background in the context of Pulsar magnetospheres \cite{Garbrecht:2018akc} and polarization of EM waves passing through the scalar clouds around BHs~\cite{Plascencia:2017kca}. The other purpose of this work is to explore thoroughly the possible instabilities associated to axionic or other scalar-type couplings to the Maxwell sector occuring in some dark matter models.

In summary, we find that the presence of electric charge or rotation leads to the appearance of new BH solutions with nontrivial axion hair. Axionic of scalar-type couplings to the Maxwell sector are also found to trigger strong instabilities that affect specially superradiant clouds around BHs.

We adopt geometric units ($G=c=1$) throughout, and a ``mostly plus'' signature. 

\section{Setup} \label{sec:setup}
There are many possible and viable DM candidates~\cite{Klasen:2015uma}. We will focus our attention
on what are perhaps the best motivated extensions of the Standard Model and of General Relativity, which include
a massive (and real) scalar $\Psi$ with possible axionic couplings to a vector (through the coupling constant $k_{\rm a}$) and scalar couplings to the Maxwell invariant through a coupling constant $k_{\rm s}$,
\beq
\label{eq:MFaction}
{\cal L}&=&\frac{R}{k}- \frac{1}{4} F^{\mu\nu} F_{\mu\nu} - \frac{1}{2} g^{\mu\nu} \p_{\mu}\Psi\p_{\nu} \Psi-\frac{\mu^{2}_{\rm S}}{2} \Psi\Psi \nonumber \\
&-& \frac{k_{\rm a}}{2} \Psi \,^{\ast}F^{\mu\nu} F_{\mu\nu}-\frac{\left(k_{\rm s}\Psi\right)^p}{4} F^{\mu\nu} F_{\mu\nu} \,,
\eeq
with $p=1,2$ being popular choices~\cite{Stadnik:2015kpa,Olive:2007aj}. Thus, depending on the parity transformation of the (pseudo)scalar, coupling to the Maxwell sector is realized through ${\bf E}\cdot {\bf B}$ (pseudo-scalar) or ${\bf E}^2- {\bf B}^2$ (scalar) invariant. We will study these couplings separately. Other couplings, such as $\nabla_\mu\Psi\nabla_\nu\Psi F^{\mu\nu}$ and $\nabla_\mu\Psi\nabla_\nu\Psi A^{\mu}A^\nu$
are possible, but we will not consider these here.
The mass of the scalar $\Psi$ is given by $m_S = \mu_S \hbar$, $F_{\mu\nu} \equiv
\na_{\mu}A_{\nu} - \na_{\nu} A_{\mu}$ is the Maxwell tensor and
$\,^{\ast}F^{\mu\nu} \equiv \frac{1}{2}\eps^{\mu\nu\rho\si}F_{\rho\si}$
is its dual. We use the definition $\epsilon^{\mu\nu\rho\si}\equiv
\frac{1}{\sqrt{-g}}E^{\mu\nu\rho\si}$ where $E^{\mu\nu\rho\si}$ is
the totally anti-symmetric Levi-Civita symbol with $E^{0123}=1$.
The quantities $k_{\rm a}, k_{\rm s}$ are constants.
We get the following equations of motion for the theory above:
%
\begin{subequations}
\label{eq:MFEoMgen}
\begin{align}
\label{eq:MFEoMScalar}
&\left(\nabla^{\mu}\nabla_{\mu} - \mu^{2}_{\rm S} \right) \Psi =\frac{k_{\rm a}}{2} \,^{\ast}F^{\mu\nu} F_{\mu\nu}+\frac{p\,k_{\rm s}^p\Psi^{p-1}}{4}F^{\mu\nu} F_{\mu\nu}\,,\\
\label{eq:MFEoMVector}
&\nabla^{\nu} \left(1+k_{\rm s}^p\Psi^p\right)F_{\mu\nu} = - 2 k_{\rm a} \,^{\ast}F_{\mu\nu} \nabla^{\nu} \Psi\,,\\
\label{eq:MFEoMTensor}
&\frac{1}{k} \left( R_{\mu\nu} - \frac{1}{2} g_{\mu\nu} R \right) = \frac{1}{2}\left(1+k_{\rm s}^p\Psi^p\right) F_{\mu}{}^{\rho} F_{\nu\rho}\nonumber\\
&- \frac{1}{8} g_{\mu\nu}\left(1+k_{\rm s}^p\Psi^p\right) F^{\rho\sigma} F_{\rho\sigma}
+ \frac{1}{2} \nabla_{\mu}\Psi \nabla_{\nu} \Psi \nonumber\\
&- \frac{1}{4} g_{\mu\nu} \left( \nabla^{\rho} \Psi \nabla_{\rho}\Psi + \mu^{2}_{\rm S}\Psi\Psi \right) - \frac{k_{\rm a}}{4} \Psi g_{\mu\nu} \,^{\ast} F^{\rho\sigma} F_{\rho\sigma}\,.
\end{align}
\end{subequations}
%
If $\Psi$ is the QCD axion,
\be
\frac{\sqrt{\hbar}}{k_{\rm a}}\sim 10^{15}\left(\frac{10^{-5}\,{\rm eV}}{\mu_S}\right)\,{\rm GeV}\,.
\ee
The inverse energy parameter $k_{\rm s}$ is tightly constrained for $p=1$ but much less so for 
$p\geq2$~\cite{Stadnik:2015kpa}. Astrophysical BHs can probe scalar fields with masses in the range  $\sim[10^{-20},10^{-11}]$ eV~\cite{Brito:2017zvb}, but larger scalar field masses can in principle also be probed by hypothetical sub-solar mass primordial BHs. Depending on the formation scenario, the QCD axion with masses in the range $\sim [10^{-12},10^{-2}]$ eV is a strong dark matter candidate (see e.g.~\cite{Ringwald:2016yge}). Current experiments are especially sensitive to this mass range but most of the relevant parameter space is  not yet ruled out (see e.g. Fig. 1 in~\cite{Cardoso:2018tly}). For masses below $~\lesssim 10^{-10}$ eV and arbitrary coupling constants the parameter space remains largely unconstrained. This range is especially interesting for stringy ultralight axions which could also explain part or all of the dark matter for axions with masses down to $\sim 10^{-23}$ eV~\cite{Marsh:2015xka}. Here, we consider arbitrary coupling constants to keep the discussion as general as possible.

For all practical purposes, the right-hand-side of Eq.~\eqref{eq:MFEoMTensor} can be set to zero when the axionic coupling constant and the BH charge are is small: In natural units, the strength of a magnetic field around a source of mass $M$ can be measured defining the characteristic magnetic field $B_M=1/M$ associated to a spacetime curvature of the same order of the horizon curvature. In physical units this is given by $B_M\sim 2.4\times 10^{19} \left(M_{\odot}/M\right) {\rm Gauss}$.
We will use this approximation when looking for new BH solutions and also when performing dynamical simulations of superradiant clouds.
Although our results are only valid when $B/B_M\lesssim 1$ (when backreaction on the metric is small), this limit includes the most interesting region of the parameter space. Indeed, for astrophysical BHs, a reference value for the largest magnetic field that can be supported in an accretion disk is given by $B\sim 4\times 10^8 \left(M/M_\odot\right)^{-1/2}{\rm Gauss}$~\cite{ReesAGN} so that the approximation $B\ll B_M$ is well justified. We also neglect the backreaction of the scalar field onto the geometry, an approximation which is justified both perturbatively and numerically~\cite{Brito:2014wla,Brito:2017zvb}.

\section{Flat-space instabilities}\label{Sec.Flat-space instabilities}
The theory above shows interesting aspects even in flat space. In fact, most of the strong-field effects that we will deal with could have been guessed from a flat-space analysis. We find that non-vanishing background EM or axion fields both may trigger instabilities, but the nature and details of such instability depends on which background field one discusses: a static background electric field gives rise to instabilities of flat-space. However, to have a similar effect for background axions, one needs a time-varying
axion (or scalar).
\subsection{Homogeneous background EM field}
We start by exploring EM and scalar or axion fluctuations, determined by Eqs.~\eqref{eq:MFEoMScalar} and~\eqref{eq:MFEoMVector},
in the background of a homogeneous EM field.

\subsubsection{Axionic couplings}
Let us turn off the scalar coupling $k_{\rm s}$ for now. A vanishing scalar field and constant background electric (in standard cartesian coordinates) $\vec{E}=(0,0,E_{z})$ and magnetic field $\vec{B}=(0,B_y,0)$ solves the equations of motion.
Let's perturb this background solution by allowing a small but nonvanishing field $\Psi\sim \epsilon \eta\,e^{-i\left(\omega t- \kwave_i x^i\right)}$ and fluctuations in the vector field,
\be
\label{eq:AnsatzAmuFlatHomogeneousEM}
A_{\mu}=\left(-z\,E_z,0,0,B_y\,x\right)+\epsilon e^{-i\left(\omega t-\kwave_ix^i\right)}X_\mu\,,
\ee
where $X_\mu,\,\mu=0,\ldots,3$ are constants and $\epsilon$ is a small book-keeping parameter. The Klein-Gordon equation yields
\beq
&&2ik_{\rm a} \kwave_y\left(X_0B_y-X_1E_z\right)+2ik_{\rm a}X_2\left(\kwave_x E_z+\omega B_y\right)\nonumber\\
&+&\eta\left(\kwave^2+\mu_S^2-\omega^2\right)=0\,,\label{axion_dispersion_KG}
\eeq
where $\kwave^2\equiv \kwave_i \kwave^i$. Maxwell's equations can be used to obtain $X_0,X_1,X_2$,
\beq
X_0&=&-\frac{2ik_{\rm a}B_y \eta\kwave_y\kwave_z+X_3\omega\left(\kwave^2-\omega^2\right)}{\kwave_z\left(\kwave^2-\omega^2\right)}\,,\\
X_1&=&-\frac{2ik_{\rm a}E_z \eta\kwave_y\kwave_z-X_3 \kwave_x\left(\kwave^2-\omega^2\right)}{\kwave_z\left(\kwave^2-\omega^2\right)}\,,\\
X_2&=&\frac{2ik_{\rm a}E_z \eta\kwave_x\kwave_z+X_3\kwave_y\left(\kwave^2-\omega^2\right)}{\kwave_z\left(\kwave^2-\omega^2\right)}\,.
\eeq
Finally, substituting back in \eqref{axion_dispersion_KG} one finds the dispersion relation
\beq
&&(\kwave^2-\omega^2)(\kwave^2+\mu_S^2-\omega^2)+4k_{\rm a}^2B_y^2(\kwave_y^2-\omega^2)\nonumber\\
&-&8k_{\rm a}^2B_yE_z \kwave_x\omega-4k_{\rm a}^2E_z^2(\kwave_x^2+\kwave_y^2)=0\,.\label{flat_dispersion}
\eeq
This equation can be solved for $\omega$, with the result that an instability appears at a threshold value of the electric field.
For example, for $B_y=0, \kwave_z=0$ one gets the dispersion relation
\be
2\omega^2=2\kwave^2+\mu_S^2\pm\sqrt{\mu_S^4+16k_{\rm a}^2E_z^2\kwave^2}\,.
\ee
Thus, for $E>E_{\rm crit}$ an instability (i.e., negative $\omega^2$, such that $\omega$ has a positive imaginary component and 
the fluctuations diverge exponentially in time) appears, with
\be
E_{\rm crit}=\frac{\sqrt{\kwave^2+\mu_S^2}}{2k_{\rm a}}\,.\label{thresh_flat_axion}
\ee
The instability requires a non-homogeneous fluctuation in the axion, something that will prove important in the discussion of BH instabilities. These results are in agreement with recent studies on axion-like phenomena in magnetic materials~\cite{Ooguri:2011aa}, and predict acritical electric field scaling like the inverse of the coupling constant $k_{\rm a}$.

\subsubsection{Scalar couplings}
Let us now focus on the scalar coupling only, i.e., we set $k_{\rm a}=0, k_{\rm s}\neq0$. 
We take the same ansatz for the scalar field and vector potential given by Eq.~\eqref{eq:AnsatzAmuFlatHomogeneousEM}.

For the scalar coupling with $p=1$, a background EM field also requires a background scalar. If $B_y=E_z$, a background scalar can be avoided and we find (turning off $k_{\rm a}$) that the Klein-Gordon
and Maxwell equations yield the dispersion relation
\be
\omega^2=\kwave^2-k_{\rm s}E_z(\omega+\kwave_x)\,.
\ee
Thus, instabilities can indeed be triggered by background EM fields.

For $p=2$ we find
\beq
&&k_{\rm s}^2\left(B_y^2-E_z^2\right)+\kwave^2+\mu_S^2-\omega^2=0\,,\label{scalar_dispersion_KG}\\
&&X_0=-\frac{X_3 \omega}{\kwave_z}\,,\quad X_1=\frac{X_3 \kwave_x}{\kwave_z}\,,\quad X_2=\frac{X_3\kwave_y}{\kwave_z}\,.
\eeq
The dispersion relation for the scalar indicates that instabilities exist generically. Note, that only $k_{\rm s}^{2}$ enters the dispersion relation, and its sign is not fixed a priori. Depending on the sign of $k_{\rm s}^2$ there are indeed instabilities triggered by either the electric or the magnetic sector.

\subsection{Homogeneous background axion field}\label{sec:mink_axion}

The same analysis as above immediately shows that a constant background axion or scalar $\Psi$ (and vanishing background EM field), triggers no instability. We thus turn to time-dependent background scalars. From the structure of the Klein-Gordon equation, the configurations of interest (dark halos, boson stars, superradiant clouds, etc) are indeed time dependent. The time dependence is harmonic and, in the non-relativistic approximation, set by the boson mass, $\Psi \sim e^{\pm i \mu_{\rm S} t}$~\cite{Brito:2015oca,Khmelnitsky:2013lxt,Boskovic:2018rub}.


\subsubsection{The instability: numerical results}

\begin{figure}[htb]
\begin{center}
\begin{tabular}{c}
\includegraphics[width=8.5cm]{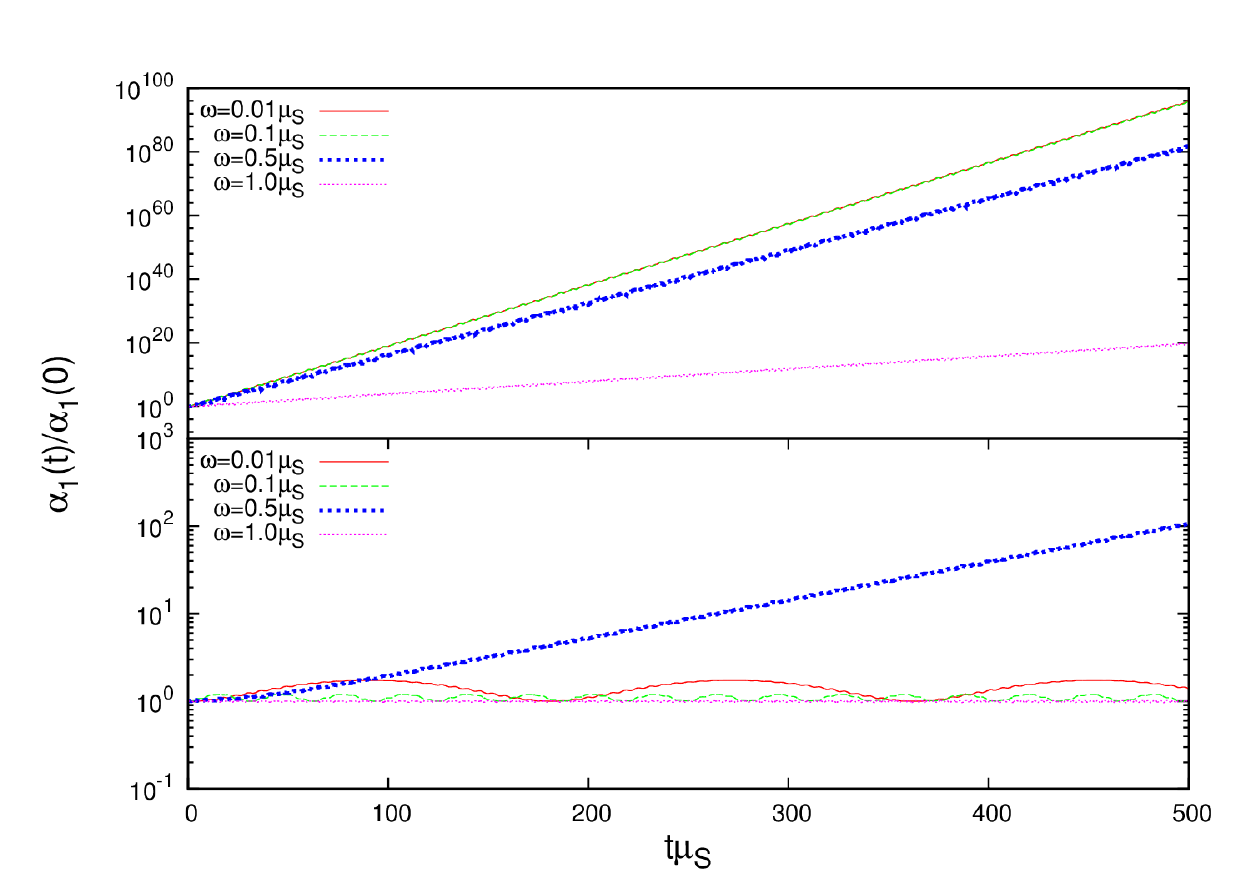}
\end{tabular}
\caption{
Time evolution of $|\alpha_{1}(\vec{\kwave},t)|$ rescaled by its initial value,
for $k_{\rm a}\kwave\Psi_0=0.5 \mu_{\rm S}$ (upper panel) and $k_{\rm a}\kwave\Psi_0=0.01 \mu_{\rm S}$ (lower panel),
and for selected frequencies. Notice that the $\kwave=\mu_{\rm S}/2$ is generically excited and dominates the evolution at low couplings. Similar results hold for $\alpha_2$.
\label{time_evolution_kaxion}}
\end{center}
\end{figure}
Consider therefore a uniform, coherent oscillating background axion state described by (this analysis closely parallels the discussion in Ref.~\cite{Sen:2018cjt})
\begin{equation}
\Psi=\frac{1}{2}\left(\Psi_{0}e^{-i\mu_{\rm S}t}+\Psi_{0}^{*}e^{i \mu_{\rm S}t}\right)\,,
\label{eq.uniform coherent oscillating axion}
\end{equation}
where $\Psi_{0}$ (which we consider to be purely real, $\text{Im}\{ \Psi_0 \}=0$) determines the amplitude of the axion's oscillations, and complex conjugation is denoted with a $*$. Maxwell equations (we set $k_{\rm s}=0$ here)
\begin{eqnarray}
\nabla^{\nu}F_{\mu\nu}=-2k_{\rm a}~^{\ast}F_{\mu\nu}\nabla^{\nu}\Psi\,,
\end{eqnarray}
can be analyzed using the following ansatz,
\begin{eqnarray}
A_{\mu}(\vec{x},t)=\sum_{\vec{\kwave}} \alpha_{\mu}e^{i(\vec{\kwave}\cdot\vec{x}-\omega t)}\,,
\label{eq.expansion of A}
\end{eqnarray}
where $\omega=|\vec{\kwave}|\equiv \kwave$ and $\left(\alpha_{\mu}, \alpha_{\mu}^{*}\right)=\left(\alpha_{\mu}(\kwave,t),\alpha^{*}_\mu(\kwave,t)\right)$. When the coupling $k_{\rm a}=0$, the solutions of Maxwell equations are a sum of plane waves, and $\alpha_{\mu}$ is time-independent. There is no instability in such cases. It is the coupling to the axion or scalar that causes a possible time dependence for $\alpha_{\mu}$. We note that there is $U(1)$ symmetry gauge redundancy. To fix such gauge degree of freedom, we impose the Lorenz gauge condition
\begin{eqnarray}
\partial_{\mu}A^{\mu}=0\,.\label{Lorenz gauge}
\end{eqnarray}
Under this gauge condition, Maxwell equations are (latin letters stand for spatial indices)
\begin{eqnarray}
\ddot{\alpha}_{i}-2i\omega \dot{\alpha}_{i}-\mu_{\rm S}k_{\rm a}\epsilon_{ikl}\kwave_{k}\alpha_{l}\left(\Psi_{0}e^{-i\mu_{\rm S}t}-\Psi_{0}^{*}e^{i\mu_{\rm S}t}\right)=0\,,\nonumber \label{eq:ax_eom_lorenz}
\end{eqnarray}
where dots stand for derivatives with respect to time, and $\epsilon_{ijk}$ is the totally anti-symmetric Levi-Civita 
symbol with $\epsilon_{xyz}=1$. Let us denote the basis orthogonal to $\vec{\kwave}$ by $\vec{e}_{(1)}$ and $\vec{e}_{(2)}$.
Then, the transverse components $\alpha_{(1)}(t,\vec{\kwave})$ and $\alpha_{(2)}(t,\vec{\kwave})$ obey the following system,
\begin{align}
&&\ddot{\alpha}_{(1)}-2i\omega\dot{\alpha}_{(1)}-\mu_{\rm S}k_{\rm a}\kwave\alpha_{(2)}\left(\Psi_{0}e^{-i\mu_{\rm S}t}-\Psi_{0}^{*}e^{i\mu_{\rm S}t}\right)=0\,,\nonumber\\
&&\ddot{\alpha}_{(2)}-2i\omega\dot{\alpha}_{(2)}-\mu_{\rm S}k_{\rm a}\kwave\alpha_{(1)}\left(\Psi_{0}e^{-i\mu_{\rm S}t}-\Psi_{0}^{*}e^{i\mu_{\rm S}t}
\right)=0\,.\label{coupled1}
\end{align}

\begin{figure}[htb]
\begin{tabular}{c}
\includegraphics[width=8.5cm]{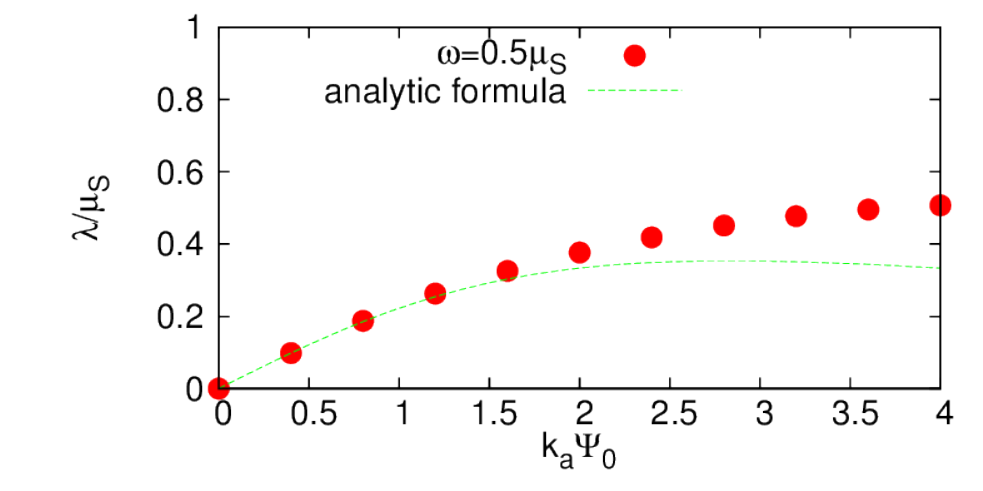}
\end{tabular}
\caption{Relation between the growth rate $\lambda/\mu_{\rm S}$ and coupling $k_{\rm a}\Psi_0$ for the $\omega=0.5\mu_{\rm S}$ mode.
Red dots are numerical values, the dashed green line is the prediction from a small-coupling expansion, described in Appendix \ref{app:RG_time_scales_2nd_ax}, and which extends the lower order prediction in the main text.
\label{relation_kaxion_lambda}}
\end{figure}

Fig.~\ref{time_evolution_kaxion} shows the time evolution of the strength $|\alpha_{(1)}|$ of the vector potential for the following initial data,
\begin{eqnarray}
\alpha_{(I)}(t=0)=\epsilon,~~\dot{\alpha}_{(I)}(t=0)=0\,,~~(I=1,2)\label{init_data_Minkowski}
\end{eqnarray}
We consider both small and moderate coupling $k_{\rm a}\Psi_0$.
For small couplings, our numerical results show that there is an instability, $\alpha_{(I)}\sim e^{\lambda t}$ whose rate peaks at $\omega \sim \mu_S/2$, and for which (see Fig. \ref{relation_kaxion_lambda})
\be
\lambda\sim \frac{1}{2}\mu_S k_{\rm a}\Psi_0\,.\label{rate_axion0}
\ee
These results are compatible with previous predictions~\cite{Sen:2018cjt}, and are not qualitatively affected by the initial conditions.
On the other hand, for strong couplings, other (unstable) modes may also be excited.
%
\subsubsection{Connection to the Mathieu equation\label{sec:Mathieu_1}}

Let us go back to Eq. \eqref{eq:ax_eom_lorenz} and introduce the ansatz
\be
\label{eq:deffcty}
y_{i}=e^{-i\omega t}\alpha_{i}\,.
\ee
Now, we project the equations to a circular $\bm{e}_{\pm}$, instead of a linear polarization basis, with $i\vec{\kwave}\times\bm{e}_{\pm}=\kwave\bm{e}_{\pm}$ and $\bm{y}=y_{\omega}\bm{e}_{\pm}$.

With a trivial time translation and re-scaling to a dimensionless time coordinate $T=\mu_S t$, the wavefunction $y_\omega$ obeys the equation\footnote{We should note that Eq.~\eqref{eq:Mathieu0} can be obtained directly from a circular polarization ansatz in the Coulomb, instead of the Lorenz gauge. This is shown in Appendix~\ref{app:Coulomb_axion}.}
\begin{equation} \label{eq:Mathieu0}
\partial^2_T y_{\omega}+\Big(\frac{\omega^2}{\mu_S^2}-2\Psi_0 k_{\rm a}\frac{\kwave}{\mu_S} \cos{T} \Big)y_{\omega}=0\,.
\end{equation}
In other words, we find that our problem is completely reduced to the well-known Mathieu equation! Thus, the complexity of having to deal with coupled (albeit ordinary) equations in the Lorenz gauge is no longer present. Mathieu equation predicts instabilities whenever $\omega^2/\mu_S^2=n^2/4, n \in \mathbb{N}$ and $\Psi_0 k_{\rm a} \kwave \neq 0$~\cite{benderbook} or in terms of the relation between wavenumber and axion frequency,
\begin{equation}\label{eq:hom_k_instable}
\omega_\ast=\{\mu_S/2,\mu_S, 3\mu_S/2,... \}\,,
\end{equation}
in agreement with the previous numerical results (see plot) for small values of the coupling.
The Fourier-transformed vector potentials with wave numbers in the sequence \eqref{eq:hom_k_instable} will dominate the other ones. One expects (based on the properties of Mathieu functions) that the first few of these modes will dominate and that -- for small couplings -- such instabilities will be significantly pronounced~\cite{arnold_ODE_book}. Perturbative investigation of unstable solutions demonstrate that the dominant rate of the instability is given by $\lambda_\ast/\mu_S=|\Psi_0 k_{\rm a}(\omega_{\ast}/\mu_S)|$ (see Appendix \ref{app:RG_time_scales} for derivation and extension to higher orders), which for our problem reduces to Eq.~\eqref{rate_axion0}, in accordance with previous work~\cite{Tkachev:2014dpa,Sen:2018cjt, Hertzberg:2018zte} and our numerical results. We thus recover in a considerably different way the same scaling for the instability rates of the coupled axion-EM system. In summary, a homogenous background of axions is an unstable configuration with a growth rate that scales linearly with the axion strength, for small couplings. We will find below in Section~\ref{sec:Time_domain_studies} that this instability has a curved spacetime analog, even when the axion field is strongly inhomogeneous, and has the form of a condensate around spinning BHs.

\subsection{Homogeneous background scalars} \label{sec:mink_scalar_coupling}
%
\begin{figure*}[htb]
\begin{tabular}{cc}
\includegraphics[width=8.5cm]{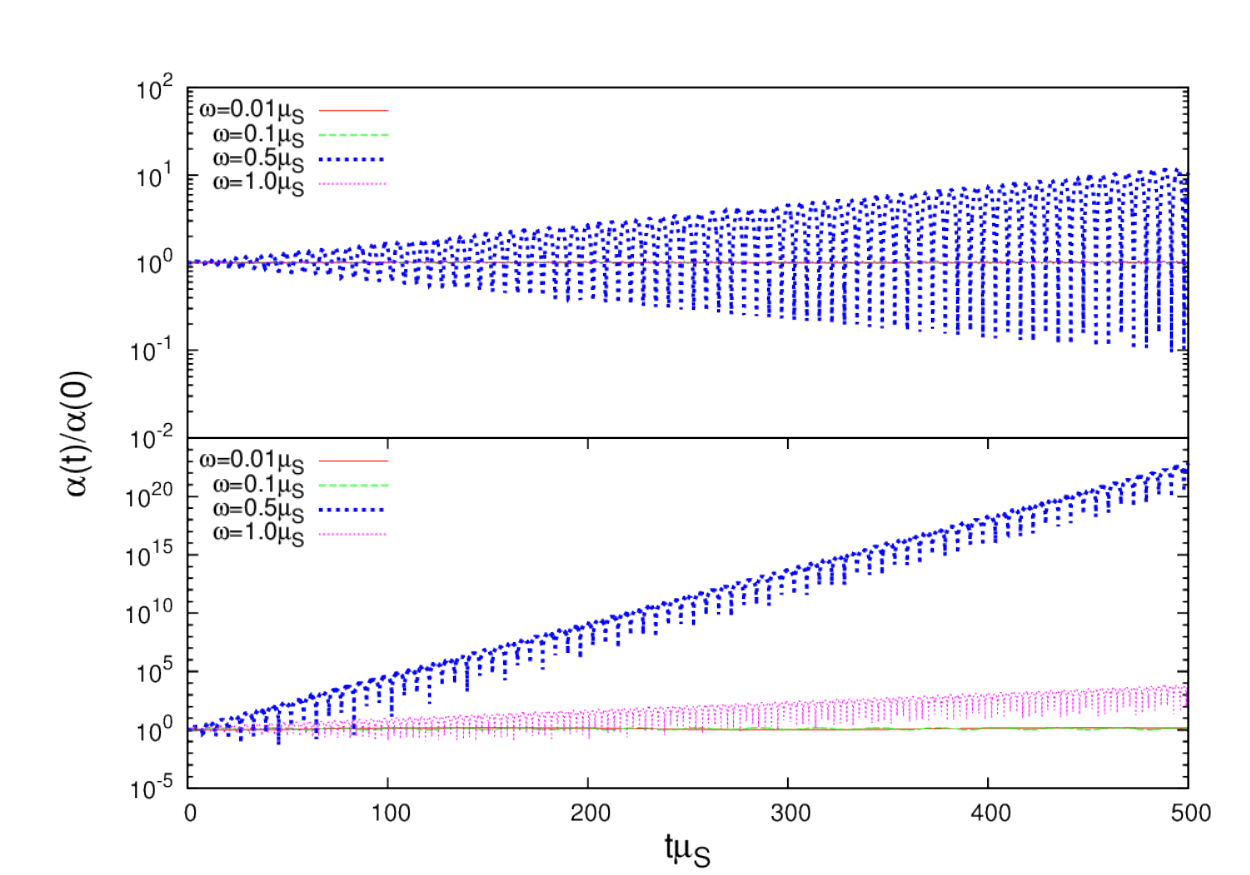}&
\includegraphics[width=8.5cm]{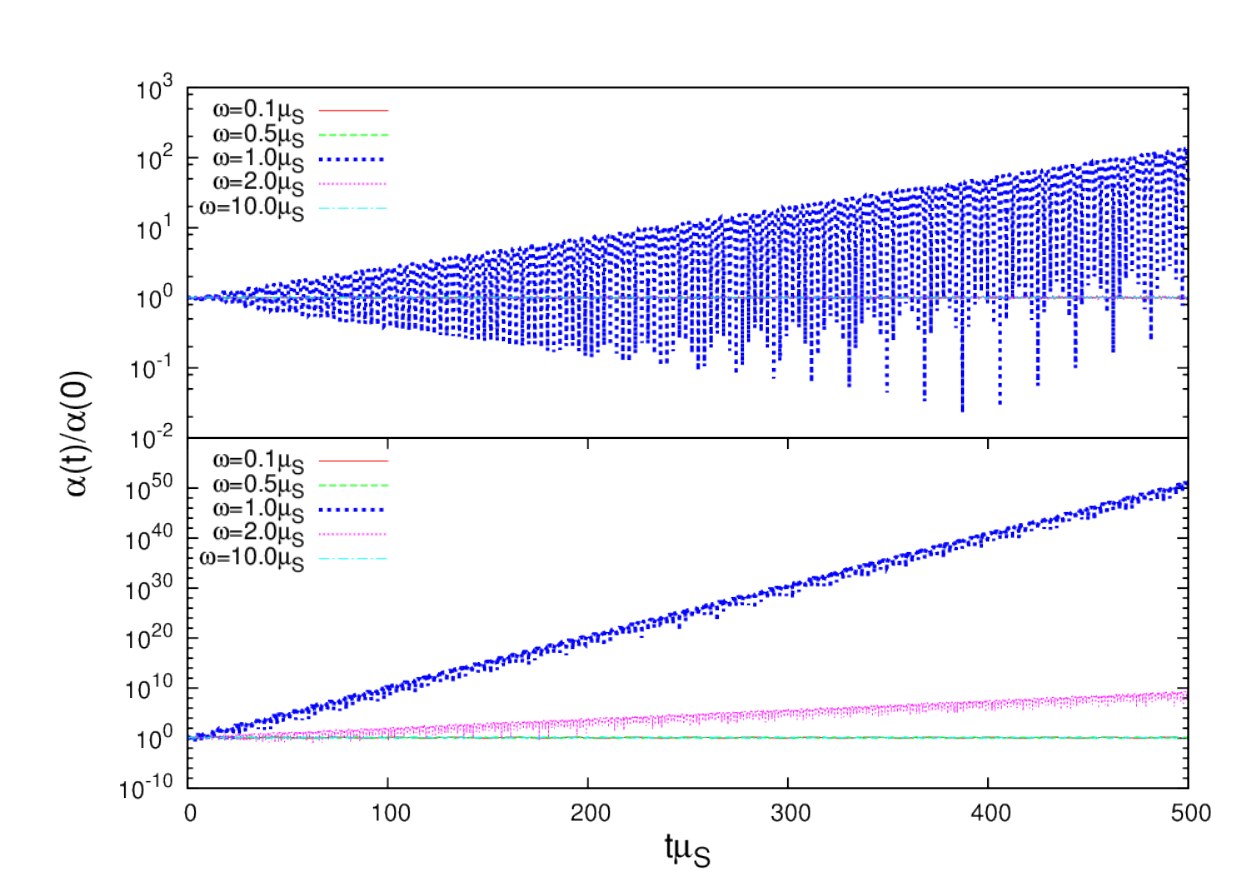}
\end{tabular}
\caption{Time evolution of the vector field. 
{\bf Left:} Scalars with $p=1$ and coupling strength $k_{\rm a}\kwave\Psi_{0}=0.005\mu_{\rm S}$ (upper panel) and $k_{\rm a}\kwave\Psi_{0}=0.1\mu_{\rm S}$ (lower panel) for selected frequencies.
The dominant mode continues to be that with frequency $\omega=0.5\mu_S$, as for axionic couplings.
{\bf Right:} Scalars with $p=2$ and coupling $k_{\rm a}\kwave\Psi_{0}=0.005\mu_{\rm S}$ (upper panel) and $k_{\rm a}\kwave\Psi_{0}=0.2\mu_{\rm S}$ (lower panel). Notice how the dominant mode now has frequency $\omega=\mu_S$.
\label{time_evolution_kscalar02_p}
}
\end{figure*}
\begin{figure*}[htb]
\begin{tabular}{cc}
\includegraphics[width=8.5cm]{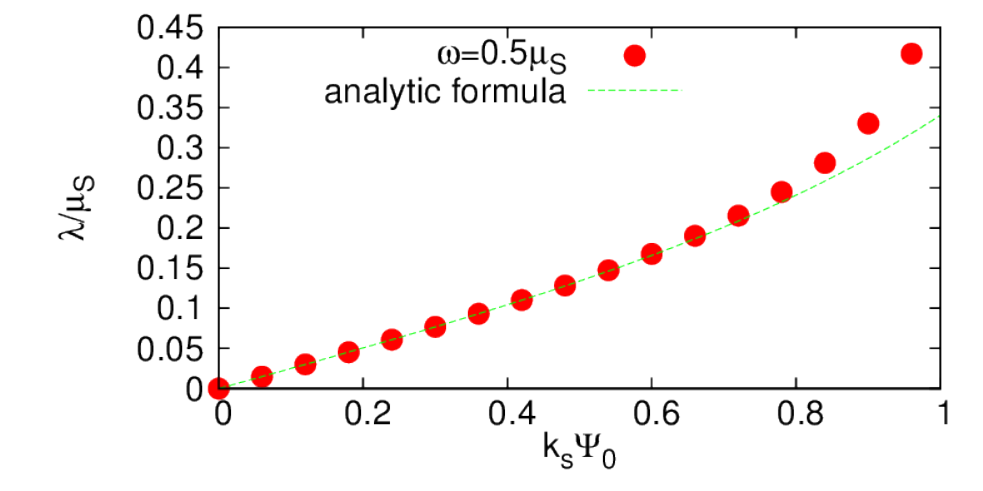}
&\includegraphics[width=8.5cm]{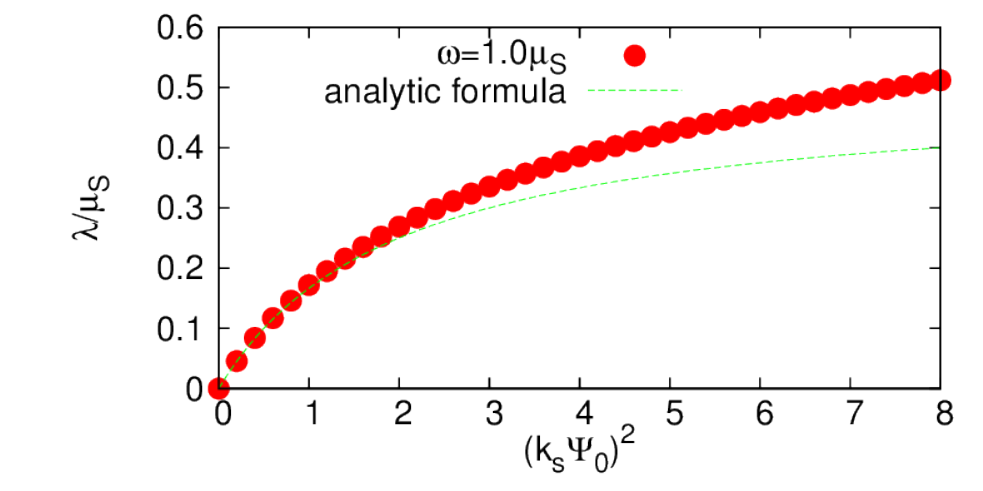}
\end{tabular}
\caption{The relation between the growth rate and the coupling constant for $p=1$ (left panels) and $p=2$ (right panels). Red dots are full numerical values, the dashed green line
is our perturbative description of the rate, described in Appendix \ref{app:RG_time_scales_2nd_sc}. \label{relation_kscalar_lambda_p}}
\end{figure*}
Let us apply the above analysis to scalar coupling case. Maxwell equations with scalar coupling are
\begin{eqnarray}
\nabla_{\mu}\left(1+k_{\rm s}^p\Psi^{p}\right)F^{\mu\nu}=0\,.
\end{eqnarray}
%

\subsubsection{Numerical results for the instability}
We take again the ansatz~\eqref{eq.uniform coherent oscillating axion} and \eqref{eq.expansion of A} in the Lorenz gauge, Eq.\ref{Lorenz gauge}), to render the final equations sufficiently simple. 

For $p=1$, we find 
\begin{eqnarray}
&0&=\left(1+k_{\rm s}\Psi_0 e^{-i\mu_{\rm S}t}/2+k_{\rm s}\Psi_0^{*} e^{i\mu_{\rm S}t}/2\right)
\left(-\ddot{\alpha}_{i}+i2\omega\dot{\alpha}_{i}\right)\nonumber\\
&-&ik_{\rm s}\mu_{\rm S}\left(\Psi_0 e^{-i\mu_{\rm S}t}/2-\Psi_0^{*}e^{i \mu_{\rm S}t}/2
\right)\left(-\dot{\alpha}_{i}+i\omega\alpha_{i}+i \kwave_{i}\alpha_{t}\right)\,.\nonumber
\end{eqnarray}
In order to extract the transverse mode, we introduce the projection operator $\mathcal{P}_{ij}=\delta_{ij}-\frac{\kwave_{i} \kwave_{j}}{\vec{\kwave}^{2}}$. Then, the evolution equation for transverse mode is
%
\begin{eqnarray}
&&\left(1-k_{\rm s}|\Psi_{0}|\cos(\mu_{\rm S}t-\theta)\right)(-\ddot{\alpha}_{(I)}+i2\omega\dot{\alpha}_{(I)})\nonumber\\
&&+k_{\rm s}\mu_{\rm S}|\Psi_{0}|\sin (\mu_{\rm S}t-\theta)(-\dot{\alpha}_{(I)}+i\omega\alpha_{(I)})=0\,,\label{scalar_p1}
\end{eqnarray}
%
where $\alpha_{i}=P_{ij}\alpha_{j}$, and $\theta$ is the phase of $\Psi_{0}$.
The time evolution of the vector potential $\alpha_{(I)}$ is shown in Fig.~\ref{time_evolution_kscalar02_p} for the initial data \eqref{init_data_Minkowski}.
We find results similar to those of the axionic case:
a homogeneous background scalar field and a vanishing EM field is an unstable configuration. An instability is triggered whereby
the vector grows exponentially, and $\omega=0.5\mu_{\rm S}$ seems to be -- visibly at small couplings -- the mode driving the state when $p=1$.

The growth rates for the dominant $\omega=\mu_{\rm S}/2$ mode are shown in Fig.~\ref{relation_kscalar_lambda_p} for different coupling strength. Our results are consistent with
\be \label{eq:rate_scalar_p_1}
\lambda\sim 0.25 k_{\rm s}\mu_s\Psi_0\,,
\ee
for sufficiently small couplings.

We can apply the same analysis to the $p=2$ case. We find,
\begin{eqnarray}
&0&=\left(-\ddot{\alpha}_{(I)}+i2\omega\dot{\alpha}_{(I)}\right)\times\left(1+k_{\rm s}^2\Psi_{0}^{2}\cos^{2}(\mu_{\rm S}t-\theta)\right)\nonumber\\
&-&2k_{\rm s}^{2}\Psi_{0}^{2}\cos(\mu_{\rm S}t-\theta)\sin(\mu_{\rm S}t-\theta)\left(-\dot{\alpha}_{(I)}+i\omega\alpha_{(I)}\right)\nonumber\,.
\end{eqnarray}
The solutions of this equation are depicted in Fig.~\ref{time_evolution_kscalar02_p}, again for the initial data \eqref{init_data_Minkowski}. The dominant mode is now $\omega=\mu_{\rm S}$ and for this mode the growth rates are
\be  
\label{eq:rate_scalar_p_2}
\lambda\sim \frac{1}{4} k_{\rm s}^2\Psi_0^2\mu_S\,,
\ee
%
\subsubsection{Analytical results} \label{sec:Mink_scalar_background_analytical}
For $p=1$ and small couplings $k_{\rm s}$, the substitution $\alpha=\frac{e^{i\omega t}}{\sqrt{1-k_{\rm s}\Psi_0\cos{\mu_St}}}y_{\omega}$ allows us to re-write  Eq.~\eqref{scalar_p1} as
\be
\partial^2_T y_{\omega}+\left(\omega^2/\mu_S^2-\frac{k_{\rm s}}{2}\Psi_0\cos{T}\right)y_{\omega}=0\,,
\ee
where again $T=\mu_S t$ and we kept only the leading-order term in the coupling. Thus, we recover again Mathieu's equation. In this regime, we then easily get that the dominant mode is\footnote{These results are also derived in the Coulomb gauge in Appendix \ref{app:Coulomb_scalar}, and we compute perturbatively the rate estimates in Appendix \ref{app:RG_time_scales}.}
\be
\omega=\mu_S/2\,,\label{eq:wp1}
\ee
and a growth rate $\lambda=0.25 k_{\rm s}\mu_S \Psi_0$, in very good agreement with the numerics.

For $p=2$ and small couplings, $\alpha=\frac{e^{i\omega t}}{\sqrt{2+k_{\rm s}^2\left(1+\Psi_0^2\cos{2\mu_St}\right)}}y_{\omega}$ allows us to rewrite the relevant equation as
\be
\partial^2_T y_{\omega}+\left(\omega^2/(4\mu_S^2)-\frac{k_{\rm s}^2\Psi_0^2}{4}\cos{T}\right)y_{\omega}=0\,,\nonumber
\ee
where now $T=2\mu_S t$. The same analysis therefore gives us the small-coupling dominant mode 
\be
\omega=\mu_S\,,\label{eq:wp2}
\ee
and the rate $\lambda=\Psi_0^2k_{\rm s}^2\mu_S/4$, in good agreement with the numerical results.

\section{Instability of Reissner-Nordstr\"{o}m BHs}\label{sec:inst_RN}

\subsection{Perturbative framework}

It is easy to see that Reissner-Nordstr\"{o}m (RN) BHs are a solution to the field equations~\eqref{eq:MFEoMgen} when the scalar field vanishes and $p\geq 2$\, \footnote{For $p=1$, RN is not a solution of the field equations, since the term $F_{\mu\nu}F^{\mu\nu}\neq 0$ sources the scalar field equation~\eqref{eq:MFEoMScalar}. This case is analogous to the more familiar Einstein-Maxwell-dilaton theories. We provide a perturbative solution in the next section.}. We now wish to show that, as expected from the flat-space analysis of the previous section, this equilibrium solution of the field equations is {\it unstable} at large enough electric fields. 
We provide details in the axionic case. The scalar coupling follows through in the same way and we discuss the results below.

We consider the vector field $A_{\mu}$ and massive scalar field $\Psi$ propagating in a static, charged BH background and coupled through the axionic coupling defined in the Lagrangian~\eqref{eq:MFaction}. The BH background is described by the metric 
\be
ds^2=-f(r)dt^2+\frac{dr^2}{f(r)}+r^2d\Omega^2\,,
\ee
where $f(r)=1-2M/r+Q^2/r^2$, with $M$ and $Q$ the BH mass and charge, respectively, and the vector field satisfies $A_{\mu}dx^{\mu}=Q/r dt$.  This spacetime has an event horizon and a Cauchy horizon located at $r_{\pm}=M\pm\sqrt{M^2+Q^2}$, respectively.  

We now linearize the field equations~\eqref{eq:MFEoMgen} around this background, and expand the perturbation functions in a complete basis of spherical harmonics. In particular we linearize the metric as
\begin{equation}
 g_{\mu\nu}=g_{\mu\nu}^{(0)}+h_{\mu\nu}\,,
\end{equation}
and decompose the metric perturbations $h_{\mu\nu}$ in the Regge--Wheeler gauge:
\begin{equation}
h_{\mu\nu}=\left(\begin{array}{cc|cc}
H_0^{{l}}Y^{l m}&H_1^{l}Y^{l m}&h_0^{l}S_\theta^{l m}
&h_0^{{l}}S_\varphi^{{l m}}\\
H_1^{{l}}Y^{{l m}}&H_2^{{l}}Y^{{l m}}&h_1^{{l}}S_\theta^{{l m}}
&h_1^{{l}}S_\varphi^{{l m}}\\\hline
h_0^{{l}}S_\theta^{{l m}}&h_1^{{l}}S_\theta^{{l m}}&r^2K^{{l}}Y^{{l m}}&0\\
h_0^{{l}}S_\varphi^{{l m}}&h_1^{{l}}S_\varphi^{{l m}}&0&r^2K^{{l}}\sin^2\theta Y^{{l m}}\\
\end{array}\right)\,,\label{expmetric}
\end{equation}
where $Y^{{l m}}=Y^{{l m}}(\theta,\varphi)$ are the ordinary scalar spherical harmonics,
$(S_\theta^{{l m}},S_\varphi^{{l m}})\equiv \left({Y^{{l m}}_{,\varphi}}/{\sin\theta},-\sin\theta
Y^{{l m}}_{,\theta}\right)$ are the axial vector harmonics, and $H_{0,1,2}^{{l}},\,h_{0,1}^{{l}},\,K^{{l}}$ are functions of
$(t,\,r)$. Here a sum over the harmonic indices $l$ and $m$ is implicit.

The EM potential can be linearized in a similar way as:
\begin{equation}\label{expansion_maxwell}
\delta A_{\mu}(t,r,\theta,\varphi)= \left[
 \begin{array}{c} 0 \\ 0\\
u_{(4)}^{{l}} S_b^{{l m}}\\
 \end{array}\right]+\left[ \begin{array}{c}u_{(1)}^{{l}} Y^{{l m}}\\u_{(2)}^{{l}} Y^{{l m}} \\
 u_{(3)}^{{l}} Y_b^{{l m}}\\ \end{array}\right]\,,
\end{equation}
where $b=(\theta,\varphi)$,
$Y_b^{l m}=(Y_{,\theta}^{l m},Y_{,\varphi}^{l m})$ are the polar vector
harmonics, $S_b^{{l m}}\equiv
\left({Y^{{l m}}_{,\varphi}}/{\sin\theta},-\sin\theta
Y^{{l m}}_{,\theta}\right)$ are again the axial vector harmonics, and $u_{(1,2,3,4)}^{{l}}$ are functions of $(t,\,r)$ and where again a sum over the harmonic indices $l$ and $m$ is implicit.

Finally, the scalar field is expanded as as
\be
\Psi(t,r,\theta,\varphi)=\frac{\psi^l(t,r)}{r} Y^{l m}\,.\label{psi_decomposition}
\ee
In the following we omit the indices $l$ and $m$, because in a spherical background different $l$ and $m$ modes
completely decouple.

This system can be studied in Fourier space by performing the decomposition:
\begin{equation}
\delta \tilde{X}(t,r)=\int d\omega \delta X(\omega, r)e^{-i\omega t}\,,\label{expa}
\end{equation}
where $\delta \tilde{X}(t,r)$ denotes schematically any perturbation function. 
The perturbation variables $\delta X$ can be classified as ``polar'' or ``axial'' depending on their behavior under parity transformations. In a spherically symmetric background polar and axial perturbations always decouple.

The background electric charge induces a coupling between gravitational and EM (grav-EM) perturbations, while due the pseudo-scalar nature of the axionic coupling, the scalar field only couples to the axial sector of the EM perturbations. Thus, only the axial sector is affected by the presence of the scalar field, while the polar sector is fully described by the usual grav-EM perturbations of the RN metric, which have been widely studied in the literature (see e.g. Ref.~\cite{Chandra}).

Let us focus then on the axial sector. We define the Regge--Wheeler function through $h_1=r\psi_{\rm RW}f(r)^{-1}$ and  $h_0=i\,f(r)\left(r\psi_{\rm RW}\right)'/\omega$ and perform the linear transformations (from $\psi_{\rm RW}, u_4$ to $Z_\pm$):
\beq
u_4&=&-\frac{Z_{+}+Z_{-}}{l(l+1)}\,,\\
\psi_{\rm RW}&=&-i\omega\frac{\left(3M-\lambda\right) Z_{+}+\left(3M+\lambda\right) Z_{-}}{(l-1)l(l+1)(l+2)Q}\,,
\eeq
with $\lambda=\sqrt{9M^2+4Q^2(l-1)(l+2)}$.
After some algebra we find that axial perturbations can be described by a system of tree coupled ordinary differential equations, given by:
\beq
&&\frac{d^2\psi}{dr_*^2}+\left(\omega^2-V_{\psi}\right)\psi+S_{\psi}Z_{+} + S_{\psi}Z_{-}=0\label{pert_RN1}\,,\\
&&\frac{d^2Z_+}{dr_*^2}+\left(\omega^2-V_{+}\right)Z_{+}+S_{+}\psi=0\label{pert_RN2}\,,\\
&&\frac{d^2Z_-}{dr_*^2}+\left(\omega^2-V_{-}\right)Z_{-}+S_{-}\psi=0\label{pert_RN3}\,,
\eeq
where the ``tortoise'' coordinate $r_*$ is defined through the relation $dr/dr_*=f(r)$ and
\beq\label{syst_RN}
&&V_{\psi}=f(r)\left(\mu_S^2+\frac{l(l+1)}{r^2}+\frac{f'(r)}{r}\right)\,,\nonumber\\
&&S_{\psi}=f(r)\frac{2k_{\rm a}Q}{r^3}\,,\nonumber\\
&&V_{\pm}=f(r)\left(\frac{l(l+1)}{r^2}-\frac{3M}{r^3}+\frac{4Q^2}{r^4}\pm \frac{\lambda}{r^3}\right)\,,\nonumber\\
&&S_{\pm}=\pm f(r)\frac{l(l+1)k_{\rm a}Q\left(3M\pm\lambda\right)}{\lambda r^3}\,.
\eeq
For spherically symmetric perturbations, $l=0$, the scalar field decouples from the EM and gravitational perturbations, and one can conclude that the RN geometry is stable against radial perturbations. The same conclusion can be drawn when $k_{\rm a}=0$ or $Q=0$.

On the other hand, for $l\geq 1$, in analogy to the flat space analysis of the previous section~\footnote{As mentioned in the context of Eq.~\eqref{flat_dispersion}, the flat-space axionic instability around background electric field requires non-homogeneous axion fluctuations. Thus, the instability only sets in for non-symmetric modes should not come as a surprise.}, one expects the existence of unstable modes above some critical value for the axion coupling constant $k_{\rm a}$. To find these modes we look for purely decaying modes at infinity\footnote{This system also allows for more generic boundary conditions describing quasibound states. For completeness we also discuss these modes in Appendix~\ref{boundstates}.}:
\be\label{boundary_infty}
\left\{\psi,Z_+,Z_-\right\}\to 0\,,\quad {\rm as}\quad r_*\to\infty\,,
\ee
while at the horizon, $r=r_+$, we impose regular boundary conditions, i.e., purely ingoing waves described by
\be\label{boundary_horizon}
\left\{\psi,Z_+,Z_-\right\}\to \left\{\psi_0,Z_{+\,0},Z_{-\,0}\right\}e^{-i \omega r_*}\,,\quad {\rm as}\quad r_*\to -\infty\,,
\ee
where $\psi_0$ and $Z_{\pm\, 0}$ are constants.
Since the system~\eqref{pert_RN1}~--~\eqref{pert_RN3} is linear, one can fix the value of, e.g., $\psi$ at the horizon and the problem becomes a 3-dimensional eigenvalue problem for $\omega$  and $Z_\pm (r=r_+)$. We solved this system by using a 3-parameter shooting method: we shoot on the parameters $\omega$ and $Z_\pm (r=r_+)$ and integrate the field equations starting from $r=r_+$ until the boundary conditions at infinity~\eqref{boundary_infty} are satisfied. 
 
\subsection{Results}

\subsubsection{Axionic couplings}
%
\begin{figure}[htb]
\begin{tabular}{c}
\includegraphics[width=8.5cm]{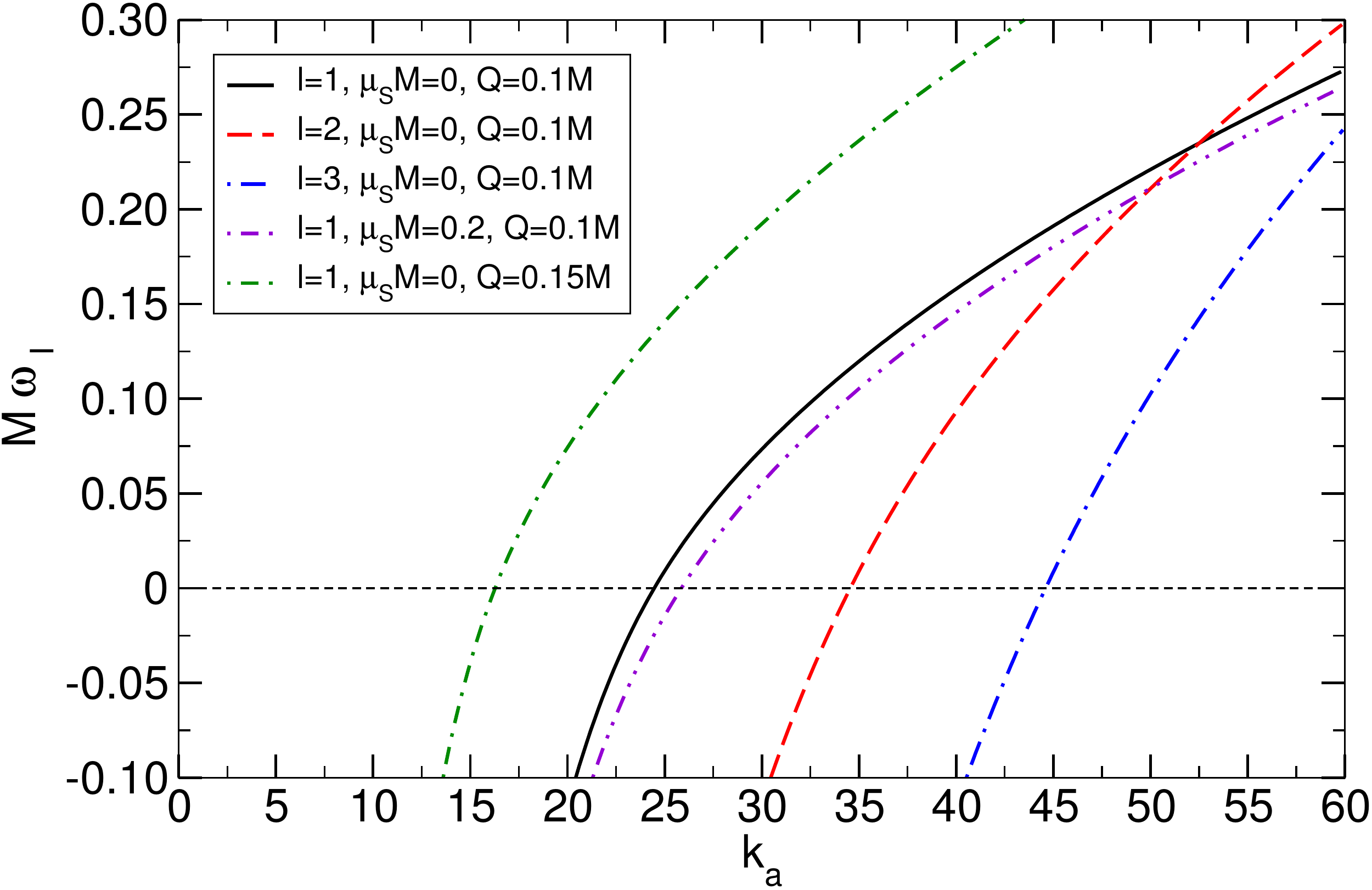}
\end{tabular}
\caption{Inverse of instability timescale, $M\omega_I$, as a function of $k_{\rm a}$ for $M\mu_S=0$, $Q=0.1M$ and different multipole numbers $l$. For comparison we also show the growth rate of the multipole $l=1$ for $M\mu_S=0$,$Q=0.15M$ and $M\mu_S=0.2$, $Q=0.1M$. For $l\geq 1$, $Q\ll M$ and $M\mu_S\lesssim 0.1$, the system becomes unstable above the critical value $\left(k_{\rm a}Q/M\right)_{\rm critical}\gtrsim 1.45+l+(M\mu_S)^{3/2}$.
\label{instability_RN}}
\end{figure}
Using the method outlined above, we find that the systems admits purely imaginary modes, $\omega=i \omega_I$, for which $\omega_I$ becomes positive above a critical value for $k_{\rm a}$. Given the ansatz~\eqref{expa} this therefore means that these modes grow exponentially in time and the system is unstable. An analogous instability was in fact shown to occur for asymptotically Anti-de Sitter RN BHs in Ref.~\cite{Donos:2011bh}. For a given $l$, $Q$ and $\mu_S$ there exist an infinite family of solutions characterized by the number of nodes and $k_{\rm a}^{\rm crit}$. This tower of solutions signals the existence of new families of BH solutions, which we will discuss shortly. For the fundamental mode, i.e., the mode with no nodes, the results are summarized in Fig.~\eqref{instability_RN} where we plot $\omega_I$ as a function of $k_{\rm a}$ for different values of $Q$, $\mu_S$ and $l$. 

We note that this instability is present even for massless scalars ($M\mu_S=0$) as shown in Fig.~\ref{instability_RN}.
We also remark that it sets in for smaller couplings $k_{\rm a}$ as we increase the BH charge and, hence, electric field.
Fig.~\ref{instability_RN} furthermore indicates that the instability timescale does not depend monotonically on the multipole number $l$
for sufficiently large coupling constants $k_{\rm a}$. For example, if the coupling is $k_{\rm a}\geq 50$, $Q=0.1M$ and $M\mu_S=0$ the $l=2$ mode is the dominant one, i.e. it is the mode that grows faster.

For small BH charges $Q\ll M$ and small mass couplings $M\mu_S\lesssim 0.1$, we find that the critical value above which the system becomes unstable for the nodeless modes is approximately given by
\be\label{crit_value}
k_{\rm a}^{\rm crit}\sim \frac{M\left(1.45+l+(M\mu_S)^{3/2}\right)}{Q}\,.
\ee
The dependence of the threshold value of $k_{\rm a}$ on the BH charge $Q$ (or in other words, on the electric field close to the horizon) is in agreement with the results from the flat-space analysis \eqref{thresh_flat_axion}. Near this critical value we find
\be
M\omega_I\sim 0.2 Q/M\left(k_{\rm a}-k_{\rm a}^{\rm crit}\right)\,.
\ee
%

\subsubsection{Scalar couplings}
%
\begin{figure*}[t]
\begin{tabular}{cc}
\includegraphics[width=8.5cm]{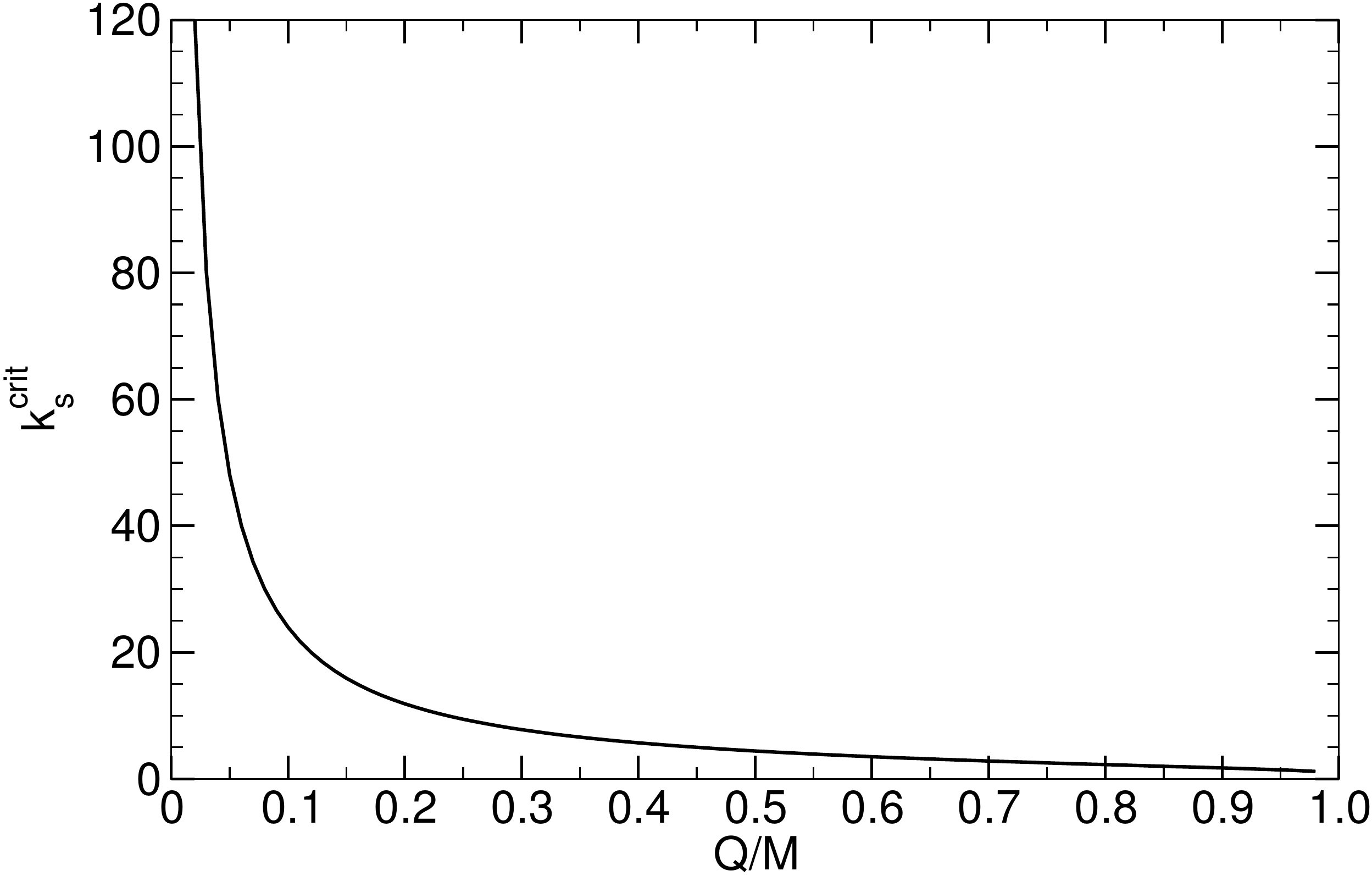}&
\hspace{-2mm}\includegraphics[width=8.3cm]{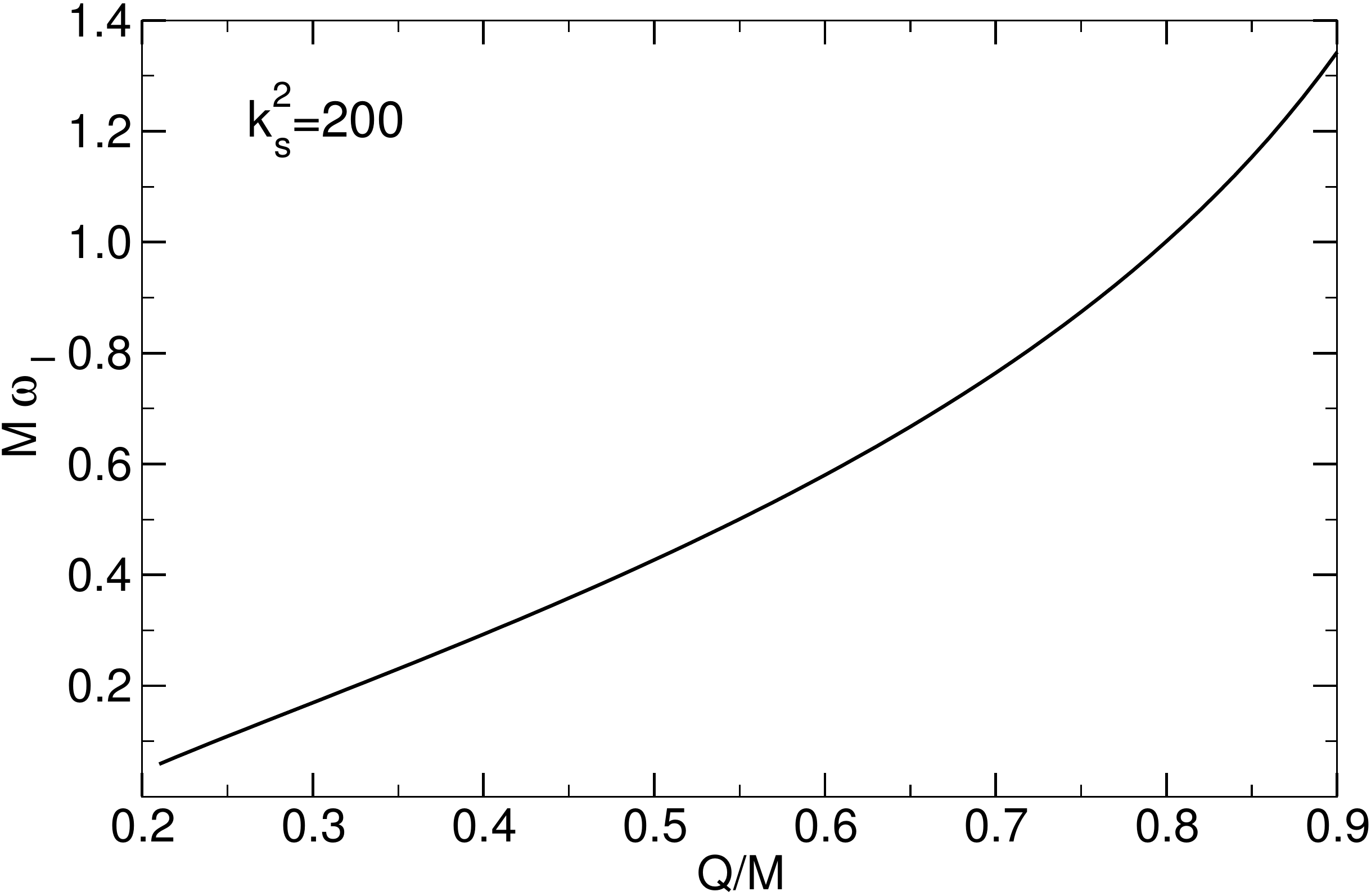}
\end{tabular}
\caption{Threshold coupling parameter as a function of BH charge, beyond which an instability sets in (left). This also represents the value of the coupling at which a new solution branchs off. For values of the coupling parameters $k_{\rm s}$ above the solid black line, RN BHs are unstable, with instability rates shown in the rightmost panel for $k_{\rm s}^2=200$ and $M\mu_S=0$.
\label{instability_RN_scalarp2}}
\end{figure*}

The results for scalar couplings arise from a similar analysis. Focusing on the $p=2$ case and on the spherically symmetric modes, we can decompose the scalar field as in \eqref{psi_decomposition} to reduce the analysis to a single ODE,
\beq
&&f^2\psi''+ff'\psi'+\left(\omega^2-V\right)\psi=0\,,\\
&&V=f\left(\frac{2Mr-(2+k_{\rm s}^2)Q^2}{r^4}+\mu_S^2\right)\,.
\eeq

For concreteness we show only the case with $\mu_S=0$, although similar conclusions can be reached for generic values of $\mu_S$. We find that there is a critical value of $k_{\rm s}$ beyond which an instability arises in a RN geometry. At this precise value of the coupling constant, $k_{\rm s}^{\rm crit}$, a new solution of the field equations is possible. In fact, we find that, for fixed $M, Q$ there is an infinite set of constants $k_{\rm s}$ for which a zero mode appears. This tower of solutions signals the existence of new families of BH solutions, analogous to the ones discussed in Refs.~\cite{Herdeiro:2018wub,Myung:2018vug}, which we will discuss at full nonlinear level shortly. We focus our attention on the smallest value of the coupling constant for which a zero mode exists, and for which the scalar has no nodes. For $p=2$, our results are summarized in Fig.~\ref{instability_RN_scalarp2}.

The threshold value is well described by $k_{\rm s}^{\rm crit}=2.4/Q$, for small charge. This dependence on the charge (or, conversely, on the electric field) is the same as the one predicted from a flat-space analysis, \eqref{scalar_dispersion_KG}. We should also stress that the scalar $\Psi$ decays as $1/r$ at large distance, thus
these BH solutions have a primary scalar hair and scalar charge, as we confirm in the next section when constructing these solutions at the full nonlinear level.

\section{New black hole solutions}

\subsection{New BH solutions for axionic couplings}

\subsubsection{End-state of the RN instability}

We showed that RN BHs are quite generically unstable when the EM field is coupled to a scalar field. The existence of a zero mode with frequency $\omega=0$ at the threshold of these instabilities suggests the existence of new BH solutions, branching-off from the RN solution for values of the coupling parameter at this threshold. As we discuss now, solutions with non-trivial scalar field indeed exist for the couplings considered here.

As shown in the previous section, for the axionic coupling the instability only exists for non-spherically symmetric axial perturbations, therefore due to the backreaction of these perturbations on the metric the end-state of this instability will most likely be an axisymmetric and stationary BH spacetime.

These solutions can then be obtained by solving a system of elliptic PDEs, which can only be solved numerically. We leave the construction of the full solutions for future work but we note that in asymptically AdS spacetimes an analogous instability and possible end-state solutions were constructed in Refs.~\cite{Donos:2011bh,Rozali:2012es,Donos:2013wia,Withers:2013loa}. 

\begin{figure}[htb]
\begin{tabular}{c}
\includegraphics[width=8.5cm]{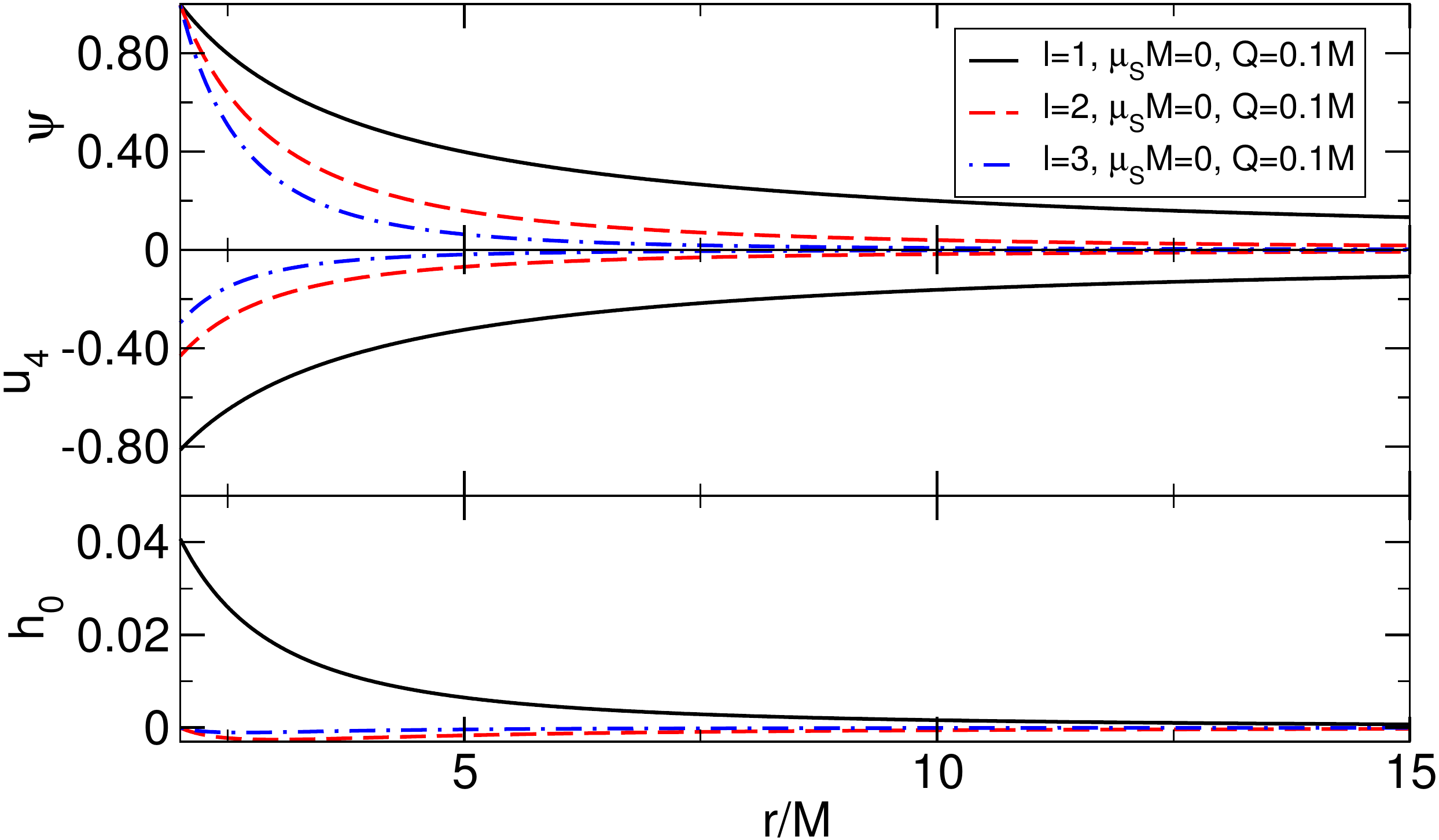}
\end{tabular}
\caption{Stationary hairy BH solutions in the presence of axionic couplings $k_{\rm a}$, obtained by linearizing around the RN solution for $Q=0.1M$ and $M\mu_S=0$. Here we plot the fundamental modes (solutions with no nodes) and we label these solutions by $k_{\rm a}^{\rm crit}$ and $l$, with $k_{\rm a}^{\rm crit}$ given approximately by Eq.~\eqref{crit_value}.
\label{RN_hairyBH}}
\end{figure}

To gain insight into these solutions, we instead construct them perturbatively by making use of the fact that at $k_{\rm a}=k_{\rm a}^{\rm crit}$ there should exist another stationary solution besides RN. Linearizing the field equations~\eqref{eq:MFEoMgen} and focusing on time-independent perturbations we find that, for $l\geq 2$, we must have $h_1=0$, while $h_0$, $\psi$ and $u_4$ are given by the system of equations:
\begin{subequations}\label{hairy_RNaxion}
\beq
&&\frac{d^2\psi}{dr_*^2}-f(r)\left(\mu_S^2+\frac{l(l+1)}{r^2}+\frac{f'(r)}{r}\right)\psi\nonumber\\
&&-f(r)\frac{2l(l+1)k_{\rm a}Q}{r^3}u_{4}=0\label{hairy_RN1}\,,\\
&&\frac{d^2u_4}{dr_*^2}-f(r)\frac{l(l+1)}{r^2}u_4-f(r)\frac{2k_{\rm a}Q}{r^3}\psi\nonumber\\
&&-f(r)\frac{Q}{r^2}h'_0+f(r)\frac{2Q}{r^3}h_0=0\label{hairy_RN2}\,,\\
&&h''_0=\frac{2}{r^2}h_0+\frac{(l-1)(l+2)}{r^2f(r)}h_0+\frac{4Q}{r^2}u'_4\label{hairy_RN3}\,,
\eeq
\end{subequations}
where a prime denotes differentiation with respect to $r$.
For $l=1$ this can be simplified to: 
\begin{subequations}\label{hairy_RNaxion_l1}
\begin{align}
&\frac{d^2\psi}{dr_*^2}-f(r)\left(\mu_S^2+\frac{2}{r^2}+\frac{f'(r)}{r}\right)\psi-f(r)\frac{4k_{\rm a}Q}{r^3}u_{4}=0\,,\\
&\frac{d^2u_4}{dr_*^2}-f(r)\left(\frac{2}{r^2}+\frac{4Q^2}{r^4}\right)u_4-f(r)\frac{2k_{\rm a}Q}{r^3}\psi=0\,,\\
&h'_0=\frac{2}{r}h_0+\frac{4Q}{r^2}u_4\,.
\end{align}
\end{subequations}
We solved these equations by imposing regular boundary conditions at the horizon and at infinity and using the shooting method outlined above. As expected, regular solutions only exist for non-zero BH electric charge and for $k_{\rm a}=k_{\rm a}^{\rm crit}$.  Specific solutions are shown in Fig.~\ref{RN_hairyBH}.

At spatial infinity the scalar field decays as $\Psi \propto 1/r^{l+1}$, the vector field component decays as $u_4 \propto 1/r^{l}$ while the metric component decays as $h_0 \propto 1/r^{l}$ for $l \geq 2$ and  $h_0 \propto 1/r^2$ for $l=1$. Therefore these solutions have the peculiar property that frame-dragging effects are present even though their ADM angular momentum vanishes. The non-zero frame-dragging is instead due to the backreaction of the magnetic field induced by the axion on the metric. 

As a final remark, we should note that when obtaining equations~\eqref{hairy_RNaxion_l1} for $l=1$ we have integrated out the solution that decays as $h_0 \propto 1/r$ and $u_4 \propto 1/r$ at spatial infinity. This additional solution, which exist for any $k_{\rm a}$, can be instead obtained from the system~\eqref{hairy_RNaxion} with $l=1$. For $k_{\rm a}=0$, this solution is none other than the Kerr-Newman solution expanded up to first-order in the BH spin, while for $k_{\rm a}\neq 0$ this corresponds to a modified Kerr-Newman geometry as we discuss below.

\subsubsection{Axions around rotating charged BHs} \label{sec:kn} 

When the BH is spinning, one expects that the scalar acquires a non-trivial profile for any value of $k_{\rm a}$, since $\,^{\ast}F^{\mu\nu}F_{\mu\nu}\neq 0$ for non-spherically symmetric spacetimes. In fact, it is easy to check that $\,^{\ast}F^{\mu\nu}F_{\mu\nu}=0$ when $F_{\mu\nu}=0$ or when the spacetime is spherically symmetric. Thus, the Kerr geometry together with $F_{\mu\nu}=\Psi=0$ is a full nonlinear solution of the theory above. In addition, for $k_{\rm a}=0$, Kerr-Newman is also a solution. It is thus natural to look for rotating, charged BHs as perturbations of the Kerr-Newman geometry.

At first order in $k_{\rm a}$\footnote{We note that when expanding in $k_{\rm a}$ with a background EM field, one is effectively considering expansions of the form $k_{\rm a} \langle A \rangle$, where $ \langle A \rangle$  is a characteristic, dimensionless and Lorentz-invariant measure of the EM field strength (e.g. $ \langle A \rangle =Q^2/M^2$ for a charged BH). In other words, strong EM fields can compensate for a ``small'' value of $k_{\rm a}$ and produce observable consequences. A similar approach was recently considered in the context of pulsar magnetospheres~\cite{Garbrecht:2018akc}.}, we find that the only nontrivial correction appears at the level of the scalar, and to first order in rotation is described by (for $\mu_S=0$ and $k_{\rm s}=0$)\footnote{This equation can be obtained from the system \eqref{hairy_RNaxion} with $l=1$ by considering the expansion parameter to be proportional to the dimensionless BH spin $a/M$.}:
\beq
&&\Psi=k_{\rm a} a\, \eta(r)\cos\theta\,,\label{KN_hair1}\\
&&(r^2-2Mr+Q^2)\eta''+2(r-M)\eta'-2\eta=-\frac{Q^2}{r^3}\,.
\eeq
Regular solutions exist for any nonzero value of the charge and can be found analytically after requiring regularity at the horizon and spatial infinity:
\be
\eta =\frac{1}{2}\left(\frac{2}{r_+}-\frac{1}{r}-\frac{\log\left[\frac{(r-r_-)^2}{r^2}\right](r_+ + r_- -2r )}{2 r_+ r_-}\right)\label{KN_hair2}\,,
\ee
where $r_{\pm}=M\pm\sqrt{M^2-Q^2}$. These solutions decay as $~1/r^2$ at infinity and up to second-order in the BH charge are given by $\eta=Q^2/(8Mr^2)+\mathcal{O}(Q^4)$.

To summarize, when rotation is turned on, one expects two distinct families of solutions with axion hair. One family of solutions exists for any value of $k_{\rm a}$ and that -- to first order in the BH spin and in the axion coupling -- is given by Eq.~\eqref{KN_hair1} with radial profile~\eqref{KN_hair2}; a second family which is the rotating generalization of the solution discussed above, which we expect to branch-off the solution~\eqref{KN_hair1} at $k_{\rm a}\sim k_{\rm a}^{\rm crit}$ for small rotation. 
 
\subsubsection{An extended Wald solution} \label{sec:wald_main} 

Aside from charged solutions, one should expect non-trivial axion configurations around non-spherically symmetric BHs with non-zero magnetic fields, since for this case one also has $\,^{\ast}F^{\mu\nu}F_{\mu\nu}\neq 0$. In fact, it has been shown by Wald~\cite{Wald:1974np} that, neglecting backreaction, Kerr BHs immersed in a homogeneous magnetic field $B$ aligned with the BH axis of symmetry allows for an exact analytical solution of Maxwell's equations:
\beq
A_{\mu}=\frac{1}{2}B(m_{\mu}+2a k_{\mu}),
\eeq
where $k^{\mu}=(1,0,0,0)$ and $m^{\mu}=(0,0,0,1)$ are the two Killing vectors that the Kerr metric admits. This field would lead to the BH accreting surrounding charge in the accretion disk and the interstellar medium. Therefore BHs would acquire a charge in those enviroments and be described by a Kerr-Newman spacetime, with a total vector potential given by~\cite{Wald:1974np}
\beq
\tilde{A}_{\mu}=\frac{1}{2}B(m_{\mu}+2a k_{\mu})-\frac{1}{2}qk_{\mu},
\eeq
with $q=Q/M$ and $Q$ is the accumulated BH charge. At equilibrium, the  BH charge-to-mass-ratio is given by $q=2B a$. We can therefore  analyse two different cases: (i) the BH is uncharged, and there is a net flow of charge from the surrounding medium, (ii) the BH is charged, but there is no net flow of charge from the surrounding medium. Recent estimates for supermassive BH in the Galactic center suggest that rotationally-induced charge is stable with respect to the discharging processes from the surroundings of an astrophysical plasma \cite{Zajacek:2018ycb}. Let us then focus on the second (equilibrium) case and estimate the importance of the induced charge on the background geometry. Using the limit for a maximal astrophysically realistic magnetic field from Section \ref{sec:setup}, we find $q \leq  10^{-11} a/M$,  i.e. the geometry is still well described by the Kerr metric.

Hence, we here consider a Kerr spacetime with the vector potential of the form\footnote{In this subsection we work in Boyer-Lindquist coordinates (Appendix \ref{App:Kerr_coordinates}) and suppress BL subscript.}
\begin{align}
A^{\rm Wald}_{\mu} = \frac{1}{2}& Bg_{\mu \nu}m^{\nu}=
\frac{B\sin^{2}\theta}{2\Sigma} \left(-2 a M r, 0, 0, \F \right)\,,\label{eq:WaldSol}
\end{align}
where $\F$ is a metric function given in Eq.~\eqref{eq:KerrBLfcts} and
we refer the reader to Appendix \ref{App:WaldSolution} for details of the Wald solution. 

Let us now consider, instead of Maxwell's equations, the generalized axionic equations~\eqref{eq:MFEoMgen}. For $k_{\rm a}=0$, Wald's solution is a solution to the problem, together with a vanishing scalar field. Thus, we are interested in a first-order (in $k_{\rm a} B^2M^2$) production of axions, as a consequence of the EM background. The dominant term describing
the axionic field is the equation 
\beq 
\left(\nabla^{\mu}\nabla_{\mu} - \mu^{2}_{\rm S} \right) \Psi = &
          \frac{1}{2}k_{\rm a} \,g^{\alpha \mu} g^{\beta \nu} ~ \,^{\ast}F^{(0)}_{\mu \nu} F^{(0)}_{\alpha \beta},\label{eq:KG_Wald}
\eeq       
where $F^{(0)}_{\mu\nu}$ denotes the Maxwell tensor corresponding to Wald's solution. 
Using Eq. \eqref{eq:WaldSol} we find, to fifth order in the spin $\tilde{a}=a/M$,
\begin{widetext}
\beq
g^{\alpha \mu} g^{\beta \nu} ~ \,^{\ast}F^{(0)}_{\mu \nu} F^{(0)}_{\alpha \beta}&=&-\frac{12aB^2M\cos\theta\sin^2\theta}{r^2}+\frac{4a^3B^2M\cos\theta\sin^2\theta\left(2r-M+\cos(2\theta)(M+5r)\right)}{r^5}\nonumber\\
&-&\frac{2a^5B^2M\cos^3\theta\sin^2\theta\left(-10M+r+\cos2\theta(10M+21r)\right)}{r^7}\,.
\eeq
\end{widetext}

One can now expand the left-hand side of Eq. \eqref{eq:KG_Wald} order by order in the spin, with $\Psi=\Phi_1 \tilde{a}+\Phi_2 \tilde{a}^2+\ldots$. To first order in rotation (and for $\mu_S=0$) one gets 
\beq
&&\frac{\partial}{\partial\theta}\left(\sin\theta \frac{\partial\Phi_1}{\partial\theta}\right)+\sin\theta\frac{\partial}{\partial r}\left((r^2-2Mr+a^2) \frac{\partial\Phi_1}{\partial r}\right) =\nonumber\\
&&=-6k_{\rm a}B^2M^2\cos\theta\sin^3\theta\,,
\eeq
and similar equations for higher order terms, each of which can be solved with an expansion in spherical harmonics. 
Finally, to first order in $k_{\rm a} M^2 B^2$ and fifth order in the spin for massless ``axions'' we find
\begin{widetext}
\be
\Psi=k_{\rm a}B^2M\left[\cos\theta\left(\frac{3a}{2}+\frac{a^3}{2r^2}\right)-\cos^3\theta\left(\frac{a}{2}+\frac{a^3}{r^2}+\frac{a^5}{2r^4}\right)
+\cos^5\theta\left(\frac{a^3}{2r^2}+\frac{a^5}{r^4}\right)-\cos^7\theta\frac{a^5}{2r^4}\right]\,.
\ee
\end{widetext}
This field will in turn contribute to the background EM field, via Eq.~\eqref{eq:MFEoMVector}, but as a second order 
(in $k_{a}$) effect.

Let us now briefly comment on the scenario in which conditions for the superradiant instability are met (see Appendix~\ref{sec:App_gravitational_atom}) \textit{and} the conditions for the EM field instability (discussed in Section~\ref{sec:Time_domain_studies}) are not satisfied. Then, $\Psi^{(0)}$ can be found by the expansion in $\alpha_{g}^2=(M\mu_S)^2$ \cite{Baumann:2018vus}. The contribution of the magnetic field will appear, to first order in $k_{\rm a}$, at the fine-structure level (in $\alpha_g$), since the dominant contribution of the driving term in Eq.~\eqref{eq:KG_Wald} is $a M/r^2 \sim \tilde{a} \alpha_{g}^4$. Even though it is a subdominant effect, it could potentially be important for the consistent calculation of the level mixing in binary systems~\cite{Baumann:2018vus}, for large magnetic fields and/or axion-photon coupling constant.

\subsection{New BH solutions for scalar couplings}

\subsubsection{End-state of the RN instability}

As already shown for the axionic couplings, the instability of the RN geometry discussed in Section~\ref{sec:inst_RN} suggests the existence of new BH solutions, branching-off from the RN solution\footnote{We note that the fact that the theory we consider has hairy BH solutions was already pointed out in Ref.~\cite{Gubser:2005ih}, although they were not explicitly constructed. In Ref.~\cite{Gubser:2005ih} only BHs with magnetic charge were constructed.}. We now construct those solutions for the scalar coupling described by the action~\eqref{eq:MFaction} with $k_{\rm a}=0$ but $k_{\rm s}\neq 0$.

Similar solutions were recently constructed for theories with a coupling of the form $e^{-\alpha \Psi^2}F_{\mu\nu}F^{\mu\nu}$~\cite{Herdeiro:2018wub}, with $\alpha$ a coupling constant. In particular in Ref.~\cite{Herdeiro:2018wub} it was shown that the scalarized BH solutions are stable against spherically symmetric perturbations and found strong evidence that the solutions are the end-state of the instability by performing fully non-linear evolutions of the instability. In fact, when $\alpha\ll 1$ and $k_{\rm s}\ll 1$ the coupling of Ref.~\cite{Herdeiro:2018wub} is equivalent to the quadratic coupling that we here study, hence all the conclusions should carry through. Here we only consider spherically symmetric spacetimes, and solutions for which the scalar field is nodeless. Solutions with nodes, and non-spherically symmetric (but static) solutions -- which correspond to unstable polar modes with $l>0$ -- have also been constructed in Ref.~\cite{Herdeiro:2018wub}. 

\begin{figure*}[t]
\begin{center}
\begin{tabular}{cc}
\includegraphics[width=8.5cm]{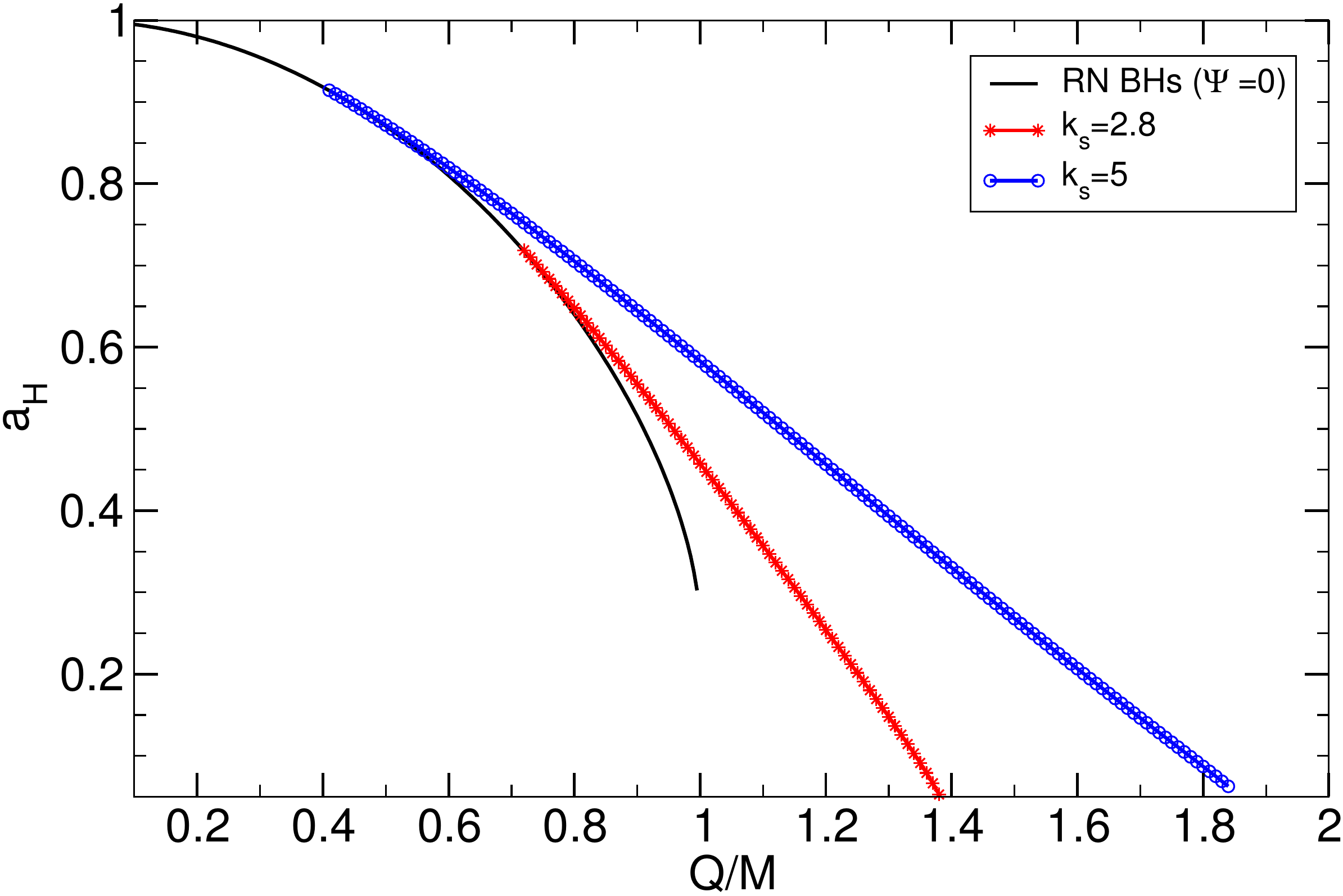}&
\includegraphics[width=8.3cm,trim={0 2 0 2},clip]{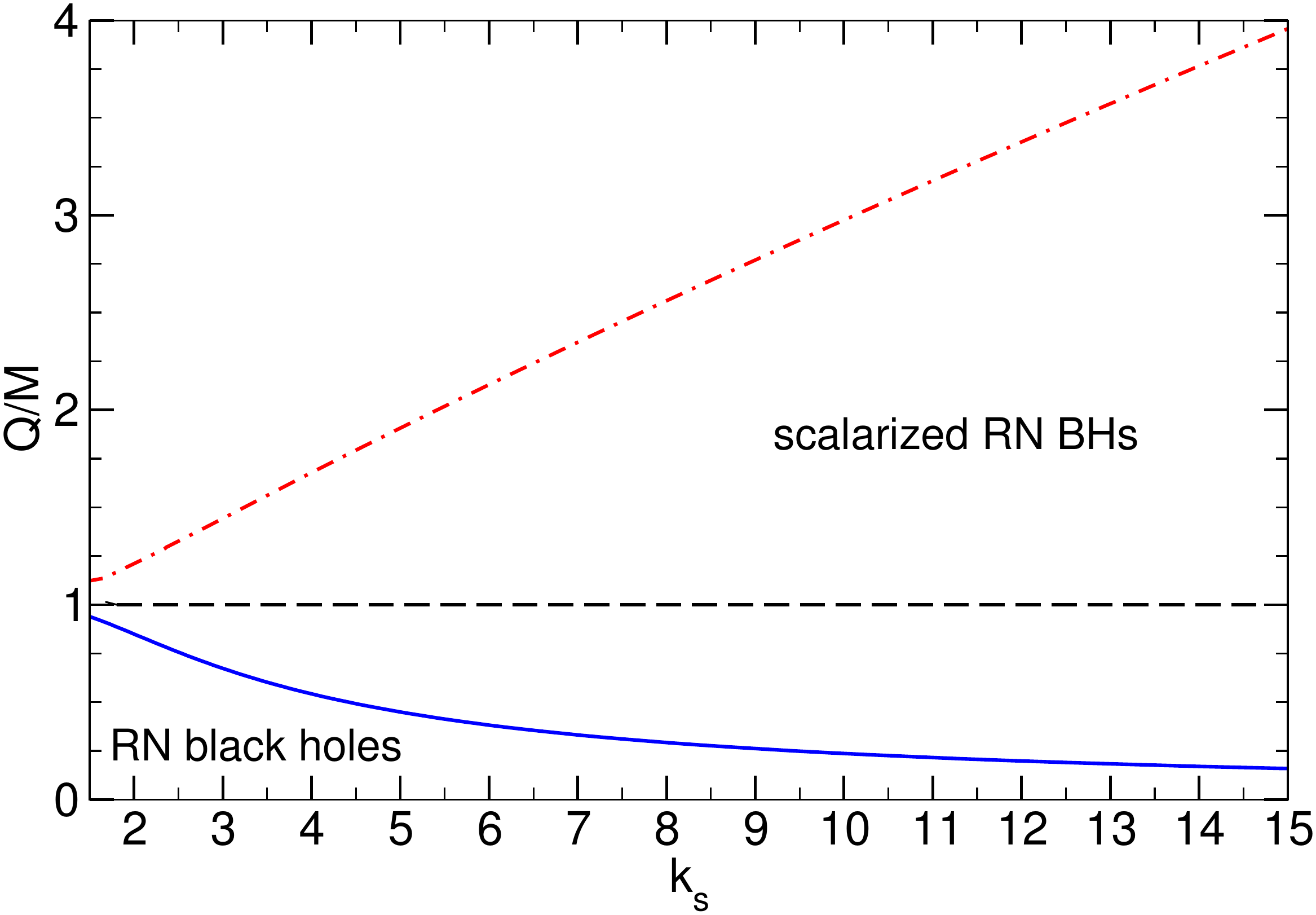}
\end{tabular}
\caption{Left panel: Dimensionless horizon area $a_{\rm H}= A_{\rm H}/(16\pi M^2)$,  with $A_{\rm H}$ the horizon area, of the nodeless scalarised BHs, for different values of $k_{\rm s}$ as a function of the BH's charge-to-mass ratio $Q/M$. For a given $k_{\rm s}$ solutions only exist above a critical $Q/M$ that can be read off from Fig.~\ref{instability_RN_scalarp2}. Numerics indicate that the scalarised solutions only exist up to maximum $Q/M$ where the horizon area vanishes. Right panel: Domain of existence of the nodeless scalarised BHs. RN BHs exist for any $k_{\rm s}$ as long as $Q/M\leq 1$, but are unstable above the solid blue line. Above that line scalarised BHs exist up to the dashed red line which marks the approximate value of $Q/M$ where the horizon area of the nodeless scalarised BHs tends to zero.
\label{RN_area}}
\end{center}
\end{figure*}

\begin{figure}[htb]
\begin{center}
\begin{tabular}{cc}
\includegraphics[width=8.5cm]{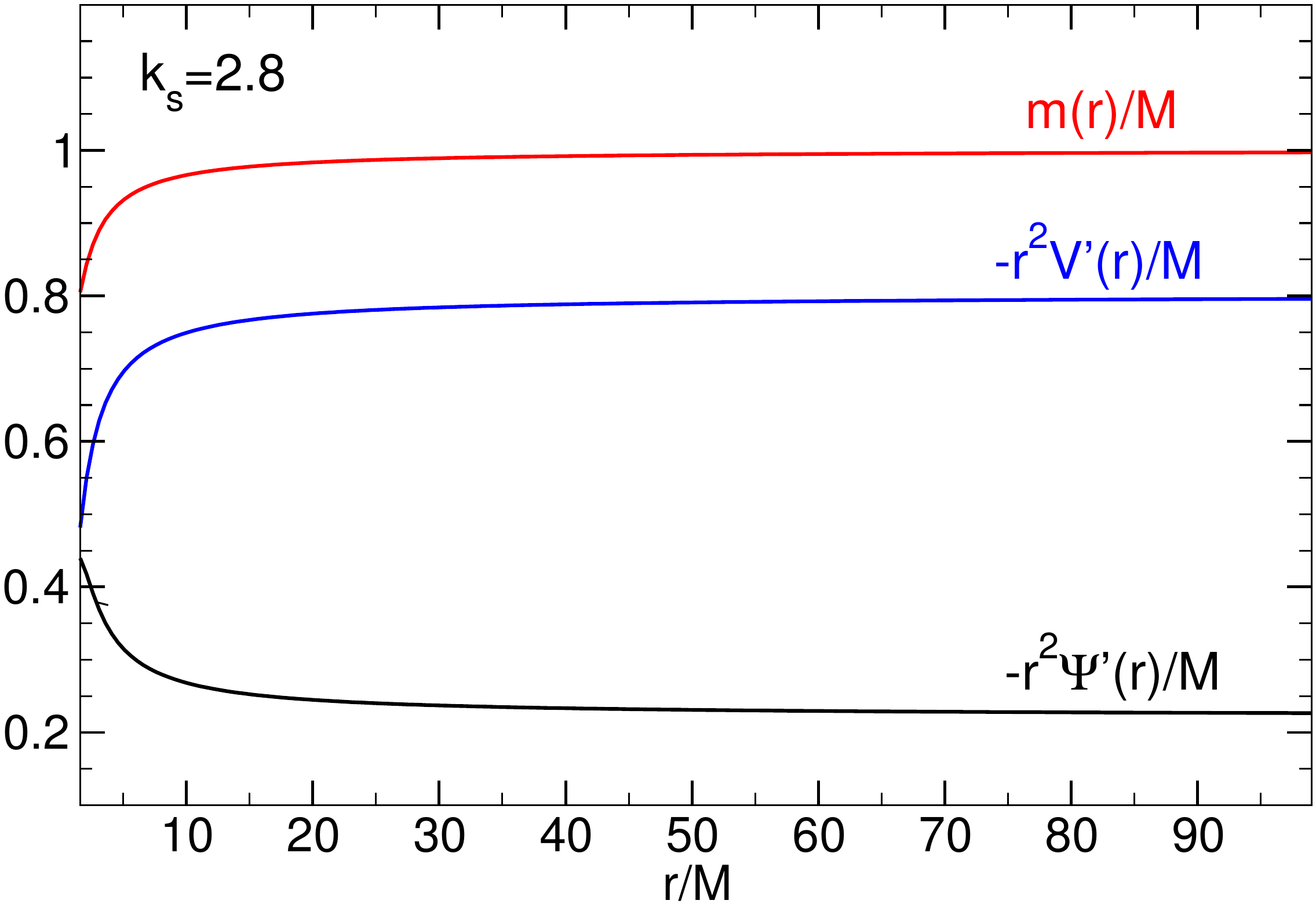}
\end{tabular}
\caption{Scalarized BH for $Q/M=0.8$ and $k_{\rm s}=2.8$. The scalar and vector fields decay as $1/r$ at spatial infinity therefore this solution carries both a scalar and an electric charge.\label{RN_sol_p2}}
\end{center}
\end{figure}

The scalar field is given by $\Psi \equiv \psi(r)$ while the most general spherically symmetric metric can be written as
\be
ds^2=-e^{-2\delta(r)}N(r)dt^2+\frac{dr^2}{N(r)}+r^2d\Omega^2\,,
\ee
where $N(r)=1-2m(r)/r$. The vector field is given by $A_{\mu}dx^{\mu}=V(r) dt$.
After substitution in the field equations~\eqref{eq:MFEoMgen}, we get (for generic values of $p$ and $\mu_S$):
\beq
&&\left(e^{-\delta}r^2 N \psi' \right)'+\frac{p\,r^2}{2}e^{\delta}k_{\rm s}^p \psi^{p-1}V'^2-e^{-\delta}r^2\mu_{S}^2\psi=0\,,\\
&& m'-\frac{r^2}{2}\left[N\psi'^2 + e^{2\delta}V'^2\left(1+k_{\rm s}^p\psi^{p}\right) + \mu_{S}^2\psi^2\right]=0\,,\\
&& \delta'+r\psi'^2=0\,,\\
&& V'=\frac{Q}{r^2}e^{-\delta}\frac{1}{1+k_{\rm s}^p\psi^{p}}\,.
\eeq
By imposing the existence and regularity of the solution across an event horizon at $r=r_H$,
in addition to regularity at infinity these equations can be solved using a standard shooting method. We refer the reader to Ref.~\cite{Herdeiro:2018wub} for more details. Our results for $p=2$ and $\mu_S=0$ are summarized in Figs.~\ref{RN_area} and~\ref{RN_sol_p2}. Similar solutions can be constructed for $p>2$ and $\mu_S\neq 0$.

The left panel of Fig.~\ref{RN_area} shows the dimensionless horizon area $a_{\rm H}= A_{\rm H}/(16\pi M^2)$, with $A_{\rm H}=4\pi r_H^2$, of the scalarized solutions as a function of the BH's charge-to-mass ratio $Q/M$. The right panel shows part of the domain of existence of the scalarized solutions. As can be seen, for a given $k_{\rm s}$, scalarized solutions only exist above a critical $Q/M$. The value at which these solutions branch-off the RN geometry agrees with the onset of the RN instability shown in Fig.~\ref{instability_RN_scalarp2}. In agreement with Ref.~\cite{Herdeiro:2018wub} we find that solutions can be overcharged $Q>M$ and for a given $k_{\rm s}$ exist up to a maximum value of $Q/M$ at which the numerics indicate that the horizon area vanishes.

In addition to the BH electric charge $Q$ and the mass $M$ these solutions have a scalar charge $Q_s$. This can be seen in Fig.~\ref{RN_sol_p2} where we show a specific scalarized solution. The scalar field can only be supported if $Q\neq 0$, i.e. $Q_s\to 0$ when $Q\to 0$, in agreement with the expectation that a Schwarzschild BH is a stable solution of the field equations.
Importantly, these results show that, for part of the parameter space, RN and scalarised solutions co-exist with the same global charges.

%

\subsubsection{Charged BH solutions for $p=1$}\label{RNp1}

For $p=1$ RN is not a solution of the field equations~\eqref{eq:MFEoMgen}. Charged BHs in this theory necessarily carry scalar hair for any value of $k_{\rm s}$ since a non-vanishing $F_{\mu\nu}F^{\mu\nu}$ will source the scalar field equation~\eqref{eq:MFEoMScalar}. This case is analogous to the more familiar Einstein-Maxwell-dilaton theories. In fact, Einstein-Maxwell-dilaton theories have a coupling of the form $e^{-2\alpha\Psi}F_{\mu\nu}F^{\mu\nu}$, with $\alpha$ as a coupling constant, and so are equivalent to the Lagrangian~\eqref{eq:MFaction} with $p=1$, $k_{\rm a}=0$ and $\mu_S=0$ when $\alpha\ll 1$ and $k_{\rm s} \ll 1$. For Einstein-Maxwell-dilaton theory, closed exact analytical BH solutions have been found~\cite{Garfinkle:1990qj}. For our specific coupling we were unable to find exact analytical solutions. However, since RN is a solution for $k_{\rm s}=0$, perturbative solutions around RN can be found in an expansion in $k_{\rm s}$. For spherically symmetric solutions, to first order in $k_{\rm s}$ and generic BH electric charge $Q$ corrections to the RN solution appear only at the level of the scalar field and are given by solving the ODE (for $\mu_S=0$ and $k_{\rm a}=0$):
\beq
&&\Psi=k_{\rm s}\psi(r)\,,\\
&&(r^2-2Mr+Q^2)\psi''+2(r-M)\psi'=-\frac{Q^2}{2r^2}\,.
\eeq
Imposing regularity at the horizon and vanishing scalar field at infinity the solution is given by
\be
\Psi =-\frac{k_{\rm s}}{2}\log\left(1-\frac{r_-}{r}\right)\,,
\ee
where $r_{-}=M-\sqrt{M^2-Q^2}$. One may easily find the first corrections to the metric and vector potential by expanding the solution around $Q=0$. In particular up to second order in $k_{\rm s}$ and fourth order $Q$ the vector potential is given by:
\be
A_\mu dx^{\mu}= \left(\frac{Q}{r}-k_{\rm s}^2\frac{Q^3}{8M r^2}\right) dt\,,
\ee
while the metric components are given by 
\beq
m(r)&=&M-\frac{Q^2}{2r}-k_{\rm s}^2Q^4\frac{r-3M}{32M^2r^2}\,,\\
\delta(r)&=&k_{\rm s}^2\frac{Q^4}{32M^2r^2}\,.
\eeq
We note that the scalar field decays as $~1/r$ at large distances and therefore these solutions carry scalar charge $Q_s$ given by $Q_s=k_{\rm s}Q^2/(4M)$.

\section{Bursts of light from scalar clouds}\label{sec:Time_domain_studies}
%
\begin{figure*}[htb]
\begin{tabular}{cc}
\includegraphics[width=8.5cm]{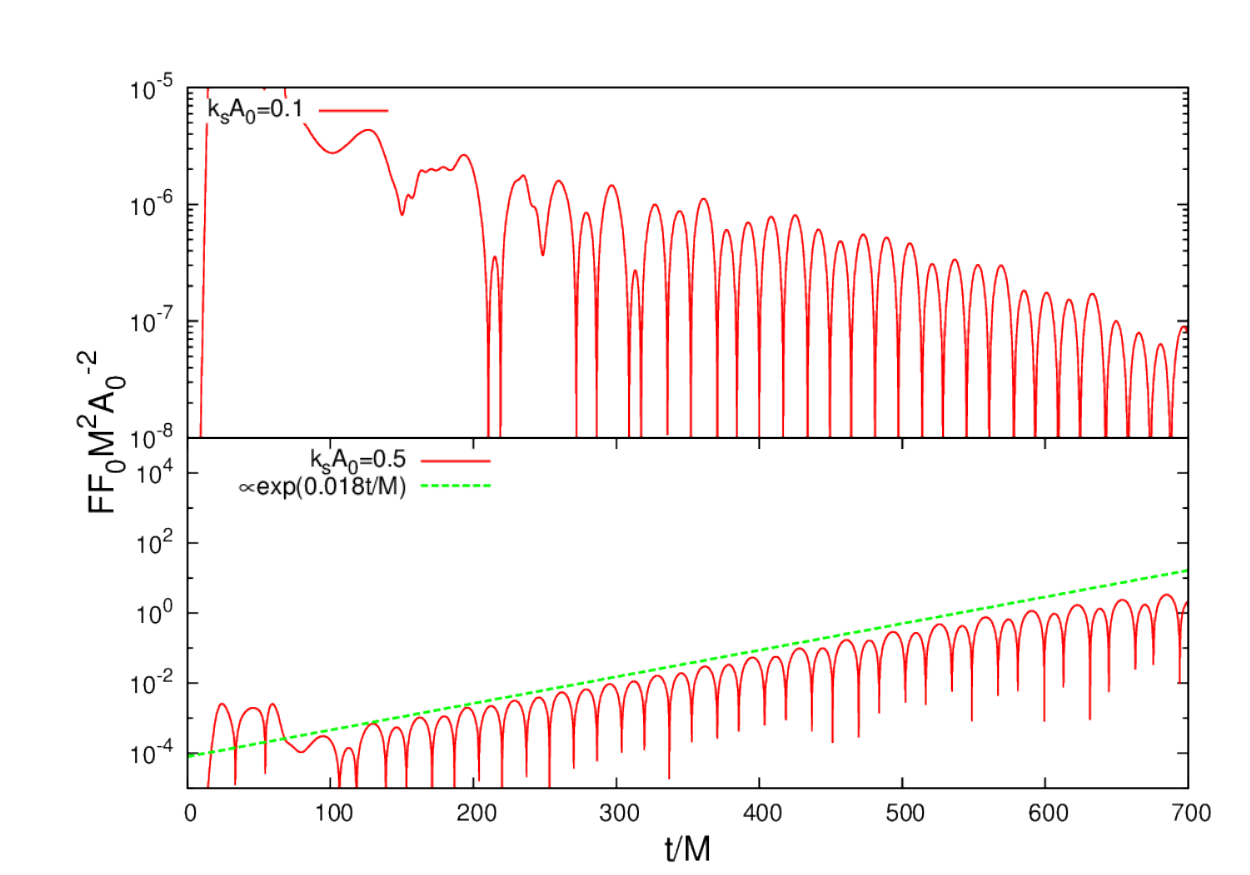}
&\includegraphics[width=8.5cm]{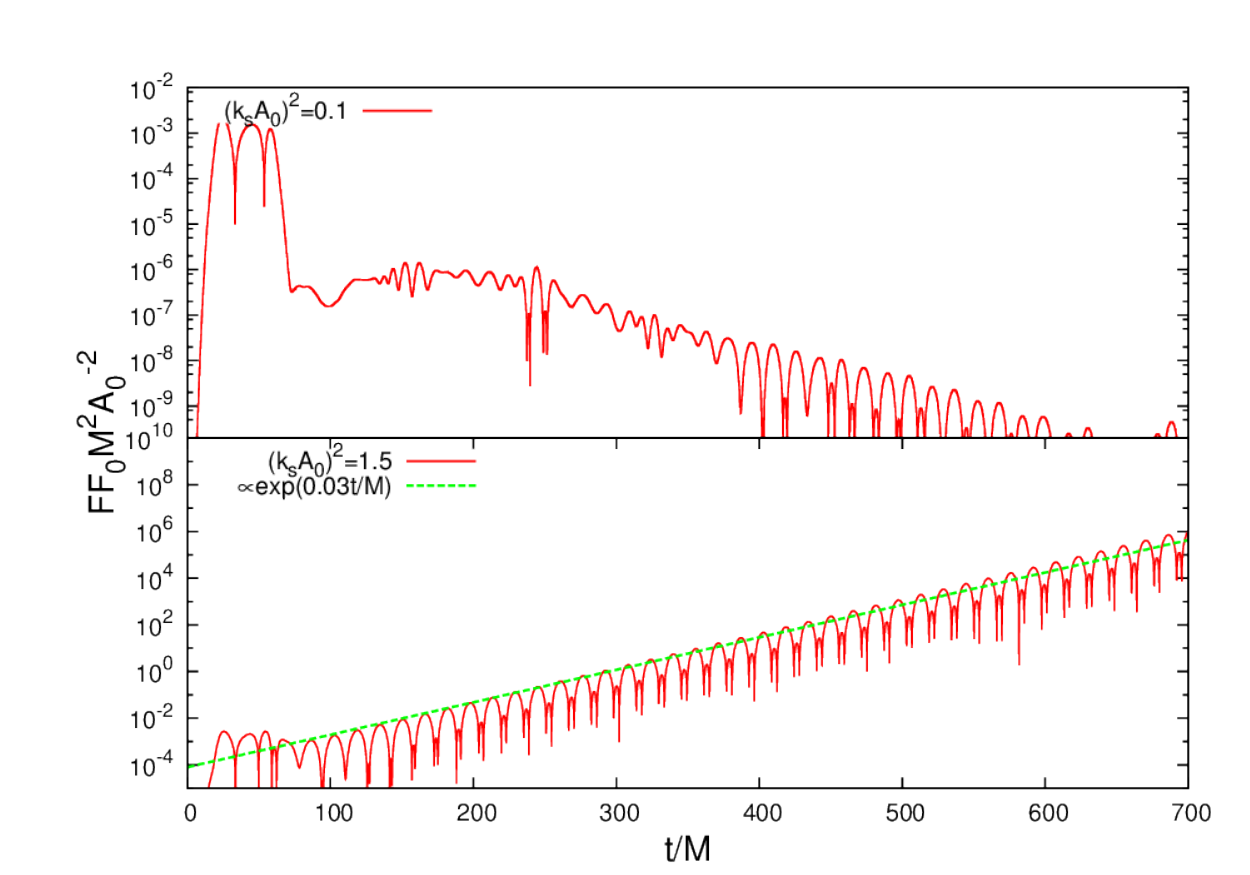}
\end{tabular}
\caption{
Time evolution of the Maxwell scalar $FF_{0}$ (see Appendix \ref{app:TDS-notation} for definition of notation) measured
at $r=20M$ for extended initial-data profiles (Eq.~\eqref{ID_extended} in Appendix \eqref{ssec:InitialData}), for $p=1,2$ (left and right panel respectively). The spacetime is flat and the scalar is not evolved. The initial data corresponds to a 
Gaussian EM field with width $w$, gaussian-centered radius $r_0$ and amplitude of $(w,r_{0},E_{0})=(5M,40M,0.001)$. In both panels the mass coupling is $\mu_{S} M=0.2$. 
\label{graph_fixed_scalar_sample_log_p}}
\end{figure*}
\begin{figure*}[htb]
\begin{tabular}{cc}
\includegraphics[width=8.5cm]{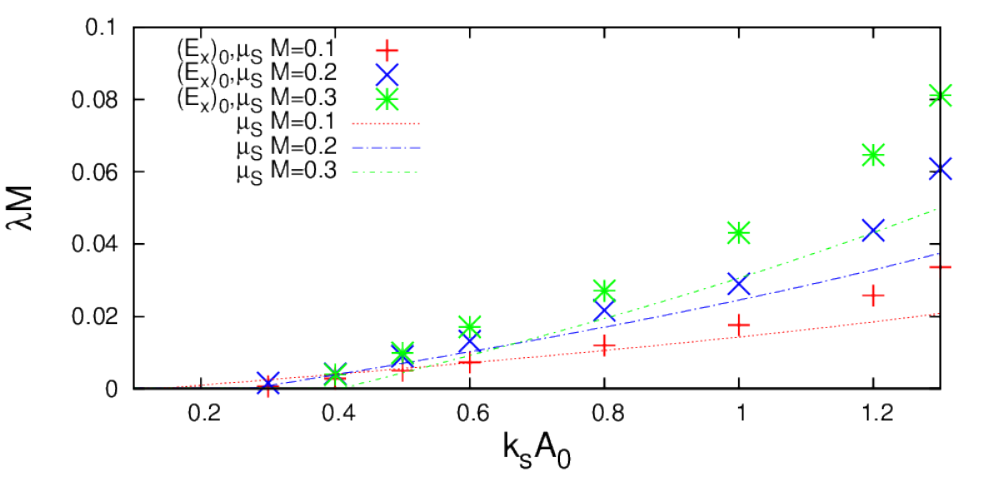}
&\includegraphics[width=8.5cm]{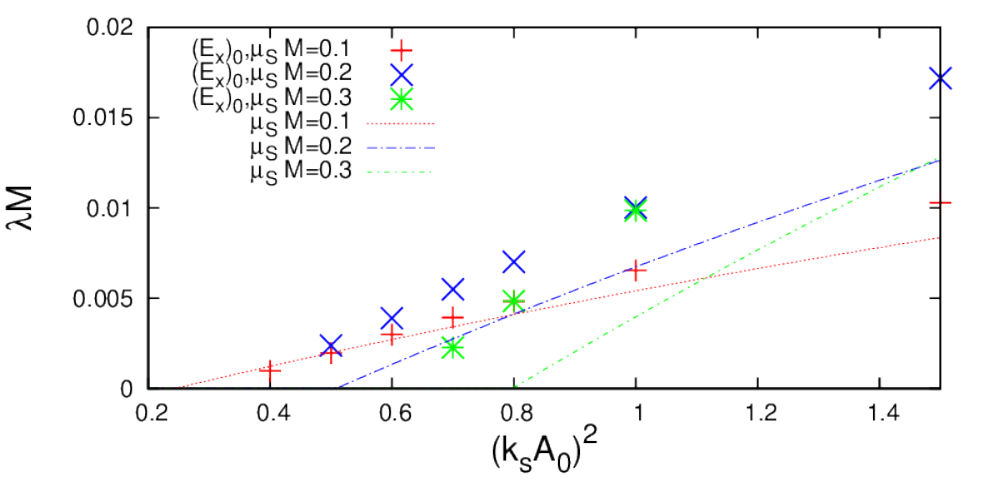}
\end{tabular}
\caption{Growth rates $M\lambda$ as a function of the coupling parameters for $p=1$ (left) and $p=2$ (right) for a Minkowski background. Crosses stand for numerically extracted rates, dashed lines are our analytical estimates, which to first order in the coupling are described by Eqs.~\eqref{ana_rate1}-\eqref{ana_rate2} (a full description of the perturbative framework can be found in Appendix \ref{app:RG_time_scales_2nd_sc}). Our results are consistent with the existence of a critical coupling below which no instability is triggered, and well described by our analytical estimates in the small coupling regime.
\label{graph_lambda_p}}
\end{figure*}

Axion and axion-like light particles -- even with negligible initial abundance -- trigger superradiant instabilities around massive, spinning BHs~\cite{Detweiler:1980uk,Cardoso:2005vk,Dolan:2007mj,East:2017ovw,East:2018glu,Brito:2015oca}. The instability extracts rotational energy away from the spinning BH and deposits it into a cloud of scalars, with a spatial extent $\sim 1/(M\mu_S^2)$~\cite{Brito:2015oca}. Over long timescales, when the mass of the cloud is sufficiently large, GW-emission becomes important, and leads to a secular spin-down of the cloud (and BH), and a consequent cloud decay. Such systems are a promising source of GWs, both as resolvable and as a stochastic background, that can be detected with current and future detectors~\cite{Arvanitaki:2010sy,Arvanitaki:2014wva,Brito:2014wla,Arvanitaki:2016qwi,Baryakhtar:2017ngi,Brito:2017wnc,Brito:2017zvb,Hannuksela:2018izj}.

Our current understanding of the evolution of superradiant instabilities and accompanying GW emission neglects the coupling to matter, expected to be very weak. Arguments based on flat-space calculations similar to those worked out in Section~\ref{Sec.Flat-space instabilities}, suggest that when the axion strength exceeds a critical value
(or in other words when the cloud extracts too much energy) an instability is triggered that might give rise to large amounts of EM 
radiation being emitted from BH systems~\cite{Rosa:2017ury}. Recently, these conjectures were shown to be true by some of us, through the evolution of Maxwell's field equations coupled to an axion field in a Kerr background.
In particular it was shown that for critical values of the coupling $k_{\rm a}\Psi_0$, EM fields are spontaneously excited in such environments, even at a classical level~\cite{Ikeda:2018nhb}. These instabilities can be indeed completely
understood in the context of classical field theory, owing to the bosonic nature of axions and photons, that allows buildup of macroscopic numbers of particles.

Here, we will extend the results of Ref.~\cite{Ikeda:2018nhb} to scalar couplings and provide some analytical understanding of the mechanism.
We refer the reader to Appendix~\ref{app.Formulation as Cauchy problem} for the numerical formulation and initial data construction.
We will always focus on a background axion or scalar field which is the product of the evolution of superradiant instabilities around spinning BHs. The growth of the scalar or axion due to superradiance is extremely slow to perform in a full nonlinear evolution; 
but see Refs.~\cite{Okawa:2014nda,Zilhao:2015tya,East:2018glu}.
Thus, our setup is that of an axion fully grown by superradiance to some value, at which point we start monitoring the coupled system of Maxwell-Klein Gordon equations, in a fixed background geometry~\footnote{In Ref.~\cite{Ikeda:2018nhb} superradiant-like growth was also monitored using a modification of the Klein-Gordon equation. The modification mimics and gives rise to superradiance, and was used in Zel'dovich's pioneering work on rotational superradiance~\cite{zeldovich1,zeldovich2,Brito:2015oca}. This study showed that the physical results did not change with respect to frozen-superradiance evolutions. However, while the results of Ref.~\cite{Ikeda:2018nhb} are compelling and lead to periodic bursts of EM radiation, an alternative scenario is still not ruled out: that if the superradiant-growth timescale is extremely large the signal is less burst-like than observed in Ref.~\cite{Ikeda:2018nhb}.\label{footnoteX}} following the approach of Ref.~\cite{Witek:2012tr}.
The initial data for the vector consists of a small azimuthal electric field of the form (see Appendix~\ref{ssec:InitialData} for further details)
\be
E^{\varphi}=E_0e^{-(r-r_0)^2/w^2}\,,\quad{\rm axions}\,,
\ee
and
\be
E^{\varphi}=\frac{E_0}{1+k^p_{\rm s}\Psi^p}e^{-(r-r_0)^2/w^2}\Theta(\theta)\,,\quad{\rm scalars}\,,\label{Eq. initial data Ephi in text}
\ee
The $\Theta$ profile is shown in Appendix~\ref{ssec:InitialData}, and is used for completeness, since all our results
are initial-data independent at the qualitative level.

\subsection{Flat space analysis}
Our main purpose is to show that EM instabilities may arise when axions or scalars exist, and they couple strongly to the Maxwell field. We will do this by order of complexity. Here, we artificially use a Minkowski background and we fix the scalar field to have the profile appropriate for clouds around BHs~\cite{Brito:2014wla},
\begin{equation}
\Psi=A_0 r M \mu_S^2 \exp{\Big(-\frac{1}{2}rM\mu_S^2\Big)}\cos{(\phi-\omega t)}\sin{\theta}\,.\label{eq:211psi}
\end{equation}
Here, $A_0$ is amplitude of the field. For further details we refer the reader to Appendix \ref{sec:App_gravitational_atom},
and Refs.~\cite{Brito:2014wla} and~\cite{Ikeda:2018nhb}. The purpose is to show that an instability exists even in this setup, but now with a critical threshold. In other words, the results worked out in Section~\ref{Sec.Flat-space instabilities} in a Minkowski background generalize, except that the non-homogeneous nature of the scalar or axion results in a critical coupling below which no instability is triggered. These results were reported for axions in our recent Letter~\cite{Ikeda:2018nhb}.

We solved the (Maxwell) evolution equation of the EM field with a fixed scalar field for $p=1, 2$ and mass couplings $\mu_S M=0.1,0.2,0.3$,
where $M$ is the BH mass that supports the solution~\eqref{eq:211psi}. Our results are summarized in Figs.~\ref{graph_fixed_scalar_sample_log_p}~--~\ref{graph_lambda_p}, and are consistent with the results we obtained for axion couplings recently~\cite{Ikeda:2018nhb}.

The novel feature with respect to the homogeneous background scalar of Section~\ref{sec:mink_axion} is the existence of a critical coupling $k_{\rm s} A_0$ below which no instability occurs. This is apparent in Fig.~\ref{graph_fixed_scalar_sample_log_p} for both $p=1$ and $p=2$ (we stress that the results for axionic couplings can be found in Ref.~\cite{Ikeda:2018nhb}). The critical value is estimated below with simple analytical arguments. If the coupling $k_{\rm s} A_{0}$ is smaller than the threshold, the EM field dissipates away. This feature is induced by finite-size effects of the scalar cloud, as we argue below. 

At large enough couplings, all initial conditions lead eventually to an instability (and exponential growth of the EM field), examples are shown in the bottom panels of Fig.~\ref{graph_fixed_scalar_sample_log_p}. 
The growth rate depends very weakly on the initial data and on the coordinate at which the EM field is extracted. A closer inspection of instability rates allows us to estimate the critical coupling value. The rates are shown in Fig.~\ref{graph_lambda_p} for different couplings, which also strongly indicates the existence of a critical threshold.

\subsection{Kerr BHs}
%
\begin{figure*}[htb]
\begin{tabular}{cc}
\includegraphics[width=8.5cm]{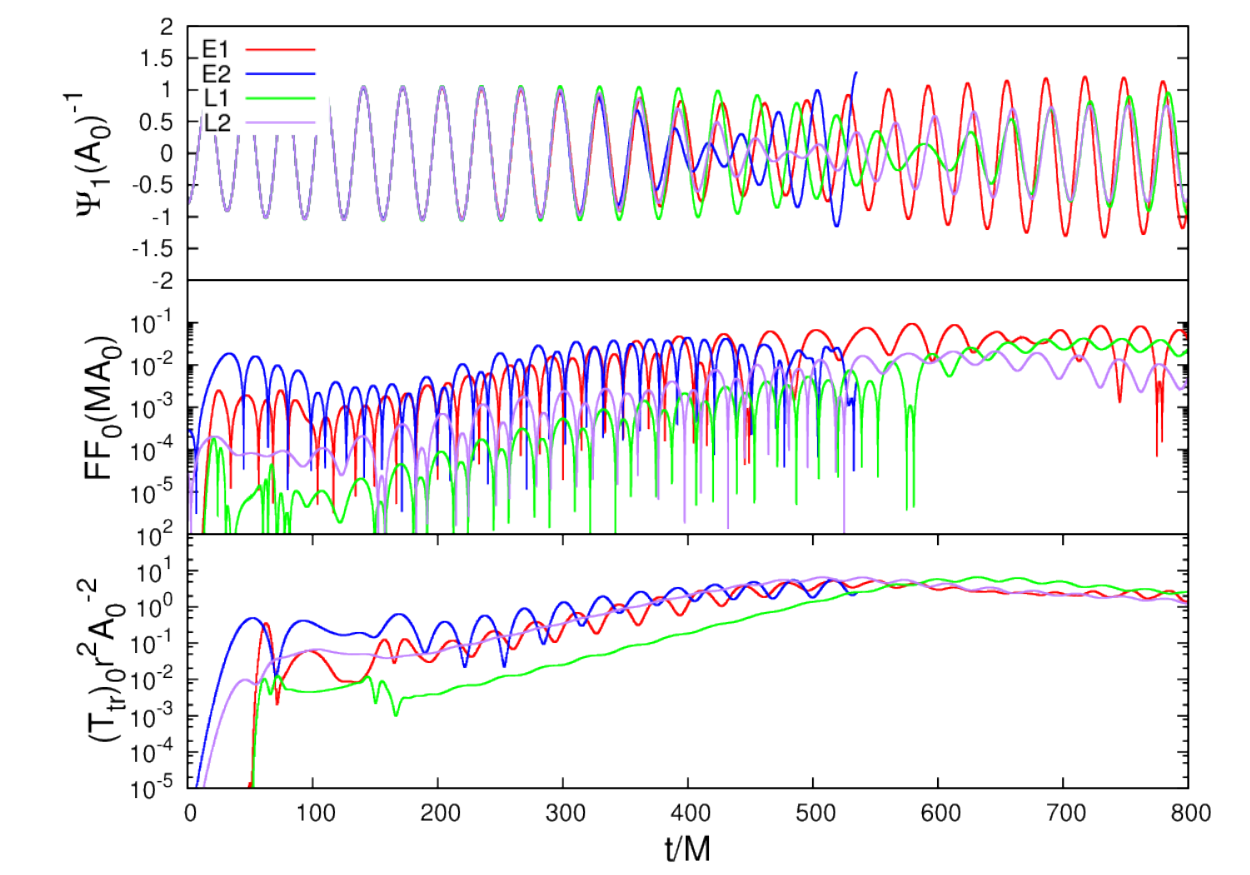}&
\includegraphics[width=8.5cm]{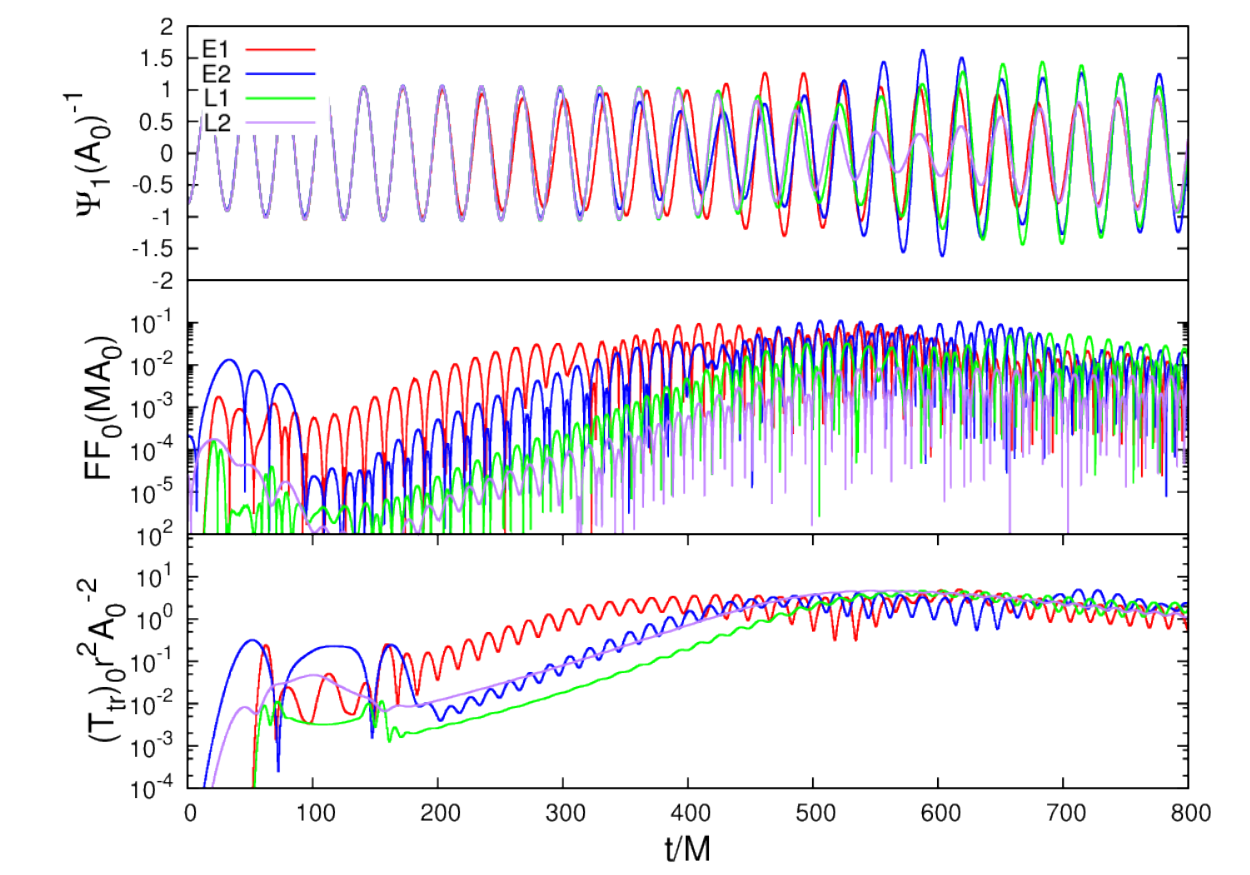}
\end{tabular}
\caption{
Time evolution of the dipolar component of the scalar field, $\Psi_{1}$ (top panel, extracted at $r=20M$, see Appendix \ref{app:TDS-notation} for notation), of the monopolar component of the Maxwell invariant $FF_{0}$ (middle panel, extracted at $r=20M$), and $(T_{tr})_{0}$ (bottom panel, extracted at $r=100M$) for $p=1$ (left panels) and $p=2$ (right panels). The mass of the scalar is $\mu_S M=0.2$, the coupling $k_{\rm s}A_{0}=0.5,k_{\rm s}^2A_{0}^2=1.0$ ($p=1,2$ respectively), and the spin parameter is $a=0.5M$.
The initial data is either of the extended (E) or localized (L) type as defined in Eqs.~\ref{Eq.Initial data of Er}, \eqref{Eq.Initial data of Etheta}, \eqref{Eq.F profile} and \eqref{ID_localized}, and described by a Gaussian centered at $r_0=40M$ and an amplitude of $E_0=10^{-3}$. For $E_1,\,L_1$ the gaussian width is $5M$, for $E_2,\,L_2$ it is $20M$.
\label{graph_a05_mu02_ks05_p1_phi_FF_Sr}
}
\end{figure*}
We have just discussed the exponential growth of an EM field around a ``frozen'' axionic or scalar cloud in a flat space background.
These results suggest that when the effective coupling is larger than a threshold value, the EM field may grow exponentially -- fed by the axionic cloud, which itself grew through super-radiance and extracted its energy from the spinning BHs. Here, we confirm this scenario with a fully numerical simulation around Kerr BHs
(the geometry is kept fixed, but the coupled Maxwell-scalar system is evolved; following the approach of~\cite{Witek:2012tr}).
We refer the reader to Appendix \ref{App:Kerr_coordinates} for notation on the Kerr metric representations, 
and summarize the formulation and initial data in Appendix \ref{app.Formulation as Cauchy problem}.

We solved the evolution equations for $M\mu_S=0.2, a=0.5M$ (we also studied higher spins, the results are qualitatively the same), 
the results are summarized in Fig.~\ref{graph_a05_mu02_ks05_p1_phi_FF_Sr} for scalar couplings with $p=1$ and $p=2$.
As expected from the previous flat-space analysis, for small enough couplings any small EM disturbance dissipates away, and the profile of the axionic or scalar cloud is basically undisturbed. On the other hand, when the coupling is larger than a threshold, the EM field grows exponentially. As shown in Fig.~\ref{graph_a05_mu02_ks05_p1_phi_FF_Sr}, for large couplings an instability is indeed triggered. Because the instability acts to produce $k\sim \mu_S/2$ vector fluctuations (for $p=1$), at the nonlinear level these backreact on the scalar field, producing transient clumps of scalar field on these scales. This translates into an increase of the scalar, when observed sufficiently close to the BH, as seen in the upper panels of Fig.~\ref{graph_a05_mu02_ks05_p1_phi_FF_Sr}. On long timescales, the instability extracts energy from the scalar cloud and eventually lowers the effective coupling to sub-threshold values, leading to a now stable cloud. On even longer timescales, superradiance will grow the scalar to super-threshold values and the cycle begins again, as demonstrated in the axion scenario in Ref.~\cite{Ikeda:2018nhb} (see footnote \ref{footnoteX}). Similar effects have been found for scalar condensates with a self-interacting potential, but in the absence of couplings to the EM sector~\cite{Yoshino:2012kn,Yoshino:2015nsa}. 

We would like to highlight a potential issue with the scalar couplings in general, and that clearly shows up when $p=1$.
When the effective coupling $k_{\rm s}\Psi$ is of order unity, the kinetic term (left hand side of Eq.~\eqref{eq:MFEoMVector}) for the vector field can vanish and the system becomes strongly coupled. The evolution in such case is ill-defined. In particular, we find for example that we cannot evolve $E2$ (see definition in caption of Fig.~\ref{graph_a05_mu02_ks05_p1_phi_FF_Sr}) in Fig.~\ref{graph_a05_mu02_ks05_p1_phi_FF_Sr} past $t=520M$, for this reason.
It is possible that the dynamics of the gravity sector (neglected in this work) cure such anomalies,
for example by producing BHs close to the threshold. Another possibility is that coupling to fermions will ensure that Schwinger-type creation works to prevent the EM field to ever approach such large values. The calculation of the time evolution near the strong coupling is beyond the purpose of our paper.

\subsection{A simple analytic description of the results} 
%
\begin{figure}
\centering
\includegraphics[width=\columnwidth]{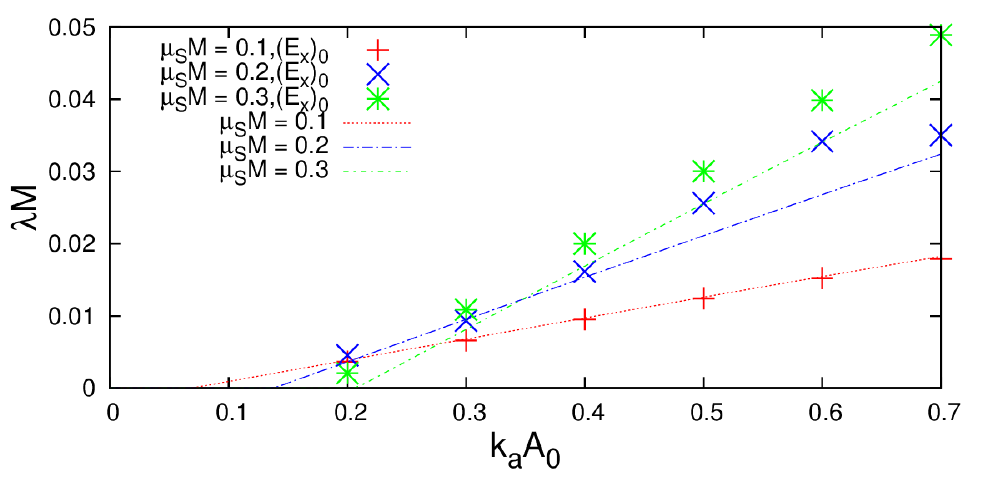}
\caption{
Instability rates for axionic couplings, in the presence of a background axion described by the cloud \eqref{eq:211psi}, for a Minkowski background. The analytical estimates for the instability rate $\lambda$ for axionic couplings, as given to first order by Eq.~\eqref{ana_rate0} (dashed lines, full expansion is described in Appendix \ref{app:RG_time_scales_2nd_ax}), are compared with the numerical results of Ref.~\cite{Ikeda:2018nhb} (crosses). We find good agreement between our analytic estimates and numerical data for small mass ($M\mu_{S}\sim0.1$) or small axion couplings.
\label{fig:Mink_axion_rates}}
\end{figure}
Compared to the flat space analysis from Section \ref{sec:mink_axion}, the (localized) axion ``cloud'' configuration introduces one more timescale in the problem, that of the time $d$ needed for photons to leave the axion configuration, where $d$ is a measure of the configuration size. Thus, there is another rate in the problem,
\begin{equation}
\lambda_\gamma \sim \frac{1}{d}\,.\label{eq:inst_rate2}
\end{equation}
If $\lambda_\gamma>\lambda_\ast$, with $\lambda_\ast$ being the estimate of the EM field instability rate for the homogeneous condensate, photons leave the configuration before the instability ensues and the effective rate of the instability is zero. In the other extreme, we can approximate the rate of the dominant instability by \cite{Hertzberg:2010yz,Hertzberg:2018zte}
\begin{equation}\label{eq:inst_rate3}
\lambda \approx \lambda_\ast\,\langle \Psi \rangle-\lambda_\gamma, 
\end{equation}
where $\langle \Psi \rangle$ is some estimate of the average value of the axion field, to be implemented in the expression obtained for the homogeneous case (see Eq.~\eqref{rate_axion0}):
\begin{equation}
\lambda_\ast\,\langle \Psi \rangle    \approx \frac{1}{2}k_{\rm a}\langle \Psi \rangle\mu_{\rm S}\,.
\end{equation}
Although the rate estimates were derived in flat space, the system under consideration is mostly in a weak field regime and we expect that they will provide a good description even in the context of instabilities around Kerr BHs.

Let us therefore consider the dominant mode in the gravitational atom \eqref{eq:211psi}, frozen and embedded in Minkowski spacetime, and estimate the instability rate. For the measure of $d$ we use the full-width-at-half-maximum (FWHM) of the function \eqref{eq:211psi}
\begin{equation}
d \approx \frac{4.893}{M\mu_{\rm S}^2}. 
\end{equation}
For the estimate of the field value we take the radial mean of the field on the FWHM and maximal contribution from the harmonic part of the function
\begin{equation}
\langle \Psi \rangle \approx (1/d)\int_{\text{FWHM}}|\Psi(r)|dr \approx 0.592 A_0\,.\label{ana_rate0}
\end{equation}
These estimates are in very good agreement with the results from the Letter~\cite{Ikeda:2018nhb} as can be seen in Fig.~\ref{fig:Mink_axion_rates}.
In addition, a cutoff coupling below which no instability arises shows up naturally, explaining all the previous numerical results. 
In summary, a very simple and elegant analytic formula explains most of the results that we observe numerically.

The results above were worked out for the axionic coupling, but the underlying physics and mechanism remains the same for scalar-type couplings. Accordingly, we expect that the rate of the dominant instability is given by Eq. \eqref{eq:inst_rate3}. Using Eq.~\eqref{eq:rate_scalar_p_1}, we find,
\begin{equation}
\lambda_\ast\,\langle \Psi \rangle_{p=1} \approx \frac{1}{4}k_{\rm s}\langle \Psi \rangle\mu_{\rm S}\,,\label{ana_rate1}
\end{equation}
for $p=1$. Similarly, using Eq.~\eqref{eq:rate_scalar_p_2} we find for $p=2$,
\begin{equation}
\lambda_\ast\,\langle \Psi \rangle_{p=2}  \approx \frac{1}{4}\left(k_{\rm s}\langle \Psi \rangle\right)^2\mu_{\rm S}\,.\label{ana_rate2}
\end{equation}
These estimates are shown together with numerical data in Figs.~\ref{graph_lambda_p}. 
Notice how such a simple estimate agrees very well with the full numerical evolution
in the small coupling regime where the perturbative approximation is valid.

\subsection{A note on plasma effects} \label{sec:plasma}
%
\begin{figure}[htb]
\begin{tabular}{cc}
\includegraphics[width=8.5cm]{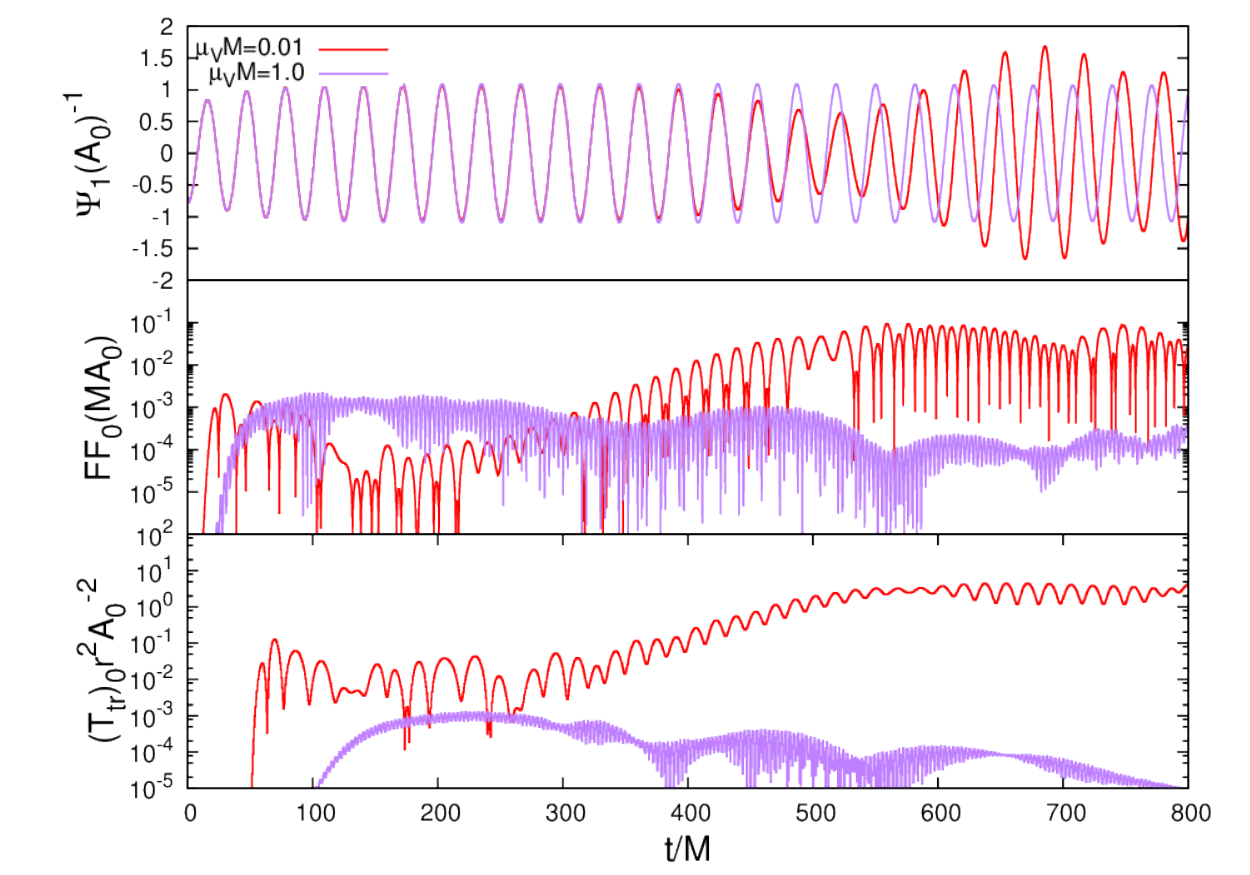}
\end{tabular}
\caption{
Time evolution of the massive scalar -- massive vector field system around a Kerr BH with $p=2$ and coupling $k_{\rm s}^{2}A_{0}^{2}=1.0$, in which the initial data is an extended profile with $(r_{0},w,E_{0})=(40M, 5M,0.001)$. Here the scalar mass is $\mu_{\rm S}M=0.2$ and the scalar cloud is evolving around a Kerr BH with $a=0.5M$. The notation for the $y-$axis is explained in Appendix \ref{app:TDS-notation}.\label{graph_Proca_a05_mu02_ks05_p2_phi_FF_Sr_E1}
}
\end{figure}

Thus far, the system was assumed to evolve in a vacuum environment, when in reality the universe is filled with matter.
We will approximate all this matter by plasma. The influence of plasma on axion-photon conversion has been discussed for superradiant axions~\cite{Sen:2018cjt, Rosa:2017ury}, but also in other contexts~\cite{Hertzberg:2018zte, Carlson:1994yqa}. EM wave propagation through plasma is described by the modified dispersion relation \cite{JacksonED}
\begin{equation}
\kwave^2=\omega^2-\frac{\omega^2_{\rm plasma} \omega}{\omega+i \nu},
\end{equation}
where $\nu$ is the collision frequency and
\begin{equation}
\omega_{\rm plasma}=\frac{4 \pi e^2 n_{e}}{m_{e}},
\end{equation}
is the plasma frequency; $m_e$, $e$ and $n_e$ are the mass, charge and the concentration of the free electrons, respectively. Conceptually, it is helpful to consider two limiting cases - collisional ($\omega \ll \nu$; appropriate in the context of plasma in the accretion disks) and collisionless ($\omega \gg \nu$; in the context of the interstellar matter or a thin accretion disk). 
\subsubsection{Collisionless plasma}
EM waves in this limit have a modified dispersion relation that is equivalent to providing a photon with a mass $\mu_{\rm V}=\omega_{\rm plasma}$. For high $\mu_{\rm V} \geq (1/2)\mu_{\rm S}$, decay processes become kinematically forbidden\footnote{The decay process is $a\to \gamma +\gamma$ (for a Lagrangian with a $\Psi F^2$ term), so if the photon has an (effective) mass $\mu_{\rm V}$, in order for the decay to be energetically favourable, we should have $\mu_{\rm S} \geq 2\mu_{\rm V}$.}.
For interstellar matter \cite{Carlson:1994yqa}
\begin{equation}
\omega_{\rm plasma}=\sqrt{\frac{n_e}{0.03\text{cm}^{-3}}}(6.4  \,\times\, 10^{-12} \rm{eV}),
\end{equation}
the plasma frequency is below  the range of the QCD axion mass and some of the ultra-light axions [see comments around Eq.~\eqref{eq:ax_mass_range}]. Hence, one can expect EM instability not to be quenched on the primordial and lower range of the stellar BHs mass spectrum.

We have numerically modeled this scenario as an scalar-Proca system. The time domain study is summarized in Fig.~\ref{graph_Proca_a05_mu02_ks05_p2_phi_FF_Sr_E1}, and confirms this picture: when the (effective) mass of the vector field is larger than the axion mass, the burst is suppressed. We have checked that this suppression effect occurs in all our initial data and also for axionic couplings. 

We will now provide some analytical control over these results by concisely reproducing some of the results of Ref.~\cite{Sen:2018cjt} and also expanding them to the scalar coupling case. Consider the dimensionless Mathieu equation, that describes parametric resonances of the EM field in the background of the homogeneous axion/scalar condensate (Section \ref{sec:mink_axion} and \ref{sec:mink_scalar_coupling}) in the form
\begin{equation}  \label{eq:mathieu_general}
\partial^2_T y+(\Upsilon+2 \epsilon \cos{T})y=0\,,
\end{equation}
where $y$ is defined in Eq.~\eqref{eq:deffcty} for axions, and in Section \ref{sec:Mink_scalar_background_analytical} for scalar couplings. Here, $\Upsilon=(\kwave^2+\mu^2_{\rm V})/\mu_S^2$ for axions and $p=1$ scalars, and $\Upsilon=(\kwave^2+\mu^2_{\rm V})/(2\mu_S)^2$ for $p=2$ scalars. Also, $\epsilon=-\Psi_0 k_{\rm a}(\kwave/\mu_S),\,(1/2) \left(k_{\rm s}\Psi_0/2\right)^p$ for axions and scalars, respectively. As we showed already, the dominant instabilities occur for $\Upsilon=1/4$. It is well known that the critical stability curves on the $\Upsilon-\epsilon$ (Ince-Strutt) diagram are given by \cite{benderbook}
\begin{equation}
\label{eq:AvsEps}
\Upsilon=\frac{1}{4} + \eps+\mathcal{O}(\eps^2)\,.
\end{equation}
Inserting the appropriate $\Upsilon$ and $\eps$, we can find the values of the parameters for which~\eqref{eq:AvsEps}, 
a quadratic equation in $\kwave$, has real solutions. First, consider the axion case where $\eps=\eps(\kwave)$. 
We find the critical plasma frequency 
\begin{equation}
\omega^{\rm crit}_{\rm plasma}=\frac{1}{2}\mu_S \sqrt{1+(\Psi_0 k_{\rm a})^2}\,,\quad {\rm axions}
\end{equation}
in agreement with Ref. \cite{Sen:2018cjt}. For scalar couplings we find 
\begin{eqnarray}
\omega^{\rm crit}_{\rm plasma}&=&\mu_S\sqrt{\frac{1}{4} + k_{\rm s}\Psi_0}\,,\quad p=1\,,\\
&=&\mu_S\sqrt{1 +\frac{1}{2} (k_{\rm s}\Psi_0)^2}\,,\quad p=2\,.
\end{eqnarray}
Our results are in qualitative agreement with this prediction. One can also straightforwardly find corrections to the instability rate, induced by the effective mass. For example, in the axion case\footnote{Note that for the most part of the parameter space, plasma corrections will be smaller than the $k_{\rm a}$ corrections, discussed in Appendix \ref{app:RG_time_scales}.}
\begin{equation} \label{eq:ax_rate_plasma}
\lambda_{\rm a}=\frac{1}{2}\Psi_0 k_{\rm a}\mu_{\rm S} \sqrt{1-4\left(\frac{\omega_{\rm plasma}}{\mu_{\rm S}}\right)^2}+\mathcal{O}(k^2_{\rm a}).
\end{equation}

The preceding analysis neglects the time-dependence of the plasma distribution, in particular it neglects also the backreaction by the cloud on the plasma. Although the full problem is outside the scope of this work, we note that the arguments of Ref.~\cite{Hertzberg:2018zte} suggest that non-harmonic time dependence would not jeopardize parametric resonances as long as the scalar mass is much smaller than the plasma frequency. However, the time-periodic background of real scalars can drive matter resonantly in peculiar configurations oscillating with the (multiple of) scalar mass \cite{Boskovic:2018rub, Ferreira:2017pth}. This behaviour would manifest itself as an additional harmonic term in the Mathieu equation, similar to the ones in Appendix \ref{app:RG_time_scales}, and is expected to modify, but not eliminate the instability. The driving of plasma by the cloud could also deplete the plasma from the central regions~\cite{Ferreira:2017pth}, leading to a smaller effective photon mass and therefore to a more efficient instability.

\subsubsection{Collisional plasma}
Using estimates from Ref. \cite{Sen:2018cjt} we find
\begin{equation}
\nu  \sim  \Big( \frac{M}{10 M_{\odot}} \Big)^{-\frac{5}{8}} 10^{-6}\text{eV}\,,
\end{equation}
for the collision frequency and 
\begin{equation}
\omega_{\rm plasma} \sim \Big( \frac{M}{10 M_{\odot}} \Big)^{\frac{1}{2}} 10^{-3}\text{eV}\,,
\end{equation}
for the plasma frequency in the inner rims of the accretion disk around BHs. For BHs larger than $M \sim 10^{-3} M_{\odot}$, $\omega_{\rm plasma}>10^{-5} \text{eV}$ and the axion decay is forbidden in all of the parameter range interesting in a BH-superradiance context.

However, one should also consider the geometry of the problem. Accretion disks are planar structures (when thin), immersed in a ``spheroidal'' scalar cloud. The EM field enhancement can originate in the space external to the accretion disk (there is a limitation from interstellar matter there, discussed in the previous subsection). Such waves can lead to Ohmic heating of the disk or disperse it through the radiation pressure. The quantitative analysis of this would probably depend on the geometry of the initial fluctuations. We should also note that the estimates of the peak luminosity from Ref.~\cite{Rosa:2017ury} (which are even lower than the ones estimated numerically in Ref.~\cite{Ikeda:2018nhb}; see below) indicate that the radiation pressure (if EM instability ensues) would blow away the surrounding plasma.

\subsubsection{$e^{+}e^{-}$ plasma}

Besides astrophysical plasma large electric fields can lead to Schwinger $e^{+}e^{-}$ pair production. It was argued, at the order-of-magnitude level, that such plasma can indeed be created and reach large enough densities (and consequently critical $\omega_{\rm plasma}$) to block EM bursts~\cite{Rosa:2017ury}. Subsequently, $e^{+}e^{-}$ annihilations would drive the plasma density down and restart the process again.

An adequate treatment of this phenomenon is beyond the scope of this work, but here we show that our numerical estimates are consistent with such high EM fields. We work with estimates from Ref.~\cite{Ikeda:2018nhb} for the axion coupling, $\mu_{\rm S} M=0.2$ and $k_{\rm a}A_{0} \approx 0.3-0.4$, where the peak luminosity was given by
\begin{eqnarray}
\frac{dE}{dt}=5.0\times 10^{-6}\left(\frac{M_{\rm S}}{M}\right)\frac{c^{5}}{G}\,,
\end{eqnarray}
with $M_{\rm S}$ representing the cloud mass (for the relation between $A_0$, $M$ and $M_{\rm S}$ see Ref.~\cite{Brito:2014wla}). The Maxwell invariant is
\begin{eqnarray}
F^{\mu\nu} F_{\mu\nu} \sim \frac{ 1}{V}\frac{dE}{dt}t_{\rm p}\,,
\end{eqnarray}
where $V \sim \langle r \rangle^3 \sim 125 \, r_c^3$ ($r_c$ is the first Bohr radius, see Appendix \ref{sec:App_gravitational_atom}) is the cloud volume and $t_{\rm p} \sim 500 M$ \cite{Ikeda:2018nhb} is the time when the luminosity plateau develops. Taking $|\bm{B}| \sim |\bm{E}|$, one finds
\begin{eqnarray}
|\bm{E}|\sim 10\, E_c \Big( \frac{M}{M_\odot}\Big )^{-1} \sqrt{\frac{M_{\rm S}}{M}}\,,
\end{eqnarray}
where $E_c = 1.3 \,\times\, 10^{18}  \, \text{V}/\text{m}$ is the critical Schwinger field.

\section{Discussion}
Extensions of the standard model of particle physics where a (ultralight) 
scalar or pseudoscalar $\Psi$ couples to photons,
are very ``natural'' to consider. These include terms of the form $\Psi ^{\ast}F_{\mu\nu}F^{\mu\nu}$
or $\Psi^p F_{\mu\nu}F^{\mu\nu}$, with $p$ an integer. Such terms have been considered widely in the literature
in the context of dark matter physics and cosmology, and in fact axionic-type couplings are elegant resolutions of the strong CP problem. Our purpose here was to discuss what looks like an important gap in our understanding: effects of axionic or scalar couplings in strong-gravity situations.

We have shown that such couplings lead to instabilities of homogeneous configurations and in fact seem to lead to violations
of BH uniqueness results in General Relativity (see Ref.~\cite{Cardoso:2016ryw} for a discussion on these results and observational tests). In particular, for large enough couplings, charged RN BHs are unstable and spontaneously acquire a nontrivial scalar profile. In fact, unlike in Yang-Mills theory (where coloured BHs are unstable), it is plausible that the hairy solutions that we discussed at a perturbative level are stable (given that GR solutions are unstable, and a time evolution will in principle drive them to a hairy one). Thus, there are two possible solutions for a fixed set of physical parameters, despite the instability of one such solution.

Ultralight scalars or axions generically lead to superradiant instabilities of spinning BHs: on relatively short timescales the scalar field extracts a sizeable amount of the BH rotational energy and deposits it in a non axi-symmetric ``condensate''~\cite{Brito:2015oca}. Since the scalar field is time-dependent, this massive condensate emits gravitational waves, a mechanism which can be used to place impressive constraints on the scalar mass~\cite{Pani:2012vp,Brito:2013wya,Brito:2017zvb}.
In the presence of axionic or scalar-type couplings, a new channel onto which the system can radiate exists. Thus, one might rightfully worry that the superradiant instability responsible for the growth of the condensate might be quenched. What we showed, here and in a previous Letter~\cite{Ikeda:2018nhb} is that indeed such couplings will -- if large enough {\it and} in the absence of plasma -- give rise to periodic bursts of low-frequency EM radiation (but see footnote \ref{footnoteX}). These bursts deplete the condensate of a fraction of the scalar until superradiance becomes effective again. Thus, they effectively limit the maximum amount of energy that superradiance can extract before the EM burst. Our numerical results are supported by simple analytical estimates which cover in a concise and simple way all the phenomena we study, from flat to curved spacetime.

In practice, the interstellar plasma can influence the realization of the bursts. For collisionless plasma (interstellar medium and thin accretion disks), we find that the plasma would not affect significantly the existence of bursts for axions or scalars more massive than $\mu_{\rm S} \sim 10^{-12} \text{eV}$.

\begin{acknowledgments}
We would like to thank the anonymous referee for useful comments.
The authors acknowledge financial support provided under the European Union's H2020 ERC 
Consolidator Grant ``Matter and strong-field gravity: New frontiers in Einstein's theory'' grant 
agreement no. MaGRaTh--646597. 
H.W.~acknowledges financial support provided by the Royal Society University Research Fellowship
UF160547 and the Royal Society Research Grant RGF\textbackslash R1\textbackslash 180073.
She also acknowledges partial support by grants FPA-2016-76005-C2-2-P, AGAUR SGR-2017-754.
This project has received funding from the European Union's Horizon 2020 research and innovation programme under the Marie Sklodowska-Curie grant agreement No 690904.
The authors would like to acknowledge networking support by the GWverse COST Action CA16104, ``Black holes, gravitational waves and fundamental physics.''
We acknowledge financial support provided by FCT/Portugal through grant PTDC/MAT-APL/30043/2017.
We acknowledge the SDSC Comet and TACC Stampede2 clusters through NSF-XSEDE Award Nos. PHY-090003.
We thankfully acknowledge the computer resources at Marenostrum IV, Finis Terrae II and LaPalma and the technical
support provided by the Barcelona Supercomputing Center via the PRACE grant Tier-0 PPFPWG, and via the BSC/RES grants
AECT-2017-2-0011, AECT-2017-3-0009 and AECT-2018-1-0014.
Computations were performed on the ``Baltasar Sete-Sois'' cluster at IST, and XC40 at YITP in Kyoto University.
\end{acknowledgments}

\appendix

\section{Flat-space instabilities in the Coulomb gauge}\label{app:Coulomb}

\subsection{Axion coupling} \label{app:Coulomb_axion}

The results in Section~\ref{sec:Mathieu_1} arise also in a simpler version of the relevant equations, if one changes to Coulomb gauge. 
In the Coulomb gauge, $(\bm{\nabla} \bm{A}=0)$, the space component of Maxwell's equations reduces to $\nabla^2\bm{A}-\partial^2_t \bm{A}-\nabla (\partial_t A_0)=-\bm{j}$ \cite{JacksonED}. Using Helmholtz theorem we can decompose $\bm{j}=\bm{j}_l+\bm{j}_t$, with $\nabla \times \bm{j}_l=0$ (longitudinal component) and $\nabla\bm{j}_t=0$ (transverse component). Finally, the time component of Maxwell's equations gives $\partial^2_t \bm{A}-\nabla^2\bm{A}=\bm{j}_t$. Notice that the effective current sourced by non-relativistic axions (where $|\nabla \Psi| \ll |\partial_t \Psi| $) is irrotational.
From Eq. \eqref{eq:MFEoMVector} in the Coulomb gauge,
\begin{equation} \label{eq:max_ax_ed_Mink}
\partial^2_t\bm{A}-\bm{\nabla}^2 \bm{A}+2k_{\rm a} \partial_t \Psi \bm{\nabla} \times \bm{A} =0\,.
\end{equation}
The momentum space representation of the previous equation shows that the fluctuations of the Fourier-transformed vector potential $\bm{A}_{\bm{\kwave}}$ are described by
\begin{equation} \label{eq:Mathieu_mod}
\partial^2_t\bm{A}_{\kwave}+\kwave^2\bm{A}_{\kwave}+ik_{\rm a}\bm{\kwave} \times \int\frac{d^3\bm{\kwave}'}{(2\pi)^3}\partial_t{\Psi}_{\bm{\kwave}-\bm{\kwave}'}\bm{A}_{\bm{\kwave}'}=0\,.
\end{equation}

Consider the homogeneous axion field $\Psi=\Psi_0 \cos{(\mu_S t)}$. As shown in Ref.~\cite{Hertzberg:2018zte}, in a circular polarization representation  $\bm{A}_{\bm{\kwave}}=\sum_{\lambda}y_{\bm{\kwave}}\bm{\xi}^{(\lambda)}_{\bm{\kwave}}+\text{c.c.}\,$ the vectors ${\xi}^{(\lambda)}_{\bm{\kwave}}$ decouple and we are left with Eq. \eqref{eq:Mathieu0}, after the variable change $\mu_{\rm S} t=T+\pi/2$. We thus recover in one go Mathieu's equation for this problem.

\subsection{Scalar coupling} \label{app:Coulomb_scalar}

For a non-relativistic scalar filed, Maxwell's equations in the Coulomb gauge and a Minkowski background reduce to
\begin{equation}\label{eq:max_sc_coupl_Mink_hom}
\partial_{\mu}F^{\mu \nu}=-g_p(t)F^{0 \nu}\,,
\end{equation}
where
\begin{equation} 
g_p(t)=\frac{pk_{\rm s}(k_{\rm s}\Psi)^{p-1}\partial_t{\Psi}}{1+(k_{\text{s}}\Psi)^{p}}\,,
\end{equation}
and the RHS of Eq. \eqref{eq:max_sc_coupl_Mink_hom} is a well-defined\footnote{This is the case in general, when the RHS of Eq. \eqref{eq:max_sc_coupl_Mink_hom} is proportional to $(\partial_{\mu}\Psi)F^{\mu \nu}$.} current $j^{\nu}$.

The general comments on the Coulomb gauge from the previous section also apply to this case. Equation \eqref{eq:max_sc_coupl_Mink_hom} reduces to
\begin{equation}
\partial^2_t\bm{A}+g_p(t)\partial_t\bm{A}-\bm{\nabla}^2 \bm{A}=0.
\end{equation}
Note that here $\bm{j}_{t}=g_p(t)\partial_t \bm{A}$ and $\bm{j}_{l}=-g_p(t)\nabla A_0$. Fourier transforming this equation we obtain
\begin{equation} \label{eq:max_scalar_FT_Mink}
\partial^2_t\bm{A}_{\kwave}+\kwave^2\bm{A}_{\kwave}+\int\frac{d^3\bm{\kwave}'}{(2\pi)^3}g^{(p)}_{\bm{\kwave}-\bm{\kwave}'}(t)\partial_t\bm{A}_{\bm{\kwave}'}=0,
\end{equation}
with $g^{(p)}_{\bm{\kwave}}(t)$ being the Fourier transform of $g_{p}(t)$.

If we consider an homogeneous scalar field $\Psi=\Psi_0 \cos{(\mu_S  t)}$ and decouple the polarization vectors, with $\bm{A}_{\bm{\kwave}}=\sum_{\lambda}y_{\bm{\kwave}}\bm{\xi}^{(\lambda)}_{\bm{\kwave}}+\text{c.c.}\,$, the previous equation reduces to 
\begin{equation} \label{eq:max_scalar_FT_Mink_hom}
\partial^2_t y_{\bm{\kwave}}+\kwave^2y_{\bm{\kwave}}+g_p(t)\partial_t y_{\bm{\kwave}}=0.
\end{equation}
The form of this equation is similar to the Ince equation \cite{Hartono:2004} and we can use a change of variables of the form
\begin{equation} 
y_{\bm{\kwave}}=\exp{\Big(- \frac{1}{2}\int^{t}_{0} g_p(t')dt' \Big)}f_{\bm{\kwave}}
\end{equation}
to obtain an equation\footnote{Note that: (i) $g_p(0)=0$; (ii) the conversion factor between $f_{\bm{\kwave}}$ and $y_{\bm{\kwave}}$ is harmonic and can not change the conclusions regarding stability.} of the Hill type
\begin{equation} 
\partial^2_t f_{\bm{\kwave}}+(\kwave^2+W_p(t)) f_{\bm{\kwave}}=0,
\end{equation}
where we defined
\begin{equation} \label{eq:Wp_scalar}
W_p(t)=-\frac{1}{2}\partial_t g_p-\frac{1}{4} g^2_p.
\end{equation}
We see that $W_p(t)$, i.e. the harmonic term that drives the instability, scales to leading order as $k^p_{\rm s}$.
To the lowest order in $k_{\rm s}$ this equation reduces to the Mathieu equation for both $p=1$:
\begin{equation} 
W_1(t)=\frac{1}{2}\mu_S^2 k_{\rm s}\Psi_0 \cos{(\mu_S  t)}
\end{equation}
and the $p=2$ case:
\begin{equation} 
W_2(t)=\mu_S^2 (k_{\rm s}\Psi_0)^2 \cos{(2\mu_S  t)}.
\end{equation}



\section{Timescales for the instabilities of the flat-space homogeneous configurations}\label{app:RG_time_scales}

\subsection{First order results} \label{app:RG_time_scales_1st}

Here we derive the instability timescales, using a perturbative expansion in a small coupling parameter $\epsilon(k_{\rm a},k_{\rm s},\Psi_0)$, to be defined latter. We expect that for specific values of $\omega$, given by Eqs.~\eqref{eq:hom_k_instable}, \eqref{eq:wp1} and \eqref{eq:wp2}, regular perturbation theory breaks down.  However, one can start from the regular perturbation theory, see how instabilities buildup and use multi-scale \cite{benderbook} or dynamical renormalization group (DRG) \cite{Chen:1995ena} methods to regularize the problem. We will here use the later approach. As instabilities from a homogeneous axion field in flat space are exactly described by the Mathieu equation and for scalar couplings approximately, to leading order in $k_{\rm s}$, we will here derive the first-order result for the leading instability rate of the Mathieu equation. 

We will consider Mathieu equation in the form of Eq. \eqref{eq:mathieu_general}, with $\mu_{\rm V}=0$.
%
%
Instabilities arise when $\Upsilon=n^2/4, n \in \mathbb{N}$. We here focus on the dominant instability $\sqrt{\Upsilon}=1/2$ and denoth the subscript of the function that governs dominant time-dependence of the vector potential with $\sqrt{\Upsilon}$. At zeroth order in $\epsilon$, the solution is given by $y_{1/2}^{(0)}=Ae^{i(1/2)T}+\text{c.c}$. The differential equation for the first order correction is
\begin{equation}
\partial^2_T y_{1/2}^{(1)}+\Big(\frac{1}{2} \Big)^2 y_{1/2}^{(1)}= - \Big(
 A^{\ast}  e^{i(1/2)T} +  A  e^{i(3/2)T} \Big) +\text{c.c.}
\end{equation}
and the full solution is given by
\beq\label{eq:app_Mathieu_1st_bare}
y_{1/2}&=& \Big( A-\frac{1}{2}A^{\ast} \epsilon \Big)  e^{i(1/2)T} +\epsilon \Big(i
 A^{\ast}  e^{i(1/2)T}(T-T_0) +  \nonumber \\
&&  \frac{1}{2}   A  e^{i(3/2)T} \Big) +\text{c.c.}\,,
\eeq
where $T_0$ is some arbitrary time where we imposed initial conditions. Higher-order terms will build up a secular terms of the form $(T-T_0)^m$, where $m$ is the order of the expansion, in the limit $m \to \infty$ giving exponential growth. However, this behaviour invalidates our perturbative expansion. 

The DRG approach is based on the insight that the invalidation of the regular perturbation theory is a consequence of the big interval between $T_0$ and $T$ \cite{Chen:1995ena, Galley:2016zee}. In order to remedy this problem, we declare the parameters of the solution in Eq. \eqref{eq:app_Mathieu_1st_bare} as ``bare'' and rewrite them as the renormalized ones:
\be \label{eq:renormalized_amplitude}
A(T_0)=Z(T_0, \tau)A(\tau) \,, \,Z(T_0, \tau)=1+\sum^{\infty}_{n=1}a_n\epsilon^n.
\ee
Next, we expand $T-T_0=T-\tau+\tau-T_0$ and choose $a_1$ (``counter-term'') in such way to cancel secular terms $\propto (\tau-T_0)$. The renormalized solution has the form
\begin{align}\label{eq:app_Mathieu_1st_renor}
y_{1/2}&=\Big(A(\tau)-\frac{1}{2}A^{\ast}(\tau) \epsilon \Big)   e^{i(1/2)T}  \\
& + \epsilon \Big(i
 A^{\ast}(\tau)  e^{i(1/2)T}(T-\tau)  +\frac{1}{2}   A (\tau) e^{i(3/2)T} \Big) +\text{c.c.}\,, \nonumber 
\end{align}
with $a_1 A(\tau)=-iA^{\ast}(\tau)(\tau-T_0)$. Arbitrariness of $\tau$ leads to the RG equation 
\begin{equation} \label{eq:RGeq}
\frac{\partial A(t_0)}{\partial \tau}=0.
\end{equation}
Working consistently at the $\epsilon^1$ order and decomposing $A(\tau)=X(\tau)+iY(\tau)$, we find
\begin{equation}
\frac{\partial^2 X }{\partial \tau^2}-\epsilon^2 X(\tau)=0,
\end{equation}
i.e.
\begin{equation}
X(\tau)=e^{\pm \epsilon \tau},
\end{equation}
with $\partial_{\tau} X=\eps Y$. Finally, we choose $\tau=T$ as the ``observational'' time and conclude that the instability rate, to first order in $\epsilon$ is $\lambda=\epsilon$, for (dimensionless) Mathieu equation. Substituting appropriate $\eps$  and rescaling back to the physical time [see definitions of $\sqrt{\Upsilon}$ and $\eps$ below Eq. \eqref{eq:mathieu_general}], we obtain the estimates consistent with the numerical results in Eqs.\eqref{rate_axion0}, \eqref{eq:rate_scalar_p_1} and \eqref{eq:rate_scalar_p_2}.

\subsection{Higher order results}

\subsubsection{Axion coupling} \label{app:RG_time_scales_2nd_ax}

Instabilities for the homogeneous configurations with the axion couplings are exactly described by the Mathieu equation. Hence, we will here obtain solution of the Mathieu equation to the second order, using DRG. Differential equation for the second order contribution is of the form
\beq
\partial^2_T y_{1/2}^{(2)}+\Big(\frac{1}{2} \Big)^2 y_{1/2}^{(2)} &=& - (2 \cos{T})y_{1/2}^{(1)}\,.
\eeq
DRG procedure is analogous to the first order case. We will consider only leading order harmonics with $T/2$  as they will give dominant contribution to the instability. Note that the term of the form $(T-\tau)(\tau-T_0)$  will self-consistently cancel, sign of the renormalizability of the differential equation \cite{Galley:2016zee}. Second-order coefficient in Eq. \eqref{eq:renormalized_amplitude} is
\beq
a_2 A(\tau)=-iA(\tau)(\tau-t_0)-\frac{A(\tau)}{2}(\tau-t_0)^2\,.
\eeq
From the RG equation \eqref{eq:RGeq} we obtain $\lambda^{(R,2)}_{\rm a}=\epsilon \sqrt{1-\eps^2}$. As we should trust this solution to the order of $\mathcal O (\eps^3)$, we perform Pad\'e resummation of the results \cite{benderbook}. As the perturbative result $ \lambda^{(R,2)}_{\rm a}$ is even function, we used the first non-trivial approximant $(2,1)$ and the final rate estimate is
\beq
\lambda_{\rm a}=\frac{\mu \epsilon}{1+\frac{1}{2}\epsilon^2},
\eeq
with appropriate $\epsilon$ defined below below Eq. \eqref{eq:mathieu_general}. This results gives very good description of the numerical data in both Sections \ref{sec:mink_axion} and \ref{sec:Time_domain_studies}, as shown on Fig. \ref{relation_kaxion_lambda} and Fig. \ref{fig:Mink_axion_rates}.

\subsubsection{Scalar coupling} \label{app:RG_time_scales_2nd_sc}

As will become clear later, we will first consider $p=2$ case. Equation for the second order correction\footnote{We use $Y_{1/2}$ label for the higher order $p=2$ corrections with $Y^{(1)}_{1/2} \equiv y^{(1)}_{1/2}$ and $Y^{(0)}_{1/2} \equiv y^{(0)}_{1/2}$. For $p=1$ we use $J_{1/2}$ mutatis mutandis.} is
\beq
\partial^2_T Y_{1/2}^{(2)}+\Big(\frac{1}{2} \Big)^2 Y_{1/2}^{(2)} &=& - (2 \cos{T})Y_{1/2}^{(1)} \\ \nonumber
&& -W_2(T;k^{4}_{s})Y_{1/2}^{(0)}+\mathcal{O}(k^6_s)\,,
\eeq
where $W_2(T;k^{4}_{s})$ is the Taylor expansion coefficient of the function in Eq. \eqref{eq:Wp_scalar} at the order of $k^{4}_{s}$. Performing DRG as in the Appendix \ref{app:RG_time_scales_2nd_ax} we obtain $\lambda^{(R,2)}_{p=2}=2\epsilon \sqrt{(1-3\epsilon) (1-5\epsilon)}$. After $(1,1)$ Pad\'e resummation we have
\beq
\lambda_{p=2}=\frac{2\mu \epsilon}{ (4\epsilon+1)} \,,
\eeq
[$\epsilon$ is defined below Eq. \eqref{eq:mathieu_general}] as shown on Figs. \ref{relation_kscalar_lambda_p} and \ref{graph_lambda_p}.

For $p=1$ case, second order correction is governed by the equation
\beq
\partial^2_T J_{1/2}^{(2)}+\Big(\frac{1}{2} \Big)^2 J_{1/2}^{(2)} &=& - (2 \cos{T})J_{1/2}^{(1)} \\ \nonumber
&& -W_1(T;k^{2}_{s})J_{1/2}^{(0)}+\mathcal{O}(k^3_s)\,.
\eeq
For the rate estimate we obtain the same results as for the axion case. This result is clearly not a good description as the numerical results indicate (Fig. \ref{relation_kscalar_lambda_p}) that the function $\lambda_{p=1}(\eps)$ is divergent. Therefore, we go to the third-order contribution
\beq
\partial^2_T J_{1/2}^{(3)}+\Big(\frac{1}{2} \Big)^2 J_{1/2}^{(3)} &=& - (2 \cos{T})J_{1/2}^{(2)} \\ \nonumber
&& -W_1(T;k^{2}_{s})J_{1/2}^{(1)}-W_1(T;k^{3}_{s})J_{1/2}^{(0)}\\ \nonumber
&& +\mathcal{O}(k^4_s)\,.
\eeq
Renormalized rate is $\lambda_{p=1}^{(R,3)}=\eps \sqrt{1+(17/2) \eps^2+(16/3) \eps^3+(2225/144)\eps^4}$. After Pad\'e resummation at the order $(1,2)$ we obtain
\beq
\lambda_{p=1}=\frac{\mu \eps}{1-\frac{17}{4}\eps^2}\,,
\eeq
[see beneath Eq. \eqref{eq:mathieu_general} for the $\epsilon$ definition] as shown on Figs. \ref{relation_kscalar_lambda_p} and \ref{graph_lambda_p}.

\section{Axions in a RN background: long-lived configurations}\label{boundstates}

Besides the unstable modes discussed in the main text, massive fields around BHs can also form long-lived quasi-bound states (see e.g. Ref.~\cite{Brito:2015oca} and references therein). Here, we discuss the effect that the axionic coupling has on these modes. In particular, due to this coupling, we expect that the long-lived scalar cloud will in turn trigger the excitation of a long-lived signal of EM waves.

We consider the system of eqs.~\eqref{pert_RN1}--~\eqref{pert_RN3} and focus on the $l=1$ case for simplicity. For this case, the field equation for $Z_-$~\eqref{pert_RN3} completely decouples and to compute the modes of the system one can set $Z_-=0$. At the horizon we impose the usual regular boundary conditions~\eqref{boundary_horizon}. On the other hand one can check that asymptotically the most general solution behaves as:
\beq\label{bc_boundstate}
&&\psi\sim A e^{-k_S r}r^{-\frac{M\left(\mu_S^2-2\omega^2\right)}{k_S}}+B e^{k_S r}r^{\frac{M\left(\mu_S^2-2\omega^2\right)}{k_S}}\nonumber\\
&&+C\frac{2 Qk_{\rm a}}{\mu_S^2r^3} e^{i\omega r}r^{2iM\omega}+D\frac{2 Qk_{\rm a}}{\mu_S^2 r^3}e^{-i\omega r}r^{-2iM\omega}\,,\nonumber\\
&&Z_+\sim C e^{i\omega r}r^{2iM\omega}+D e^{-i\omega r}r^{-2iM\omega}\nonumber\\
&&-A\frac{4 Qk_{\rm a}}{\mu_S^2 r^3}e^{-k_S r}r^{-\frac{M\left(\mu_S^2-2\omega^2\right)}{k_S}}\nonumber\\
&&-B\frac{4 Qk_{\rm a}}{\mu_S^2 r^3} e^{k_S r}r^{\frac{M\left(\mu_S^2-2\omega^2\right)}{k_S}}\,,
\eeq
where $k_S=\sqrt{\mu_S^2-\omega^2}$. If $\Re(\omega)<\mu_S$, one can then find regular modes by imposing $B=D=0$. These modes are spatially localized states that slowly leak EM radiation to infinity due to the axionic coupling. On the other hand, if $\Re(\omega)>\mu_S$, the condition $A=C=0$ yields purely outgoing waves at infinity, and allow us to compute the quasinormal modes of the system~\cite{Berti:2009kk}.

To find the quasi-bound state modes we employ two complementary methods. We used a direct extension of the two-parameter shooting method explained in the main text: fixing the scalar field at the horizon, we shoot for $\omega$ and the value of the vector at the horizon~\eqref{boundary_horizon}, such that the quasibound-state boundary conditions are met. We also used a direct integration method that allow us to reduce the problem to a one-parameter shooting method~\cite{Rosa:2011my,Pani:2012bp,Brito:2013wya,Macedo:2018txb}, that we here describe.
This method allows to compute modes of systems with an arbitrary number of coupled equations but for concreteness let us consider the system of coupled equations~\eqref{pert_RN1} and~\eqref{pert_RN2} for $l=1$ with ingoing wave boundary condition at the horizon~\eqref{boundary_horizon}.

By imposing this boundary condition we obtain a family of solutions at infinity characterized by $2$ parameters, corresponding to the $2$-dimensional vector of the coefficients $\left\{\psi_0,Z_{+\,0}\right\}$. The quasibound-state spectrum can be computed by choosing a suitable orthogonal basis for the space of initial coefficients $\left\{\psi_0,Z_{+\,0}\right\}$. To do so we perform two integrations from the horizon to infinity and construct the $2\times 2$ matrix 
\be
\label{detS}
\bm{S_m} (\omega)=
 \begin{pmatrix}
  B^{(1)} & B^{(2)}\\
  D^{(1)} & D^{(2)}\\
 \end{pmatrix}\,,
\ee
where $B$ and $D$ are obtained from the boundary conditions at infinity~\eqref{bc_boundstate} and the superscripts denote a particular vector of the chosen basis, for example, $B^{(1)}$ corresponds to $\left\{\psi_0,Z_{+\,0}\right\}=\{1,0\}$ and $B^{(2)}$ corresponds to $\left\{\psi_0,Z_{+\,0}\right\}=\{0,1\}$. The  eigenfrequency $\omega=\omega_R+i\omega_I$ will then correspond to the solutions of
\be
\det|\bm{S_m}(\omega_0)|=0\,,
\ee
which in practice corresponds to minimizing $\det\bm{S_m}$ in the complex plane.

\begin{figure}[htb]
\begin{center}
\begin{tabular}{c}
\epsfig{file=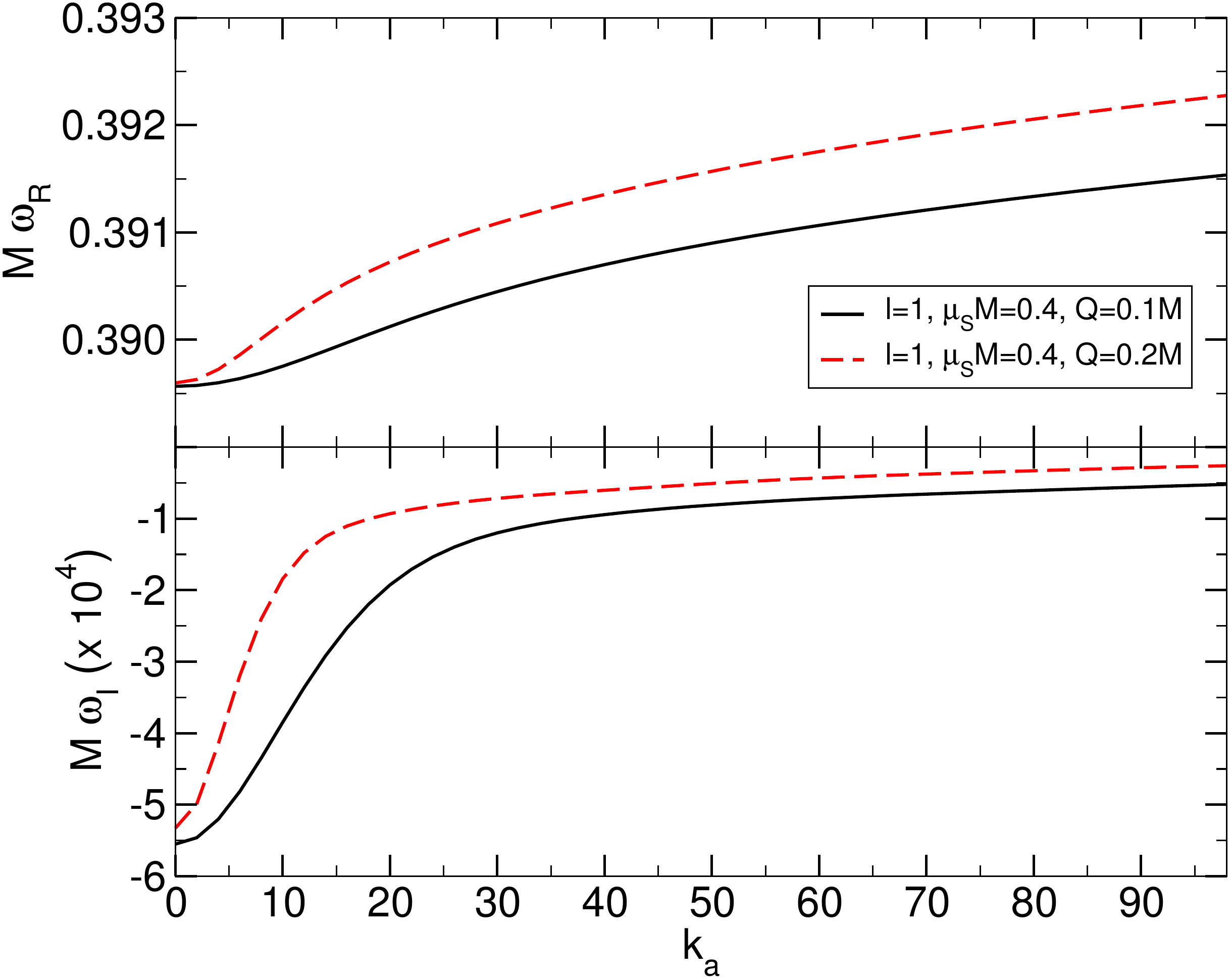,width=8.5cm,angle=0,clip=true}
\end{tabular}
\caption{Quasi-bound state fundamental eigenfrequencies, $\omega=\omega_R+i\omega_I$, for $l=1$, $M\mu_S=0.4$ and different $Q$, as a function of $k_{\rm a}$.\label{RN_boundstate}}
\end{center}
\end{figure}

 The eigenfrequencies for $M\mu_S=0.4$ as a function of $k_{\rm a}$ are shown in Fig.~\ref{RN_boundstate}. A generic conclusion of our analysis is that the coupling does not significantly affect the quasibound states, however the time-scale, $1/\omega_I$, over which these states decay slightly increases the larger the coupling $k_{\rm a}$.

%

\section{Kerr spacetime - coordinates and notation}\label{App:Kerr_coordinates}

\subsection{Boyer-Lindquist coordinates}

Kerr spacetime in Boyer-Lindquist coordinates $(t_{\rm BL},r,\theta,\varphi_{\rm BL})$ is given by \cite{Teukolsky:2014vca}
\begin{align}
\label{eq:KerrBLlineelement}
\dif s^{2} = & - \left( 1 - \frac{2 M r}{\Sigma} \right) \dif t^{2}_{\rm BL}
        - \frac{4 M r a \sin^{2}\theta}{\Sigma} \dif t_{\rm BL} \dif \varphi_{\rm BL}
\nonumber \\ &
        + \frac{\Sigma}{\Delta} \dif r^{2} 
        + \Sigma \dif \theta^{2}
        + \frac{\F}{\Sigma} \sin^{2}\theta \dif \varphi^{2}_{\rm BL}
\,,
\end{align}
where
\begin{subequations}
\label{eq:KerrBLfcts}
\begin{align}
\Delta = & r^{2} + a^{2} - 2 M r 
       = (r - r_{+})(r - r_{-} )
\,,\\
\Sigma = & r^{2} + a^{2} \cos^{2}\theta
\,,\\
\F = & \left( r^2 + a^2 \right)^{2} - \Delta a^{2} \sin^{2}\theta
\,,
\end{align}
\end{subequations}
and $a=J/M$, while $J$ and $M$ are the BH's angular momentum and mass, respectively. It is also useful to define the dimensionless rotational parameter $\tilde{a}=a/M$ and note that $0<\tilde{a}<1$ (consequence of the Cosmic Censorship Conjecture). 

The Kerr spacetime admits two horizons
\beq
r_{\pm}=M \pm \sqrt{M^2-a^2}  \,,\label{eq:horizons}
\eeq
and the angular velocity of the outer one is
\beq
\Omega_{+}=\frac{a}{2Mr_{+}} \,.\label{eq:horizons_angular}
\eeq
%
\subsection{Kerr-Schild coordinates}\label{app.Kerr-Schild coordinates}
For numerical purposes it is often more useful to use Kerr-Schild coordinates. The two coordinates systems are related via~\cite{Alcubierre:2008}
\begin{align}
\label{eq:TrafoBL2KS}
\dif t_{\rm KS} = & \dif t_{\rm BL} + \frac{2 M r}{\Delta} \dif r
\,,\quad
\dif \varphi_{\rm KS} = \dif \varphi_{\rm BL} + \frac{a}{\Delta} \dif r
\,.
\end{align}
In these ``spherical''-type Kerr-Schild coordinates the line element is 
\begin{align}
\dif s^{2} = &  - \left( 1 - \frac{2 M r}{\Sigma} \right) \dif t^{2}_{\rm KS}
                - \frac{4 M r a \sin^{2}\theta}{\Sigma} \dif t_{\rm KS} \dif \varphi_{\rm KS}
\nonumber \\ &
                + \frac{4 M r}{\Sigma} \dif t_{\rm KS} \dif r
                + \left( 1+\frac{2 M r}{\Sigma} \right) \dif r^{2}
                + \Sigma \dif \theta^{2} 
\\ &
                - 2 a \sin^{2}\theta \left( 1 + \frac{2 M r}{\Sigma} \right) \dif r \dif\varphi_{\rm KS}
                + \frac{\F}{\Sigma} \sin^{2}\theta \dif\varphi^{2}_{\rm KS}
\,.\nonumber
\end{align}
%

\section{Wald's solution}\label{App:WaldSolution}
Here we summarize the solution to Einstein-Maxwell theory derived by Wald~\cite{Wald:1974np}
that represents a rotating BH encompassed by an originally uniform magnetic field.
%
The EM field $F$ in the background of a Kerr BH is written as~\cite{Wald:1974np}
\begin{align}
\label{eqApp:WaldSolGen}
F = &   F_{10}\, \omega^{1} \wedge \omega^{0}
      + F_{13}\, \omega^{1} \wedge \omega^{3}
\non\\ &
      + F_{20}\, \omega^{2} \wedge \omega^{0}
      + F_{23}\, \omega^{2} \wedge \omega^{3}
\,,
\end{align}  
where the orthonormal tetrad is
\begin{subequations}
\begin{align}
\omega^{0} = & \sqrt{\frac{\Delta}{\Sigma}}\left( \dif t_{\rm BL} - a \sin^{2}\theta \dif\varphi_{\rm BL} \right)
\,,\\
\omega^{1} = & \sqrt{\frac{\Sigma}{\Delta}} \dif r
\,,\\
\omega^{2} = & \sqrt{\Sigma} \dif\theta
\,,\\
\omega^{3} = & \frac{\sin\theta}{\sqrt{\Sigma}} \left( \left(r^2+a^2\right)\dif\varphi_{\rm BL} - a \dif t_{\rm BL} \right)
\,,
\end{align}
\end{subequations}
and 
\begin{subequations}
\label{eq:FABWaldGen}
\begin{align}
F_{10} = & B_{0} \left[
        \frac{a r \sin^{2}\theta}{\Sigma}
        - \frac{a M \left( 1+\cos^{2}\theta \right) \left(r^2 - a^2\cos^{2}\theta \right)}{\Sigma^{2}}
        \right]
\,,\\
F_{13} = & B_{0} \frac{r \sin\theta \sqrt{\Delta} }{\Sigma}
\,,\\
F_{20} = & B_{0} \frac{ a \sin\theta \cos\theta \sqrt{\Delta} }{\Sigma}
\,,\\
F_{23} = & B_{0} \frac{\cos\theta}{\Sigma} \left[
        r^2 + a^2 - \frac{ 2 M r a^{2}\left(1+\cos^{2}\theta \right)}{\Sigma}
        \right]
\,.
\end{align}
\end{subequations}

In BL coordinates the Maxwell tensor is given by~\footnote{Note that we suppress subscripts ${\rm{BL}}$
to deflate the notation, and simply remind the reader that we refer to $(t_{\rm BL},r,\theta,\varphi_{\rm BL})$.}
\begin{align}
\label{eq:WaldMTFullDiffForm}
F = &   F_{tr}\,            \dif t_{\rm BL} \wedge \dif r
      + F_{t\theta}\,       \dif t_{\rm BL} \wedge \dif\theta
\non\\ &
      + F_{r\varphi}\,      \dif r          \wedge \dif\varphi_{\rm BL}
      + F_{\theta\varphi}\, \dif\theta      \wedge \dif\varphi_{\rm BL}
\,,
\end{align}
with components
\begin{subequations}
\label{eq:WaldMTFullComp}
\begin{align}
F_{tr}            = & F_{13} \frac{a \sin\theta}{\sqrt{\Delta}} - F_{10}
\,,\\
F_{t\theta}       = & F_{23}\, a \sin\theta - F_{20} \sqrt{\Delta}
\,,\\
F_{r\varphi}      = & F_{13} \frac{( a^2 + r^2 ) \sin\theta}{\sqrt{\Delta}} - F_{10}\, a \sin^{2}\theta
\,,\\
F_{\theta\varphi} = & F_{23} ( a^2 + r^2 ) \sin\theta - F_{20}\, a\sin^{2}\theta \sqrt{\Delta}
\end{align}
\end{subequations}
The corresponding $4$-vector potential is given by
\begin{align}
\label{eq:AvecBL4d}
A_{\mu} = & \frac{B_{0}}{2\Sigma} 
        \left( 2 a M r \left( 1 + \cos^{2}\theta \right), 0, 0,
               \sin^{2}\theta \left( \F - 4 a^{2} M r \right)
        \right)
\,.
\end{align}
The explicit expressions for the electric and magnetic fields are rather involved,
but they asymptote to
\begin{align}
\lim_{r\rightarrow\infty} E^{i} = & (0,0,0)
\,,\quad
\lim_{r\rightarrow\infty} B^{i} =   B_{0} \left( \cos\theta, 0, 0 \right)
\,,
\end{align}
at spatial infinity.
In the limit of a non-rotating BH background the fields reduce to
\begin{align}
E^{i} = & (0,0,0)
\,,\quad
B^{i} = \alpha B_{0} \left( \cos\theta, -\frac{\sin\theta}{r},0 \right)
\,,
\end{align}
where $\alpha$ is the lapse function.
The invariants of the EM field are
\begin{subequations}
\begin{align}
F^{\mu\nu}F_{\mu\nu} = & 2 \left( - F^{2}_{10} + F^{2}_{13} - F^{2}_{20} + F^{2}_{23} \right)
\,,\\
\,^{\ast}F^{\mu\nu} F_{\mu\nu} = & 4 \left( F_{13} F_{20} - F_{10} F_{23} \right)
\,.
\end{align}
\end{subequations}
The latter vanishes in limit of zero spin.

\section{Gravitational atom}  \label{sec:App_gravitational_atom}

Superradiant instability around Kerr BHs happens as long as the superradiant condition is satisfied \cite{Brito:2015oca}
\beq
\frac{\omega}{m}<\Omega_{+},
\eeq
where $\omega$ is a bosonic wave frequency and $m$ is a spherical harmonic ``quantum'' number. Since the field can ``leak'' through the horizon, solutions of the Klein-Gordon equation are quasi-bound and the field frequency is a complex number. 

In the weak-field regime (scenario dubbed as a gravitational atom) velocities and densities are small, as the cloud spreads over a large volume around the BH \cite{Brito:2014wla}. Particle dispersion relation is (for now we omit the imaginary part)  $\omega^2=\mu^2+\kwave^2$ and in the weak-field regime we expect $\omega\sim \mu$, so that, up to the second order in wave number $\kwave$,
\begin{equation} \label{eq:Nomega}
\omega=\mu+\frac{\kwave^2}{2\mu}+{\cal O}(k^4)\,.
\end{equation}
Dimensionless expansion parameter here is the group velocity of the field $v=\kwave/\mu$. In order to understand long-range behaviour, note that the field is trapped by BH gravity $(\kwave^{2}<0)$, so that we expect exponential tail $\Psi \sim e^{i \kwave r} \sim e^{-|\kwave| r}$. 

Length scale associated with the particles momentum is De Broglie wavelength $\lambda_{\text{D}} =2\pi/ |k|$ and in the near-horizon limit important scale is the gravitational radius $r_g=2M$. We will use virial theorem\footnote{Here we expect that the leading order behaviour of the weak-field gravitational potential is $\propto 1/r$.} $(2T \sim V)$ to understand dependence of the typical size of the cloud $r_c$ on $\mu$ and $M$:
\begin{equation}
\frac{|\kwave^2|}{\mu} \sim \frac{M \mu}{r_c}.
\end{equation}
De Broglie wavelength of the wave on radius $r_c$ depends on the number of modes excited as $n\lambda_{\text{D}}=2r_c \pi$. We find
\begin{equation}
r_c \sim \frac{n^2}{\mu \alpha_{g}},
\end{equation}
where $\alpha_{g}$ is the fine structure constant
\begin{equation}
\alpha_g = \frac{r_g}{\lambda_{c}},
\end{equation}
and $\lambda_{c}=1/\mu$ is the (reduced) Compton wavelength. Finally, we see that the behaviour of the real part of the spectrum is the same as for the Hydrogen atom, mutatis mutandis:
\begin{equation} \label{eq:Nomega2}
\omega=\mu \Big(1- \frac{\alpha_{g}^2}{2n^2} \Big).
\end{equation}
This equation describes leading-order behaviour of the real part of the frequency. For higher-order corrections see Ref. \cite{Baumann:2018vus}. 

Imaginary part (decay width) was analytically calculated in the weak-field limit in Ref. \cite{Detweiler:1980uk} and it implies that the dominant growth mode is $|2 \, 1 \, 1  \rangle$, where the field is described by Eq.~\eqref{eq:211psi}. In this state the density peak is located at $\langle r \rangle=5r_c$.
%
%

Relevant values of axion mass are given between \cite{Baumann:2018vus}
\begin{equation} \label{eq:ax_mass_range}
\frac{\alpha^{\rm (min)}_g}{0.07} \Big(\frac{M}{10M_{\odot}}\Big)^{-1}< \frac{\mu}{10^{-12} \text{eV}} < \frac{\alpha^{\rm (max)}_g}{0.07} \Big(\frac{M}{10M_{\odot}}\Big)^{-1}\,,
\end{equation}
with 
\begin{equation}
\alpha^{\rm (min)}_g=0.006 \Big(\frac{M}{10M_{\odot}}\Big)^{\frac{1}{9}}\,,
\end{equation}
and $\alpha^{\rm (max)}_g$ depending on $\Tilde{a}$. For example, $\alpha^{\rm (max)}_g=0.42$ for $\Tilde{a}=0.7$ and $\alpha^{\rm (max)}_g=0.19$ for $\Tilde{a}=0.8$ (see Ref.~\cite{Dolan:2007mj}). Physically, the lower limit arises from the condition that the significant growth of the cloud occurs during the age of the Universe, while the upper limit is numerically estimated from the growth rate function. For primordial BHs \cite{Rosa:2017ury}, $M/M_{\odot} \in (10^{-10},10^{-4})$ we find $\mu/(10^{-12}\text{eV}) \in (10^3,10^{12})$, while for stellar ($M/M_{\odot} \in (10^{0},10^{2})$) and supermassive ($M/M_{\odot} \in (10^{6},10^{10})$) BHs we find $\mu/(10^{-12}\text{eV}) \in  (10^{-2},10^{2})$ and $\mu/(10^{-12}\text{eV}) \in (10^{-9},10^{-5})$, respectively. Primordial BH-axion mixed dark matter scenario was considered in Ref. \cite{Rosa:2017ury}.

\section{Formulation as Cauchy problem}\label{app.Formulation as Cauchy problem}

In this Appendix, we summarize the Cauchy problem for our system, which we use in Section \ref{sec:Time_domain_studies}.

\subsection{3+1 decomposition}

The equations of motion of this system are given by Eq.~\ref{eq:MFEoMgen}.
In this work, we ignore the dynamics of gravity, and solve the Klein-Goldon equation Eq.\ref{eq:MFEoMScalar}, and Maxwell's equations coupled to a scalar field according to Eq.\ref{eq:MFEoMVector}.
In order to calculate the time evolution of this system, we apply the 3+1 decomposition to these equations with Lorenz gauge.
In this decomposition, the metric function is written as
\begin{eqnarray}
ds^{2}=-\alpha^{2}dt^{2}+\gamma_{ij}(dx^{i}+\beta^{i}dt)(dx^{j}+\beta^{j}dt),
\end{eqnarray}
where $\alpha$ is a lapse function, $\beta^{i}$ is a shift vector, and $\gamma_{ij}$ is a spatial metric.

The vector potential $A_{\mu}$ is written as
\begin{eqnarray}
A_{\mu}=\mathcal{A}_{\mu}+n_{\mu}A_{\phi}\,,
\end{eqnarray}
where $A_{\phi}=-n^{\mu}A_{\mu}$, and $\mathcal{A}_{i}=\gamma_{i}^{\mu}A_{\mu}$.
By substituting these expressions in the field equations, we get the following evolution equations for this system:
\begin{widetext}
\begin{eqnarray}
\partial_{t}\Psi&=&-\alpha\Pi+\mathcal{L}_{\beta}\Psi\,,\\
\partial_{t}\Pi&=&\alpha\left(
-D^{2}\Psi+\mu_{\rm s}^{2}\Psi+K\Pi
-2k_{\rm a}E^{i}B_{i}
+\frac{1}{2}pk_{\rm s}^{p}\Psi^{p-1}(\vec{B}^{2}-\vec{E}^{2})
\right)-D^{i}\alpha D_{i}\Psi+\mathcal{L}_{\beta}\Pi\,,\\
\partial_{t}\mathcal{A}_{i}&=&-\alpha(E_{i}+D_{i}\mathcal{A}_{\phi})-A_{\phi}D_{i}\alpha+\mathcal{L}_{\beta}\mathcal{A}_{i}\,,\\
\partial_{t}E^{i}&=
&\mathcal{L}_{\beta}E^{i}
+\alpha KE^{i}
-\alpha D_{j}\left(D^{j}\mathcal{A}^{i}-D^{i}\mathcal{A}^{j}\right)
-(D^{i}\mathcal{A}^{j}-D^{j}\mathcal{A}^{i})D_{j}\alpha\nonumber\\
&&+\alpha\left(
\frac{k_{\rm s}^{p}p\Psi^{p-1}}{1+k_{\rm s}^{p}\Psi^{p}}\left(\Pi E^{i}-\epsilon^{ijk}D_{j}\Psi B_{k}\right)
+\frac{2k_{\rm a}}{1+k_{\rm s}^{p}\Psi^{p}}\left(B^{i}\Pi +\epsilon^{ijk}E_{k}D_{j}\Psi
\right)
\right)\nonumber\\
&&+\alpha\left(
\frac{1}{1+k_{\rm s}^{p}\Psi^{p}}D^{i}Z+\frac{\mu^{2}_{\rm V}}{1+k_{\rm s}^{p}\Psi^{p}}\mathcal{A}^{i}
\right)\,,\\
\partial_{t}Z&=&\alpha\left(
D_{i}\left((1+k_{\rm s}^{p}\Psi^{p})E^{i}
\right)
+\mu_{\rm V}^{2}A_{\phi}
+2k_{\rm a}B^{i}D_{i}\Psi
\right)
-\kappa\alpha Z
+\mathcal{L}{_\beta}Z\,,
\end{eqnarray}
\end{widetext}
where $\Pi$ is the momentum conjugate of the scalar field, $E_{i}$ and $B_{i}$ are the electric field and magnetic field, defined as
\begin{eqnarray} \label{eq:DefEB}
\displaystyle E^{i}&=&\gamma^{i}_{~\mu}F^{\mu\nu}n_{\nu}\nonumber ,\\
\displaystyle B^{i}&=&\gamma^{i}_{~\mu}~^{\ast}F^{\mu\nu}n_{\nu}=-\epsilon^{ikl}D_{k}\mathcal{A}_{l}\,;
\end{eqnarray}
$A_{i}$ is a spatial component of vector field; $D_{i}$ is a covariant derivative with respect to $\gamma_{ij}\,$; $K$ is the trace of extrinsic curvature; $Z$ is a auxiliary field, which is introduced to stabilize the constraint damping mode; $\epsilon^{ijk}=-\frac{1}{\sqrt{\gamma}}E^{ijk}$ and $E^{ijk}$ is the totally anti-symmetric Levi-Civita symbol with $E^{xyz}=1$. The above equations of motion include both the axion and the scalar coupling (note that they are implemented mutually exclusively) as well as a massive photon (set at $\mu_{\rm V}=0$, except in Section \ref{sec:plasma}).
We also get the following constraint equations,
%
\begin{eqnarray}
D_{i}\left((1+k_{\rm s}\Phi^{p})E^{i}
\right)
+\mu_{\rm V}^{2}A_{\phi}
+2k_{\rm a}B^{i}D_{i}\Phi=0\,.\label{const_eq}
\end{eqnarray}

Our numerical time-evolution code is based on this formalism, expressed in Kerr-Schild type coordinates discussed below, 

\subsection{Background spacetime}
For the background spacetime, we use Kerr-Schild coordinates (see Appendix \ref{app.Kerr-Schild coordinates}).
But, in order to avoid the coordinate singularity, we use Cartesian type coordinates, which are defined by the following coordinate transformations:
\begin{eqnarray}
x&=&r\cos\varphi_{\rm KS}\sin\theta-a\sin\varphi_{\rm KS}\sin\theta\,,\\
y&=&r\sin\varphi_{\rm KS}\sin\theta+a\cos\varphi_{\rm KS}\sin\theta\,,\\
z&=&r\cos\theta\,.
\end{eqnarray}
In these coordinates, the metric can be written as
\begin{eqnarray}
ds^{2}&=&(\eta_{\mu\nu}+2Hl_{\mu}l_{\nu})dx^{\mu}dx^{\nu},
\end{eqnarray}
where $H$ and $l_{\mu}$ are defined as
\begin{eqnarray}
H&=&\frac{r^{3}M}{r^{4}+a^{2}z^{2}}\,,\\
l_{\mu}&=&\left(1,\frac{rx+ay}{r^{2}+a^{2}},\frac{-ax+ry}{r^{2}+a^{2}},\frac{z}{r}\right)\,.
\end{eqnarray}
One can then obtain the lapse function, the shift vector, spatial metric and the extrinsic curvature:
\begin{eqnarray}
\displaystyle \alpha&=&
\displaystyle \frac{1}{\sqrt{1+2H}}\,,\\
\displaystyle \beta_{i}&=&
\displaystyle 2Hl_{i}\,,\\
\displaystyle \gamma_{ij}&=&
\displaystyle \delta_{ij}+2Hl_{i}l_{j}\,,\\
\displaystyle K_{ij}&=&
\displaystyle \frac{\partial_{i}(Hl_{j})+\partial_{j}(Hl_{i})+2Hl^{k}_{\ast}\partial_{k}(Hl_{i}l_{j})}{\sqrt{1+2H}}\,.
\end{eqnarray}

\subsection{Initial Data}
\label{ssec:InitialData}
%
In order to construct the initial data, one must solve the constraint equations~\eqref{const_eq}.
For the scalar field, we use following a simple axion cloud profile as initial data:
\be
\label{Eq.axion cloud initial data}
\Psi(t,r,\theta,\phi)=A_{0}rM\mu^{2}e^{-rM\mu^{2}/2}\cos(\phi -\omega_{\rm R}t)\sin\theta\,,
\ee
where $A_{0}$ is the amplitude of the cloud, and $\omega_{R}\simeq \mu$ is a bounded-state frequency.

For axionic couplings, the constraint equation is given by
\begin{eqnarray}
D_{i}E^{i}+2k_{\rm a} B_{i}D^{i}\Psi&=&0\,.
\end{eqnarray}
The initial data that we use is given by
\begin{eqnarray}
E^{r}&=&E^{\theta}=\mathcal{A}_{i}=0,\\
E^{\varphi}&=&E_{0}(r,\theta),
\end{eqnarray}
where $E_{0}(r,\theta)$ is an arbitrary function of $r$ and $\theta$.
One can show that this profile satisfies the constraint equations.
For $E_{0}(r,\theta)$ we use a Gaussian profile\footnote{In Ref.~\cite{Ikeda:2018nhb} it was shown that the results between the ``localized'' and the ``extended'' profile (to be defined soon) do not change at a qualitative level, for axion couplings}.
\begin{eqnarray}
E_{0}(r,\theta)=E_{0}e^{-\left(\frac{r-r_{0}}{w}\right)^{2}},
\end{eqnarray}
where $E_{0}$, $r_{0}$, and $w$ are the amplitude, the peak radius and the width of the initial electric field.

For scalar couplings, the constraint equation is the following,
\be
D_{i}\left((1+k_{\rm s}^{p}\Psi^{p})E^{i}\right)=0\,.
\ee
As the initial profile, we use the following solution of the constraint equations,
\begin{eqnarray}\label{Eq.Initial data of Er}
E^{r}&=&E^{\theta}=\mathcal{A}_{i}=0\,,\label{Eq.Initial data of Er_v2}\\
E^{\varphi}&=&\frac{F(r,\theta)}{1+k_{\rm s}^{p}\Psi^{p}}\,,\label{Eq.Initial data of Etheta}
\end{eqnarray}
where $F(r,\theta)$ is an arbitrary function of $r$ and $\theta$.
In this work, we use
\begin{eqnarray}\label{Eq.F profile}
F(r,\theta)&=&E_{0}e^{-\left(\frac{r-r_{0}}{w}\right)^{2}}\Theta(\theta)\,,
\end{eqnarray}
where $E_{0}$, $r_{0}$ and $w$ are constants, which characterize the strength, the radius, and width of the Gaussian profile of the electric field.
$\Theta$ characterizes the $\theta$-dependence of the profile.
In our study, we use two different profiles for $F(r,\theta)$. The first is a simple constant value, 
\be
\Theta(\theta)=1\,,\label{ID_extended}
\ee
which we call ``extended profile'' since it is direction-independent.
The second profile we use is
\begin{equation}\label{ID_localized}
\Theta(\theta)=
\left\{\begin{array}{cc}
\sin^{4} 4\theta&~{\rm for}~0<\theta<\frac{\pi}{4}\\
0&~{\rm for}~\frac{\pi}{4}<\theta<\pi.
\end{array}\right.
\end{equation}
We term this a ``localized profile'' since it is sharply peaked along some directions only.

\subsection{Analysis tools} \label{app:TDS-notation}
To gain information about the time development,
the physical quantities extracted from the numerical simulation are the multipolar components of the physical quantities $\Psi_{i}$, $FF_{i}=(F_{\mu\nu}F^{\mu\nu})_{i}$, and $(T^{\rm EM}_{tr})_{i}$ with
\begin{eqnarray}
Z_{0}(t,r)&:=&\int d\Omega Z(t,r,\theta,\phi) Y_{00}(\theta,\phi)\,,\\
Z_{1}(t,r)&:=&\int d\Omega Z(t,r,\theta,\phi) Y_{R}(\theta,\phi)\,,
\end{eqnarray}
where $Y_{R}=\frac{1}{2}\left(Y_{1,1}+Y_{1,-1}\right)$.
\subsection{Numerical code} 
%
\begin{figure}[htb]
\begin{tabular}{cc}
\includegraphics[width=8.5cm]{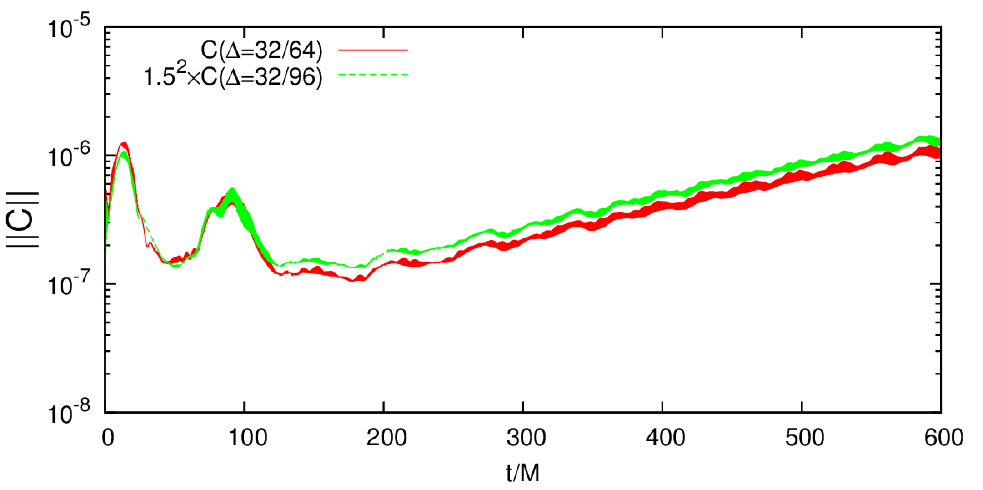}
\end{tabular}
\caption{
Time evolution of the norm of constraint. This shows second-order convergence, because of the integration scheme used to compute the norm.
\label{graph_convergence}
}
\end{figure}
We developed a numerical code which solves the time evolution problem for the scalar and EM fields under the above formulation.
Our numerical code is written in C++; we adopt a 4th order time-integration Runge-Kutta method, and 4th order discretization for the spatial direction. To calculate long term simulation, we use fixed mesh refinement.
The grid structure is layered around BHs. Furthermore, to avoid the physical singularity inside the horizon, the metric in the horizon is replaced with a smooth regular function. The numerical domain of our simulation is $-600M\leq x\leq 600M$, $-600M\leq y\leq 600M$, and $-600M\leq z\leq 600M$. Refinement level is typically 4, and the ratio of resolution between adjacent refinement level is 2.
To avoid high frequency modes which comes from boundary between adjacent refinement region, a Kreiss-Oliger artificial dissipation is added.

As a test simulation, we calculate the time evolution of extended initial data whose $(r_{0},w,E_{0})=(40M,5M,0.001)$, with $p=1$ scalar coupling, and $a=0.5M$. The time evolution of the norm of the constraint is depicted in Fig.\ref{graph_convergence}.
This evolution shows 2nd-order convergence, simply because we used a second-order-accurate integration scheme to compute the norm.

\bibliographystyle{h-physrev4}
\bibliography{axion_refs}

\begin{thebibliography}{10}

\bibitem{Peccei:1977hh}
R.~Peccei and H.~R. Quinn,
\newblock Phys.Rev.Lett. {\bf 38}, 1440 (1977).

\bibitem{Arvanitaki:2009fg}
A.~Arvanitaki, S.~Dimopoulos, S.~Dubovsky, N.~Kaloper and J.~March-Russell,
\newblock Phys.Rev. {\bf D81}, 123530 (2010), [0905.4720].

\bibitem{Bergstrom:2009ib}
L.~Bergstrom,
\newblock New J.Phys. {\bf 11}, 105006 (2009), [0903.4849].

\bibitem{Fairbairn:2014zta}
M.~Fairbairn, R.~Hogan and D.~J. Marsh,
\newblock Phys.Rev. {\bf D91}, 023509 (2015), [1410.1752].

\bibitem{Marsh:2015xka}
D.~J.~E. Marsh,
\newblock Phys. Rept. {\bf 643}, 1 (2016), [1510.07633].

\bibitem{Hui:2016ltb}
L.~Hui, J.~P. Ostriker, S.~Tremaine and E.~Witten,
\newblock Phys. Rev. {\bf D95}, 043541 (2017), [1610.08297].

\bibitem{Mirizzi:2006zy}
A.~Mirizzi, G.~G. Raffelt and P.~D. Serpico,
\newblock Lect.Notes Phys. {\bf 741}, 115 (2008), [astro-ph/0607415].

\bibitem{Hooper:2007bq}
D.~Hooper and P.~D. Serpico,
\newblock Phys.Rev.Lett. {\bf 99}, 231102 (2007), [0706.3203].

\bibitem{Detweiler:1980uk}
S.~L. Detweiler,
\newblock Phys.Rev. {\bf D22}, 2323 (1980).

\bibitem{Cardoso:2005vk}
V.~Cardoso and S.~Yoshida,
\newblock JHEP {\bf 0507}, 009 (2005), [hep-th/0502206].

\bibitem{Dolan:2007mj}
S.~R. Dolan,
\newblock Phys.Rev. {\bf D76}, 084001 (2007), [0705.2880].

\bibitem{East:2017ovw}
W.~E. East and F.~Pretorius,
\newblock Phys. Rev. Lett. {\bf 119}, 041101 (2017), [1704.04791].

\bibitem{East:2018glu}
W.~E. East,
\newblock Phys. Rev. Lett. {\bf 121}, 131104 (2018), [1807.00043].

\bibitem{Brito:2015oca}
R.~Brito, V.~Cardoso and P.~Pani,
\newblock Lect. Notes Phys. {\bf 906} (2015), [1501.06570].

\bibitem{Arvanitaki:2010sy}
A.~Arvanitaki and S.~Dubovsky,
\newblock Phys.Rev. {\bf D83}, 044026 (2011), [1004.3558].

\bibitem{Arvanitaki:2014wva}
A.~Arvanitaki, M.~Baryakhtar and X.~Huang,
\newblock Phys. Rev. {\bf D91}, 084011 (2015), [1411.2263].

\bibitem{Brito:2014wla}
R.~Brito, V.~Cardoso and P.~Pani,
\newblock Class. Quant. Grav. {\bf 32}, 134001 (2015), [1411.0686].

\bibitem{Arvanitaki:2016qwi}
A.~Arvanitaki, M.~Baryakhtar, S.~Dimopoulos, S.~Dubovsky and R.~Lasenby,
\newblock Phys. Rev. {\bf D95}, 043001 (2017), [1604.03958].

\bibitem{Baryakhtar:2017ngi}
M.~Baryakhtar, R.~Lasenby and M.~Teo,
\newblock Phys. Rev. {\bf D96}, 035019 (2017), [1704.05081].

\bibitem{Brito:2017wnc}
R.~Brito {\em et~al.},
\newblock Phys. Rev. Lett. {\bf 119}, 131101 (2017), [1706.05097].

\bibitem{Brito:2017zvb}
R.~Brito {\em et~al.},
\newblock Phys. Rev. {\bf D96}, 064050 (2017), [1706.06311].

\bibitem{Hannuksela:2018izj}
O.~A. Hannuksela, R.~Brito, E.~Berti and T.~G.~F. Li,
\newblock 1804.09659.

\bibitem{Rosa:2017ury}
J.~G. Rosa and T.~W. Kephart,
\newblock Phys. Rev. Lett. {\bf 120}, 231102 (2018), [1709.06581].

\bibitem{Sen:2018cjt}
S.~Sen,
\newblock Phys. Rev. {\bf D98}, 103012 (2018), [1805.06471].

\bibitem{Ikeda:2018nhb}
T.~Ikeda, R.~Brito and V.~Cardoso,
\newblock 1811.04950.

\bibitem{Garbrecht:2018akc}
B.~Garbrecht and J.~I. McDonald,
\newblock JCAP {\bf 1807}, 044 (2018), [1804.04224].

\bibitem{Plascencia:2017kca}
A.~D. Plascencia and A.~Urbano,
\newblock JCAP {\bf 1804}, 059 (2018), [1711.08298].

\bibitem{Klasen:2015uma}
M.~Klasen, M.~Pohl and G.~Sigl,
\newblock Prog. Part. Nucl. Phys. {\bf 85}, 1 (2015), [1507.03800].

\bibitem{Stadnik:2015kpa}
Y.~V. Stadnik and V.~V. Flambaum,
\newblock Mod. Phys. Lett. {\bf A32}, 1740004 (2017), [1506.08364].

\bibitem{Olive:2007aj}
K.~A. Olive and M.~Pospelov,
\newblock Phys. Rev. {\bf D77}, 043524 (2008), [0709.3825].

\bibitem{Ringwald:2016yge}
A.~Ringwald,
\newblock PoS {\bf NOW2016}, 081 (2016), [1612.08933].

\bibitem{Cardoso:2018tly}
V.~Cardoso {\em et~al.},
\newblock JCAP {\bf 1803}, 043 (2018), [1801.01420].

\bibitem{ReesAGN}
M.~J. Rees,
\newblock Annual Review of Astronomy and Astrophysics {\bf 22}, 471 (1984).

\bibitem{Ooguri:2011aa}
H.~Ooguri and M.~Oshikawa,
\newblock Phys. Rev. Lett. {\bf 108}, 161803 (2012), [1112.1414].

\bibitem{Khmelnitsky:2013lxt}
A.~Khmelnitsky and V.~Rubakov,
\newblock JCAP {\bf 1402}, 019 (2014), [1309.5888].

\bibitem{Boskovic:2018rub}
M.~Bo\v{s}kovi\'{c}, F.~Duque, M.~C. Ferreira, F.~S. Miguel and V.~Cardoso,
\newblock Phys. Rev. {\bf D98}, 024037 (2018), [1806.07331].

\bibitem{benderbook}
C.~M. Bender and S.~A. Orszag,
\newblock {\em Advanced Mathematical Methods for Scientists and Engineers}
  (McGraw-Hill, New York, 1978).

\bibitem{arnold_ODE_book}
V.~I. Arnold,
\newblock {\em Ordinary Differential Equations} (Springer-Verlag, Berlin,
  1992).

\bibitem{Tkachev:2014dpa}
I.~I. Tkachev,
\newblock JETP Lett. {\bf 101}, 1 (2015), [1411.3900],
\newblock [Pisma Zh. Eksp. Teor. Fiz.101,no.1,3(2015)].

\bibitem{Hertzberg:2018zte}
M.~P. Hertzberg and E.~D. Schiappacasse,
\newblock JCAP {\bf 1811}, 004 (2018), [1805.00430].

\bibitem{Chandra}
S.~Chandrasekhar,
\newblock {\em {The mathematical theory of black holes}} (, 1983).

\bibitem{Donos:2011bh}
A.~Donos and J.~P. Gauntlett,
\newblock JHEP {\bf 08}, 140 (2011), [1106.2004].

\bibitem{Herdeiro:2018wub}
C.~A.~R. Herdeiro, E.~Radu, N.~Sanchis-Gual and J.~A. Font,
\newblock Phys. Rev. Lett. {\bf 121}, 101102 (2018), [1806.05190].

\bibitem{Myung:2018vug}
Y.~S. Myung and D.-C. Zou,
\newblock 1808.02609.

\bibitem{Rozali:2012es}
M.~Rozali, D.~Smyth, E.~Sorkin and J.~B. Stang,
\newblock Phys. Rev. Lett. {\bf 110}, 201603 (2013), [1211.5600].

\bibitem{Donos:2013wia}
A.~Donos,
\newblock JHEP {\bf 05}, 059 (2013), [1303.7211].

\bibitem{Withers:2013loa}
B.~Withers,
\newblock Class. Quant. Grav. {\bf 30}, 155025 (2013), [1304.0129].

\bibitem{Wald:1974np}
R.~M. Wald,
\newblock Phys.Rev. {\bf D10}, 1680 (1974).

\bibitem{Zajacek:2018ycb}
M.~Zaja\v{c}ek, A.~Tursunov, A.~Eckart and S.~Britzen,
\newblock Mon. Not. Roy. Astron. Soc. {\bf 480}, 4408 (2018), [1808.07327].

\bibitem{Baumann:2018vus}
D.~Baumann, H.~S. Chia and R.~A. Porto,
\newblock 1804.03208.

\bibitem{Gubser:2005ih}
S.~S. Gubser,
\newblock Class. Quant. Grav. {\bf 22}, 5121 (2005), [hep-th/0505189].

\bibitem{Garfinkle:1990qj}
D.~Garfinkle, G.~T. Horowitz and A.~Strominger,
\newblock Phys. Rev. {\bf D43}, 3140 (1991),
\newblock [Erratum: Phys. Rev.D45,3888(1992)].

\bibitem{Okawa:2014nda}
H.~Okawa, H.~Witek and V.~Cardoso,
\newblock Phys. Rev. {\bf D89}, 104032 (2014), [1401.1548].

\bibitem{Zilhao:2015tya}
M.~Zilh\~{a}o, H.~Witek and V.~Cardoso,
\newblock Class. Quant. Grav. {\bf 32}, 234003 (2015), [1505.00797].

\bibitem{zeldovich1}
Y.~B. Zel'dovich,
\newblock Pis'ma Zh. Eksp. Teor. Fiz. {\bf 14}, 270 [JETP Lett. {\bf14}, 180
  (1971)] (1971).

\bibitem{zeldovich2}
Y.~B. Zel'dovich,
\newblock Zh. Eksp. Teor. Fiz {\bf 62}, 2076 [Sov.Phys. JETP {\bf 35}, 1085
  (1972)] (1972).

\bibitem{Witek:2012tr}
H.~Witek, V.~Cardoso, A.~Ishibashi and U.~Sperhake,
\newblock Phys. Rev. {\bf D87}, 043513 (2013), [1212.0551].

\bibitem{Yoshino:2012kn}
H.~Yoshino and H.~Kodama,
\newblock Prog. Theor. Phys. {\bf 128}, 153 (2012), [1203.5070].

\bibitem{Yoshino:2015nsa}
H.~Yoshino and H.~Kodama,
\newblock Class. Quant. Grav. {\bf 32}, 214001 (2015), [1505.00714].

\bibitem{Hertzberg:2010yz}
M.~P. Hertzberg,
\newblock Phys. Rev. {\bf D82}, 045022 (2010), [1003.3459].

\bibitem{Carlson:1994yqa}
E.~D. Carlson and W.~D. Garretson,
\newblock Phys. Lett. {\bf B336}, 431 (1994).

\bibitem{JacksonED}
J.~D. Jackson,
\newblock {\em {Classical Electrodynamics}} (New York: John Wiley \& Sons,
  1999).

\bibitem{Ferreira:2017pth}
M.~C. Ferreira, C.~F.~B. Macedo and V.~Cardoso,
\newblock Phys. Rev. {\bf D96}, 083017 (2017), [1710.00830].

\bibitem{Cardoso:2016ryw}
V.~Cardoso and L.~Gualtieri,
\newblock Class. Quant. Grav. {\bf 33}, 174001 (2016), [1607.03133].

\bibitem{Pani:2012vp}
P.~Pani, V.~Cardoso, L.~Gualtieri, E.~Berti and A.~Ishibashi,
\newblock Phys.Rev.Lett. {\bf 109}, 131102 (2012), [1209.0465].

\bibitem{Brito:2013wya}
R.~Brito, V.~Cardoso and P.~Pani,
\newblock Phys. Rev. {\bf D88}, 023514 (2013), [1304.6725].

\bibitem{Hartono:2004}
Hartono and A.~van~der Burgh,
\newblock J. Eng. Math. {\bf 49}, 99 (2004).

\bibitem{Chen:1995ena}
L.-Y. Chen, N.~Goldenfeld and Y.~Oono,
\newblock Phys. Rev. {\bf E54}, 376 (1996), [hep-th/9506161].

\bibitem{Galley:2016zee}
C.~R. Galley and I.~Z. Rothstein,
\newblock Phys. Rev. {\bf D95}, 104054 (2017), [1609.08268].

\bibitem{Berti:2009kk}
E.~Berti, V.~Cardoso and A.~O. Starinets,
\newblock Class.Quant.Grav. {\bf 26}, 163001 (2009), [0905.2975].

\bibitem{Rosa:2011my}
J.~G. Rosa and S.~R. Dolan,
\newblock Phys.Rev. {\bf D85}, 044043 (2012), [1110.4494].

\bibitem{Pani:2012bp}
P.~Pani, V.~Cardoso, L.~Gualtieri, E.~Berti and A.~Ishibashi,
\newblock Phys.Rev. {\bf D86}, 104017 (2012), [1209.0773].

\bibitem{Macedo:2018txb}
C.~F.~B. Macedo,
\newblock 1809.08691.

\bibitem{Teukolsky:2014vca}
S.~A. Teukolsky,
\newblock Class. Quant. Grav. {\bf 32}, 124006 (2015), [1410.2130].

\bibitem{Alcubierre:2008}
M.~Alcubierre,
\newblock {\em {Introduction to 3+1 numerical relativity }}International series
  of monographs on physics (Oxford Univ. Press, Oxford, 2008).

\end{thebibliography}

\end{document}